\newif{\ifchangetext}
\changetexttrue

\documentclass[preprint]{aastex}

\newcommand{\cmt}[1]{}
\ifchangetext

\else

\fi

\usepackage{natbib}   
\usepackage{amsmath}  
\usepackage{verbatim} 
\usepackage{enumerate}
\usepackage{amssymb}  
\usepackage{hyperref}      
\usepackage[usenames,dvipsnames]{xcolor}  
\usepackage{appendix}

\DeclareGraphicsExtensions{.pdf,.png,.jpg}

\defcitealias{Rodney:2014}{R14}
\def\R14{\citetalias{Rodney:2014}}

\defcitealias{Graur:2014}{G14}
\def\G14{\citetalias{Graur:2014}}

\newcommand{\sy}{\scriptstyle}

\def\arcsec{\ensuremath{^{\prime\prime}}}

\newcommand{\bq}{\begin{equation}} 
\newcommand{\eq}{\end{equation}}

\newcommand{\holmcnu}{$  72.04  \pm  2.67 $}
\newcommand{\holmcnosys}{$  72.04  \pm  2.56  $ km s$^{-1}$ Mpc$^{-1}$}

\newcommand{\hofnu}{$  72.25  \pm  2.51 $}
\newcommand{\hofnosys}{$  72.25  \pm  2.38  $ km s$^{-1}$ Mpc$^{-1}$}

\newcommand{\uncfnosys}{$  3.3 \% $}

\newcommand{\homwnu}{$  76.18  \pm  2.37 $}

\newcommand{\homwnosys}{$  76.18  \pm  2.17  $ km s$^{-1}$ Mpc$^{-1}$}
\newcommand{\hoand}{$  74.50  \pm  3.27  $ km s$^{-1}$ Mpc$^{-1}$}
\newcommand{\hoandnu}{$  74.50  \pm  3.27 $}
\newcommand{\hoandnosys}{$  74.50  \pm  2.87  $ km s$^{-1}$ Mpc$^{-1}$}

\newcommand{\homwfnu}{$  74.04  \pm  1.93 $}
\newcommand{\homwfnosys}{$  74.04  \pm  1.74  $ km s$^{-1}$ Mpc$^{-1}$}
\newcommand{\uncmwfnosys}{$  2.4 \% $}
\newcommand{\hoallthree}{$  73.24  \pm  1.74  $ km s$^{-1}$ Mpc$^{-1}$}

\newcommand{\hoallthreebfnu}{$ {\bf  73.24  \pm  1.74 }$}
\newcommand{\hoallthreenosys}{$  73.24  \pm  1.59  $ km s$^{-1}$ Mpc$^{-1}$}

\newcommand{\sysuncallthreenop}{$  0.71 $}
\newcommand{\uncallthree}{$  2.4 \% $}
\newcommand{\uncallthreenosys}{$  2.2 \% $}
\newcommand{\Planckdiff}{$  3.2  $}
\newcommand{\PlanckdiffTTTEEE}{$  3.4  $}
\newcommand{\WMAPdiff}{$  1.0  $}
\newcommand{\WMAPBAOdiff}{$  2.1  $}

\newcommand{\hoallfournu}{$  73.46  \pm  1.71 $}
\newcommand{\hoallfournosys}{$  73.46  \pm  1.53  $ km s$^{-1}$ Mpc$^{-1}$}

\newcommand{\hoallthreewvi}{$  71.56  \pm  2.49  $ km s$^{-1}$ Mpc$^{-1}$}
\newcommand{\hoallthreewvinu}{$  71.56  \pm  2.49 $}
\newcommand{\hoallthreenosyswvi}{$  71.56  \pm  1.52  $ km s$^{-1}$ Mpc$^{-1}$}
\newcommand{\sysuncallthreewvi}{$  2.8 \% $}
\newcommand{\uncallthreewvi}{$  3.5 \% $}

\newcommand{\hoallfwvi}{$  72.04  \pm  2.83  $ km s$^{-1}$ Mpc$^{-1}$}
\newcommand{\hoallfnosyswvi}{$  72.04  \pm  2.23  $ km s$^{-1}$ Mpc$^{-1}$}
\newcommand{\sysuncallfwvi}{$  2.4 \% $}

\newcommand{\uncallfnosyswvi}{$  3.1 \% $}
\newcommand{\hotrgbnosys}{$  71.62  \pm  1.68  $ km s$^{-1}$ Mpc$^{-1}$}
\newcommand{\hotrgbnu}{$  71.62  \pm  1.78 $}

\newcommand{\beq}{\begin{equation}}
\newcommand{\eeq}{\end{equation}}
\newcommand{\beqa}{\begin{eqnarray}}
\newcommand{\eeqa}{\end{eqnarray}}

\newcommand{\PL}{$P$--$L$}

\newcommand{\nd}{\multicolumn{1}{c}{$\dots$}}
\newcommand{\ndr}{\multicolumn{1}{r}{$\dots$}}
\newcommand{\td}{..\!\!}
\newcommand{\oh}{{\rm [O/H]}}
\newcommand{\lp}{\log P}

\newcommand{\hunit}{km\,s$^{-1}$\,Mpc$^{-1} \, $}

\usepackage{lscape}

\shorttitle{H$_0$ from a Refurbished Distance Ladder}
\shortauthors{Riess et al.}

\begin{document} 

\title{A 2.4\% Determination of the Local Value of the Hubble Constant\altaffilmark{1}}

\author{Adam G.~Riess\altaffilmark{2,3}, Lucas M.~Macri\altaffilmark{4}, Samantha L.~Hoffmann\altaffilmark{4}, Dan Scolnic\altaffilmark{2,5}, Stefano Casertano\altaffilmark{3}, Alexei V.~Filippenko\altaffilmark{6}, Brad E.~Tucker\altaffilmark{6,7}, Mark J.~Reid\altaffilmark{8}, David O.~Jones\altaffilmark{2}, Jeffrey M.~Silverman\altaffilmark{9}, Ryan Chornock\altaffilmark{10}, Peter Challis\altaffilmark{8}, Wenlong Yuan\altaffilmark{4}, Peter J. Brown\altaffilmark{4}, and Ryan J.~Foley\altaffilmark{11,12}}

\altaffiltext{1}{Based on observations with the NASA/ESA {\it Hubble Space Telescope}, obtained at the Space Telescope Science Institute, which is operated by AURA, Inc., under NASA contract NAS 5-26555.}
\altaffiltext{2}{Department of Physics and Astronomy, Johns Hopkins University, Baltimore, MD, USA}
\altaffiltext{3}{Space Telescope Science Institute, Baltimore, MD, USA; ariess@stsci.edu}
\altaffiltext{4}{George P. and Cynthia Woods Mitchell Institute for Fundamental Physics and Astronomy, Department of Physics \& Astronomy, Texas A\&M University, College Station, TX, USA}
\altaffiltext{5}{Kavli Institute for Cosmological Physics, University of Chicago, Chicago, IL, USA}
\altaffiltext{6}{Department of Astronomy, University of California, Berkeley, CA, USA}
\altaffiltext{7}{The Research School of Astronomy and Astrophysics, Australian National University, Mount Stromlo Observatory, Weston Creek, ACT, Australia}
\altaffiltext{8}{Harvard-Smithsonian Center for Astrophysics, Cambridge, MA, USA}
\altaffiltext{9}{Department of Astronomy, University of Texas, Austin, TX, USA}
\altaffiltext{10}{Astrophysical Institute, Department of Physics and Astronomy, Ohio University, Athens, OH, USA}
\altaffiltext{11}{Department of Physics, University of Illinois at Urbana-Champaign, Urbana, IL, USA}
\altaffiltext{12}{Department of Astronomy, University of Illinois at Urbana-Champaign, Urbana, IL, USA}

\begin{abstract} 
We use the Wide Field Camera 3 (WFC3) on the {\it Hubble Space Telescope (HST)} to reduce the uncertainty in the local value of the Hubble constant from 3.3\% to \uncallthree. The bulk of this improvement comes from new, near-infrared observations of Cepheid variables in 11 host galaxies of recent type Ia supernovae (SNe~Ia), more than doubling the sample of reliable SNe~Ia having a Cepheid-calibrated distance to a total of 19; these in turn leverage the magnitude-redshift relation based on $\sim300$ SNe~Ia at $z\!<\!0.15$. All 19 hosts as well as the megamaser system NGC$\,$4258 have been observed with WFC3 in the optical and near-infrared, thus nullifying cross-instrument zeropoint errors in the relative distance estimates from Cepheids. Other noteworthy improvements include a 33\% reduction in the systematic uncertainty in the maser distance to NGC$\,$4258, a larger sample of Cepheids in the Large Magellanic Cloud (LMC), a more robust distance to the LMC based on late-type detached eclipsing binaries (DEBs), {\it HST} observations of Cepheids in M31, and new {\it HST}-based trigonometric parallaxes for Milky Way (MW) Cepheids. 

We consider four geometric distance calibrations of Cepheids: (i) megamasers in NGC$\,$4258, (ii) 8 DEBs in the LMC, (iii) 15 MW Cepheids with parallaxes measured with {\it HST}/FGS, {\it HST}/WFC3 spatial scanning and/or {\it Hipparcos}, and (iv) 2 DEBs in M31. The Hubble constant from each is \hofnu, \holmcnu, \homwnu, and \hoand, respectively.  Our best estimate of H$_0=\,$ {\hoallthree} combines the anchors NGC$\,$4258, MW, and LMC, yielding a {\uncallthree}\ determination (all quoted uncertainties include fully-propagated statistical and systematic components). This value is \PlanckdiffTTTEEE $\sigma$ higher than $66.93 \pm 0.62$ \hunit predicted by $\Lambda$CDM with 3 neutrino flavors having a mass of 0.06~eV and the new {\it Planck} data, but the discrepancy reduces to \WMAPBAOdiff $\sigma$ relative to the prediction of $69.3 \pm 0.7$ \hunit based on the comparably precise combination of {\it WMAP}+ACT+SPT+BAO observations, suggesting that systematic uncertainties in cosmic microwave background radiation measurements may play a role in the tension.

If we take the conflict between {\it Planck} high-redshift measurements and our local determination of H$_0$ at face value, one plausible explanation could involve an additional source of dark radiation in the early Universe in the range of $\Delta$N$_{\rm eff}\!\approx\!0.4\!-\!1$.  We anticipate further significant improvements in H$_0$ from upcoming parallax measurements of long-period MW Cepheids. 
\end{abstract} 


\section{Introduction} 

The Hubble constant (H$_0$) measured locally and the sound horizon observed from the cosmic microwave background radiation (CMB) provide two absolute scales at opposite ends of the visible expansion history of the Universe. Comparing the two gives a stringent test of the standard cosmological model. A significant disagreement would provide evidence for fundamental physics beyond the standard model, such as time-dependent or early dark energy, gravitational physics beyond General Relativity, additional relativistic particles, or nonzero curvature. Indeed, none of these features has been excluded by anything more compelling than a theoretical preference for simplicity over complexity. In the case of dark energy, there is no simple explanation at present, leaving direct measurements as the only guide among numerous complex or highly tuned explanations.

Recent progress in measuring the CMB from {\it WMAP} \citep{Hinshaw:2013,Bennett:2013} and {\it Planck} \citep{Planck:2016} have reduced the uncertainty in the distance to the surface of last scattering ($z \sim 1000$) to below 0.5\% in the context of $\Lambda$CDM, motivating complementary efforts to improve the local determination of H$_0$ to percent-level precision \citep{Suyu:2012,Hu:2005}. Hints of mild tension at the $\sim\!2\!-\!2.5 \sigma$ level with the $3\!-\!5\%$ measurements of H$_0$ stated by \citet{Riess:2011}, \citet{Sorce:2012}, \citet{Freedman:2012}, and \citet{Suyu:2013} have been widely considered and in some cases revisited in great detail \citep{Efstathiou:2014,Dvorkin:2014,Bennett:2014,Spergel:2015,Becker:2015}, with no definitive conclusion except for highlighting the value of improvements in the local observational determination of H$_0$.

\subsection{Past Endeavors}

Considerable progress in the local determination of H$_0$ has been made in the last 25 years, assisted by observations of water masers, strong-lensing systems, SNe, the Cepheid period-luminosity (\PL) relation \citep[also known as the Leavitt law;][]{Leavitt:1912}, and other sources used independently or in concert to construct distance ladders \citep[see][for recent reviews]{Freedman:2010,Livio:2013}.

A leading approach utilizes {\it Hubble Space Telescope (HST)} observations of Cepheids in the hosts of recent, nearby SNe~Ia to link geometric distance measurements to other SNe~Ia in the expanding Universe. The SN~Ia {\it HST} Calibration Program \citep{Sandage:2006} and the {\it HST} Key Project \citep{Freedman:2001} both made use of {\it HST} observations with WFPC2 to resolve Cepheids in SN~Ia hosts. However, the useful range of that camera for measuring Cepheids, $\lesssim\!25$~Mpc, placed severe limits on the number and choice of SNe~Ia which could be used to calibrate their luminosity (e.g., SNe 1937C, 1960F, 1974G). A dominant systematic uncertainty resulted from the unreliability of those nearby SNe~Ia which were {\it photographically} observed, highly reddened, spectroscopically abnormal, or discovered after peak brightness. Only two objects (SNe 1990N and 1981B) used by \cite{Freedman:2001,Freedman:2012} and four by \cite{Sandage:2006} (the above plus SN 1994ae and SN1998aq) were free from these shortcomings, leaving a very small set of reliable calibrators relative to the many hundreds of similarly reliable SNe~Ia observed in the Hubble flow. The resulting ladders were further limited by the need to calibrate WFPC2 at low flux levels to the ground-based systems used to measure Cepheids in a single anchor, the Large Magellanic Cloud (LMC). The use of LMC Cepheids introduces additional systematic uncertainties because of their shorter mean period ($\Delta\langle\log\,P\rangle\approx 0.7$~dex) and lower metallicity \citep[$\Delta\log({\rm O}/{\rm H})=-0.25$~dex,][]{Romaniello:2008} than those found with {\it HST} in the large spiral galaxies that host nearby SNe~Ia. Despite careful work, the estimates of H$_0$ by the two teams (each with 10\% uncertainty) differed by 20\%, owing in part to the aforementioned systematic errors. 

\begin{figure}[ht]
\vspace*{170mm}
\includegraphics{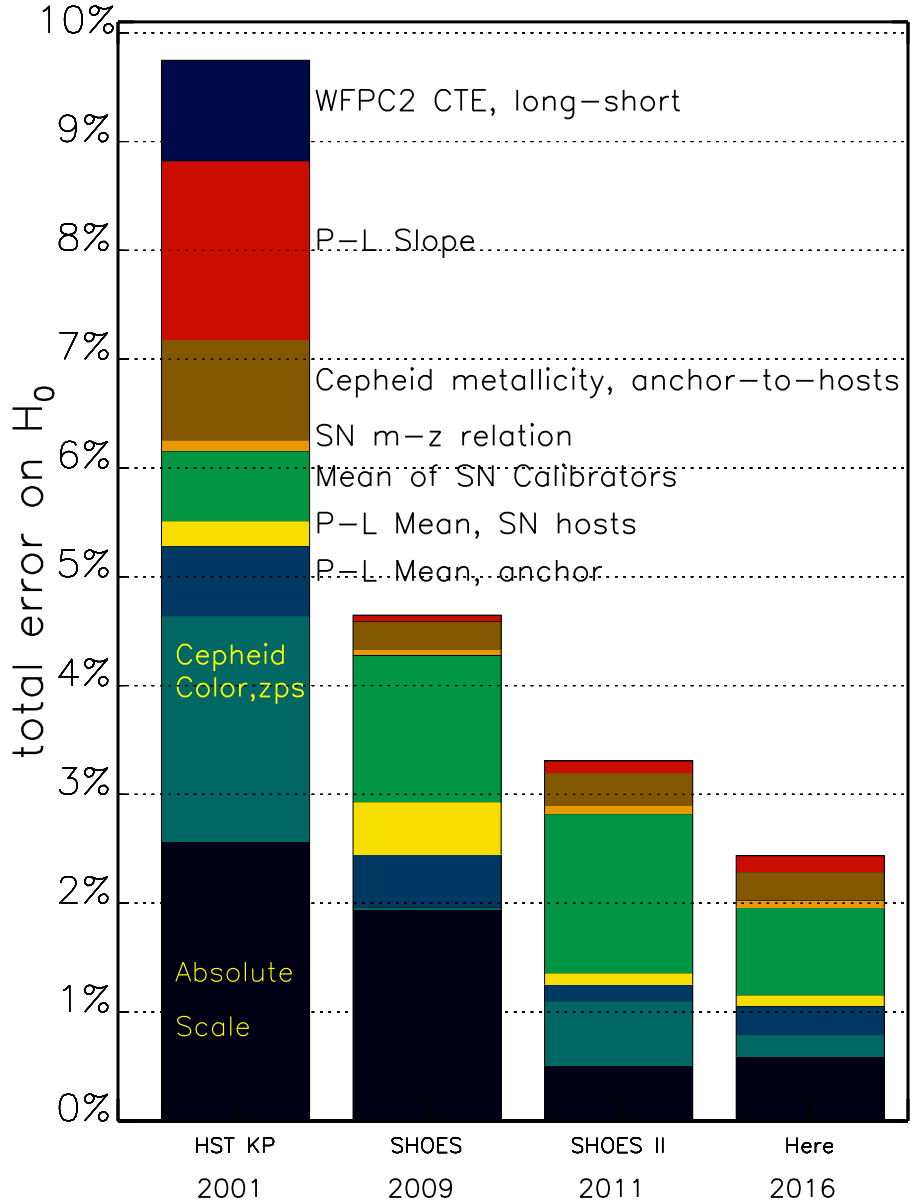}
\caption{\label{fg:errbud} Uncertainties in the determination of H$_0$. Uncertainties are squared to show their individual contribution to the quadrature sum. These terms are given in Table~\ref{tb:h0unc}.}
\end{figure}

More recently, the SH0ES (Supernovae, H$_0$, for the Equation of State of dark energy) team used a number of advancements to refine this approach to determining H$_0$. Upgrades to the instrumentation of {\it HST} doubled its useful range for resolving Cepheids (leading to an eight-fold improvement in volume and in the expected number of useful SN~Ia hosts), first with the Advanced Camera for Surveys \citep[ACS;][]{Riess:2005,Riess:2009a} and later with the Wide Field Camera 3 \citep[WFC3;][hereafter R11]{Riess:2011} owing to the greater area, higher sensitivity, and smaller pixels of these cameras. WFC3 has other superior features for Cepheid reconnaissance, including a white-light filter {(\it F350LP)} that more than doubles the speed for discovering Cepheids and measuring their periods relative to the traditional {\it F555W} filter, and a 5 square arcmin near-infrared (NIR) detector that can be used to reduce the impact of differential extinction and metallicity differences across the Cepheid sample. A precise geometric distance to NGC$\,$4258 measured to 3\% using water masers \citep[][hereafter H13]{Humphreys:2013} has provided a new anchor galaxy whose Cepheids can be observed {\it with the same instrument and filters} as those in SN~Ia hosts to effectively cancel the effect of photometric zeropoint uncertainties in this step along the distance ladder. Tied to the Hubble diagram of 240 SNe Ia \citep[now $>\!300$~SNe~Ia;][]{Scolnic:2015,Scolnic:2016}, the new ladder was used to initially determine H$_0$ with a total uncertainty of 4.7\% \citep[][hereafter R09]{Riess:2009}. R11 subsequently improved this measurement to 3.3\% by increasing to 8 the number of Cepheid distances to reliable SN~Ia hosts, and formally including {\it HST}/FGS trigonometric parallaxes of 10 Milky Way (MW) Cepheids with distance $D\!<\!0.5$~kpc and individual precision of 8\% \citep{Benedict07}. The evolution of the error budget in these measurements is shown in Figure~\ref{fg:errbud}. 

Here we present a broad set of improvements to the SH0ES team distance ladder including new near-infrared {\it HST} observations of Cepheids in 11 SN~Ia hosts (bringing the total to 19), a refined computation of the distance to NGC$\,$4258 from maser data, additional Cepheid parallax measurements, larger Cepheid samples in the anchor galaxies, and additional SNe~Ia to constrain the Hubble flow. We present the new Cepheid data in \S2 and in S.~L.~Hoffmann et al.~(2016, in prep.; hereafter H16). Other improvements are described throughout \S3, and a consideration of analysis variants and systematic uncertainties is given in \S4. We end with a discussion in \S5.

\section{HST Observations of Cepheids in the SH0ES Program}

Discovering and measuring Cepheid variables in SN~Ia host galaxies requires a significant investment of observing time on {\it HST}. It is thus important to select SN Ia hosts likely to produce a set of calibrators that is a good facsimile of the much larger sample defining the modern SN~Ia magnitude-redshift relation at $0.01\!<\!z\!<\!0.15$ \citep[e.g.,][]{Scolnic:2015,Scolnic:2016}. Poor-quality light curves, large reddening, atypical SN explosions, or hosts unlikely to yield a significant number of Cepheids would all limit contributions to this effort. Therefore, the SH0ES program has been selecting SNe~Ia with the following qualities to ensure a reliable calibration of their fiducial luminosity: (1) modern photometric data (i.e., photoelectric or CCD), (2) observed before maximum brightness and well thereafter, (3) low reddening (implying $A_V\!<\!0.5$~mag), (4) spectroscopically typical, and (5) a strong likelihood of being able to detect Cepheids in its host galaxy with {\it HST}. This last quality translates into any late-type host (with features consistent with the morphological classification of Sa to Sd) having an expectation of $D\lesssim40$~Mpc, inclination $<\!75^\circ$, and apparent size $>\!1\arcmin$.  To avoid a possible selection bias in SN~Ia luminosities, the probable distance of the host is estimated via the Tully-Fisher relation or flow-corrected redshifts as reported in NED\footnote{The NASA/IPAC Extragalactic Database (NED) is operated by the Jet Propulsion Laboratory, California Institute of Technology, under contract with the National Aeronautics and Space Administration (NASA).}. We will consider the impact of these selections in \S4. 

The occurrence of SNe~Ia with these characteristics is unfortunately quite rare, leading to a nearly complete sample of 19 objects observed between 1993 and 2015 (see Table~\ref{tb:wfc3}). Excluding supernovae from the 1980s, a period when modern detectors were rare and when suitable SNe~Ia may have appeared and gone unnoticed, the average rate of production is $\sim 1/$year. Regrettably, it will be difficult to increase this sample substantially (by a factor of $\sim 2$) over the remaining lifetime of {\it HST}. We estimate that a modest augmentation of the sample (at best) would occur by removing one or more of the above selection criteria, but the consequent increase in systematic uncertainty would more than offset the statistical gain.

Reliable SNe~Ia from early-type hosts could augment the sample, with distance estimates based on RR Lyrae stars or the tip of the red-giant branch (TRGB) for their calibration. Unfortunately, the reduced distance range of these distance indicators for {\it HST} compared to Cepheids (2.5 mag or $D\!<\!13$ Mpc for TRGB, 5 mag or $D\!<\!4$  Mpc for RR Lyrae stars) and the factor of $\sim 5$ smaller sample of SNe~Ia in early-type hosts limits the sample increase to just a few additional objects \citep[SN 1994D, SN 1980N, 1981D, and SN 2006dd with the latter 3 all in the same host;][]{Beaton:2016}, a modest fraction of the current sample of 19 SNe~Ia calibrated by Cepheids.

\begin{figure}[ht]
\vspace*{150mm}
\includegraphics{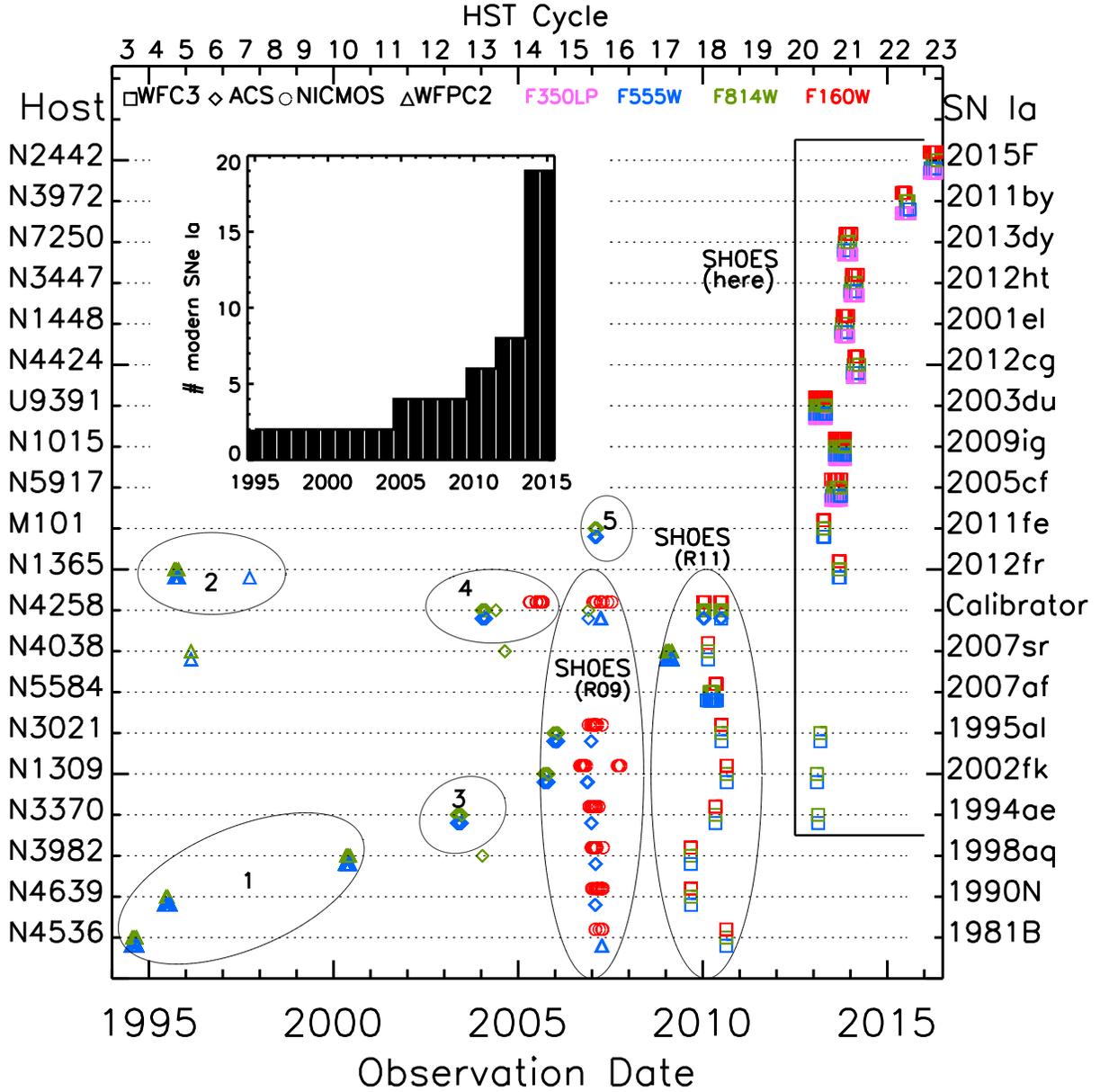}
\caption{\label{fg:hstobs} {\it HST} observations of the host galaxies of ideal SNe~Ia. The data used to observe Cepheids in 19 SN~Ia hosts and NGC$\,$4258 have been collected over 20 years with 4 cameras and over 600 orbits of {\it HST} time.  60-90~day campaigns in {\it F555W} and {\it F814W} or in {\it F350LP} were used to identify Cepheids from their light curves with occasional reobservations years later to identify Cepheids with $P>60$~d. Near-IR follow-up observations in {\it F160W} are used to reduce the effects of host-galaxy extinction, sensitivity to metallicity, and breaks in the \PL\ relation. Data sources: (1) {\it HST} SN~Ia Calibration Project, \citet{Sandage:2006}; (2) {\it HST} Key Project, \citet{Freedman:2001}; (3) \citet{Riess:2005}; (4) \citet{macri06}; and (5) \citet{Mager:2013}.}
\end{figure}

Figure~\ref{fg:hstobs} shows the sources of the {\it HST} data obtained on every host we use, gathered from different cameras, filters, time periods, {\it HST} programs and observers. All of these publicly available data can be readily obtained from the Mikulski Archive for Space Telescopes (MAST; see Table~\ref{tb:wfc3}). The utility of the imaging data can be divided into two basic functions: Cepheid discovery and flux measurement. For the former, a campaign using a filter with central wavelength in the visual band and $\sim\!12$ epochs with nonredundant spacings spanning $\sim\!60$--90 days will suffice to identify Cepheid variables by their unique light curves and accurately measure their periods \citep{madore91,saha96,Stetson:1996}. Revisits on a year timescale, although not required, will yield increased phasing accuracy for the longest-period Cepheids. Image subtraction can be very effective for finding larger samples of variables \citep{Bonanos:2003}, but the additional objects will be subject to greater photometric biases owing to blends which suppress their amplitudes and chances of discovery in time-series data \citep{Ferrarese:2000}.

Flux measurements are required in order to use Cepheids as standard candles for distance measurement and are commonly done with {\it HST} filters at known phases in optical ({\it F555W}, {\it F814W}) and NIR ({\it F160W}) bands to correct for the effects of interstellar dust and the nonzero width in temperature of the Cepheid instability strip. We rely primarily on NIR ``Wesenheit'' magnitudes \citep{madore82}, defined as
\bq m_H^W = m_{H} - R \, (V\!-\!I), \label{eq:wh} \eq 
\noindent where $H=F160W$, $V=F555W$, $I=F814W$ in the {\it HST} system, and $R\equiv A_H/(A_V\!-\!A_I)$. 
We note that the value of $R$ due to the correlation between Cepheid intrinsic color and luminosity
is very similar to that due to extinction \citep{Macri:2015}, so the value of $R$ derived for the latter effectively also reduces the intrinsic scatter caused by the breadth of the instability strip.  However, to avoid a distance bias, we include only Cepheids with periods above the completeness limit of detection (given in H16) in our primary fit.  (In future work we will use simulations to account for the bias of Cepheids below this limit to provide an extension of the Cepheid sample.)

In {\it HST} observations, Cepheid distances based on NIR measurements have somewhat higher statistical uncertainties than those solely based on optical photometry owing to the smaller field of view, lower spatial resolution, and greater blending from red giants.  However, as characterized in \S4.2,  this is more than offset by increased robustness to systematic uncertainties (such as metallicity effects and possible breaks in the slope of the \PL\ relation) as well as the reduced impact of extinction and a lower sensitivity to uncertainties in the reddening law. The latter is quantified by the value of $R$ in Equation~\ref{eq:wh}, ranging from 0.3 to 0.5 at $H$ depending on the reddening law, a factor of $\sim4$ lower than the value at $I$. At the high end, the \cite{Cardelli:1989} formulation with $R_V=3.3$ yields $R=0.47$. The \citet{Fitzpatrick:1999} formulation with $R_V=3.3$ and 2.5 yields $R=0.39$ and $R=0.35$, respectively. At the low end, a formulation appropriate for the inner Milky Way \citep{Nataf:2016} yields $R=0.31$. We analyze the sensitivity of H$_0$ to variations in $R$ in \S4. 

\begin{deluxetable}{llrrlr}
\tablecaption{Cepheid Hosts Observed with {\it HST}/WFC3\label{tb:wfc3}}
\tabletypesize{\small}
\tablewidth{0pc}
\tablehead{\colhead{Galaxy} & \colhead{SN~Ia} & \multicolumn{2}{c}{Exp.~time (s)} &\colhead{Prop IDs} & \colhead{UT Date$^c$}\\
\colhead{} & \colhead{} & \colhead{NIR$^a$} & \colhead{Opt.$^b$} &\colhead{} &\colhead{}}
\startdata
M101$^d$  &2011fe & 4847& 3776& 12880      &2013-03-03\\
N1015     &2009ig &14364&39336& 12880      &2013-06-30\\
N1309$^d$ &2002fk & 6991& 3002& 11570,12880&2010-07-24\\
N1365$^d$ &2012fr & 3618& 3180& 12880      &2013-08-06\\
N1448     &2001el & 6035&17562& 12880      &2013-09-15\\
N2442     &2015F  & 6035&20976& 13646      &2016-01-21\\
N3021$^d$ &1995al & 4426& 2962& 11570,12880&2010-06-03\\
N3370$^d$ &1994ae & 4376& 2982& 11570,12880&2010-04-04\\
N3447     &2012ht & 4529&19114& 12880      &2013-12-15\\
N3972     &2011by & 6635&19932& 13647      &2015-04-19\\
N3982$^d$ &1998aq & 4018& 1400& 11570      &2009-08-04\\
N4038$^d$ &2007sr & 6795& 2064& 11577      &2010-01-22\\
N4258$^d$ &Anchor &34199& 6120& 11570      &2009-12-03\\
N4424     &2012cg & 3623&17782& 12880      &2014-01-08\\
N4536$^d$ &1981B  & 2565& 2600& 11570      &2010-07-19\\
N4639$^d$ &1990N  & 5379& 1600& 11570      &2009-08-07\\
N5584     &2007af & 4929&59940& 11570      &2010-04-04\\
N5917     &2005cf & 7235&23469& 12880      &2013-05-20\\
N7250     &2013dy & 5435&18158& 12880      &2013-10-12\\
U9391     &2003du &13711&39336& 12880      &2012-12-14  

\enddata
\tablecomments{(a) Data obtained with WFC3/IR and {\it F160W}. (b) Data obtained with WFC3/UVIS and {\it F555W}, {\it F814W}, or {\it F350LP} used to find and measure the flux of Cepheids. (c) Date of first WFC3/IR observation. (d) Includes time-series data from an earlier program and a different camera --- see Fig. 2.}
\end{deluxetable}

\subsection{Cepheid Photometry}

\begin{figure}[ht]
\vspace*{150mm}
\includegraphics{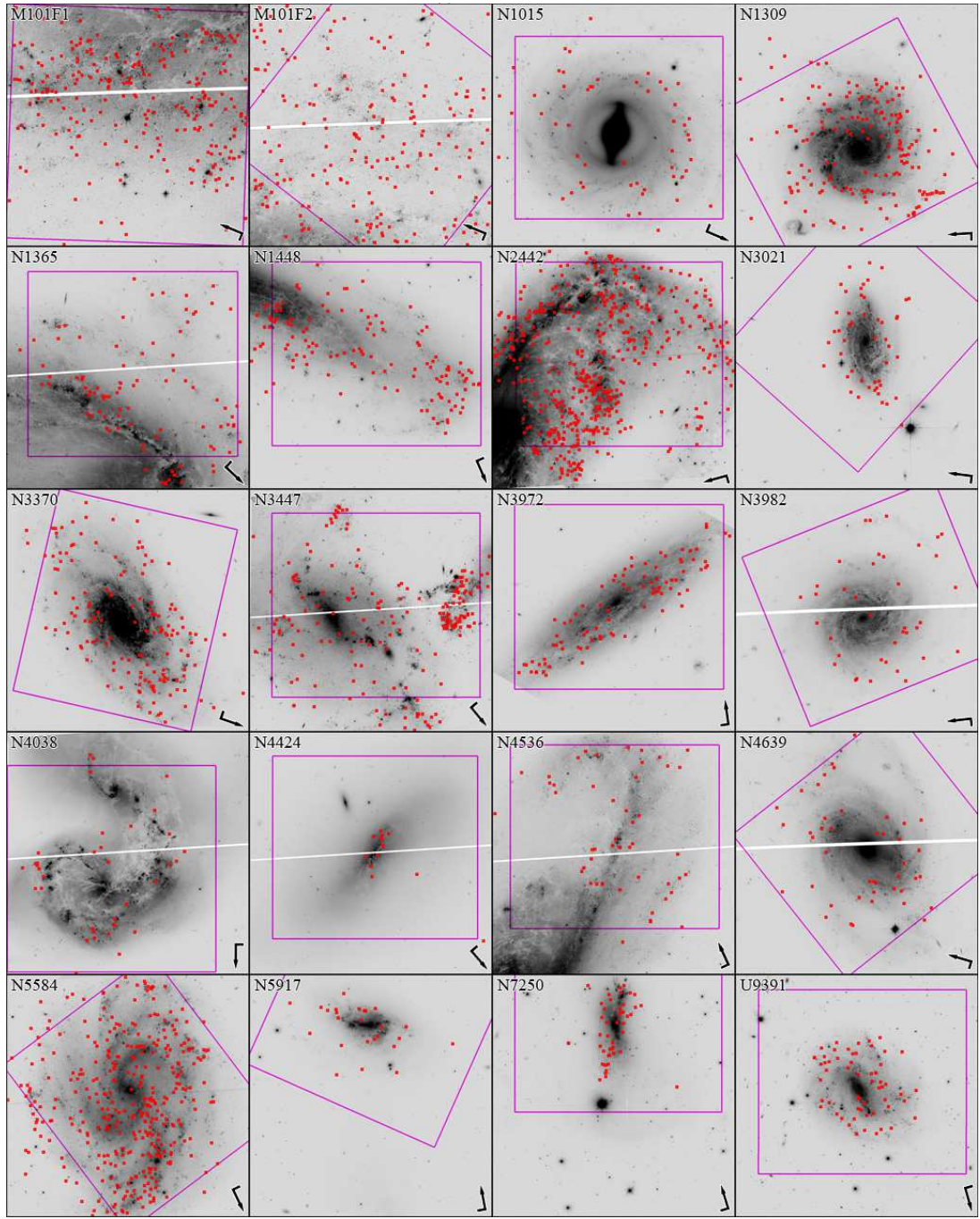}
\caption{\label{fg:hstfovs} Images of Cepheid hosts.  Each image is of the Cepheid host indicated.  The magenta outline shows the field of view of WFC3/IR, $2\farcm 7$ on a side. Red dots indicate the positions of the Cepheids. Compass indicates North (long axis) and East (short axis).}
\end{figure}

\begin{figure}[ht]
\vspace*{150mm}
\includegraphics{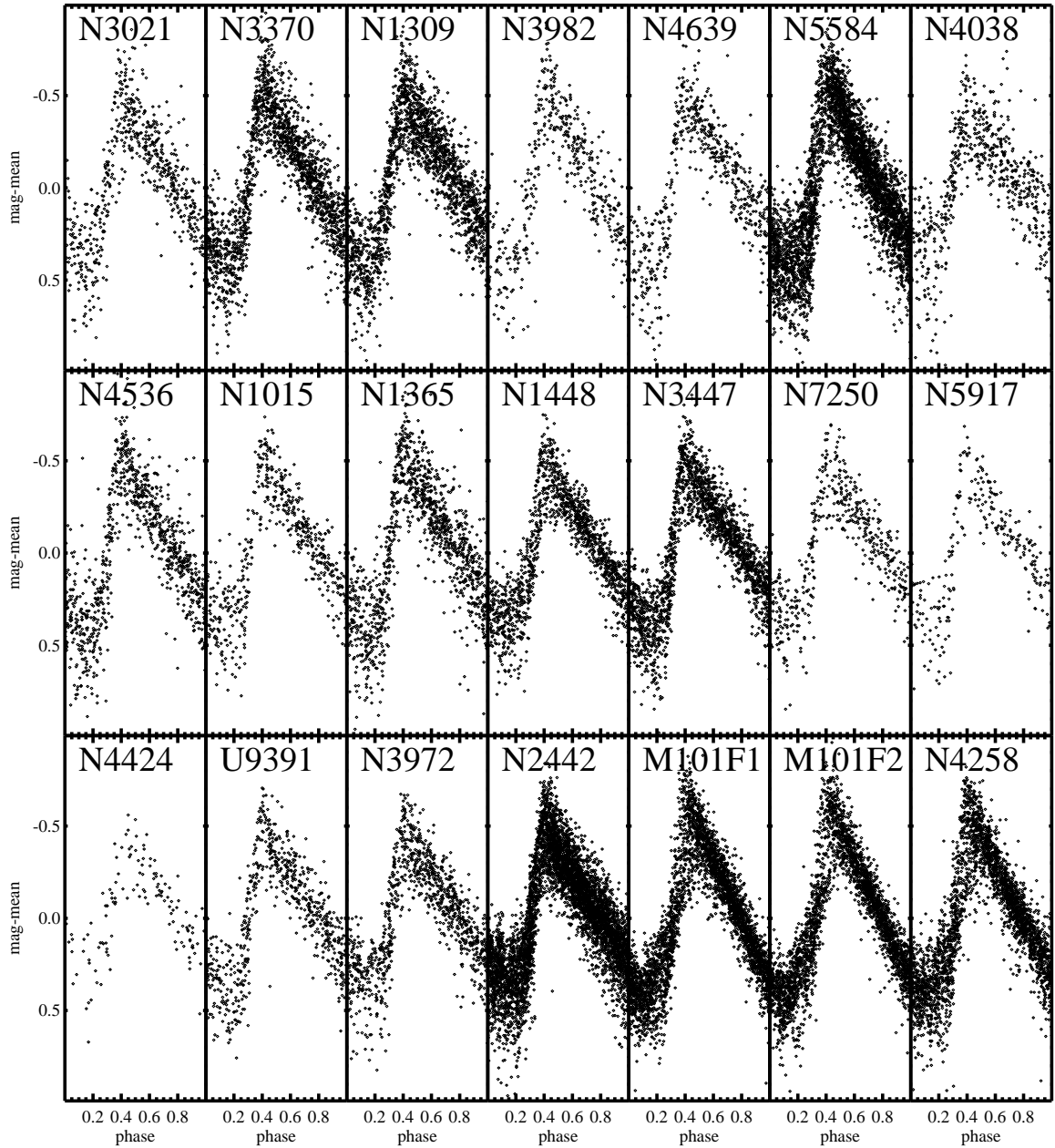}
\caption{\label{fg:cmplc} Composite visual ({\it F555W}) or white-light ({\it F350LP}) Cepheid light curves.  Each {\it HST} Cepheid light curve with $10<P<80$ days is plotted after subtracting the mean magnitude and determining the phase of the observation.  Two fields (F1 and F2) are shown for M101.}
\end{figure}

The procedure for identifying Cepheids from time-series optical data (see Table~\ref{tb:wfc3} and Figure~\ref{fg:hstobs}) has been described extensively \citep{saha96,Stetson:1996,Riess:2005,macri06}; details of the procedures followed for this sample are presented by H16, utilize the DAO suite of software tools for crowded-field PSF photometry, and are similar to those used previously by the SH0ES team. The complete sample of Cepheids discovered or reanalyzed by H16 in these galaxies (NGC$\,$4258 and the 19 SN~Ia hosts) at optical wavelengths contains 2113 variables above the periods for completeness across the instability strip (with limits estimated using the {\it HST} exposure-time calculators and empirical tests as described in that publication). There are 1566 such Cepheids in the 19 SN~Ia hosts within the smaller WFC3-IR fields alone. The positions of the Cepheids within each target galaxy are shown in Figure~\ref{fg:hstfovs}. For hosts in which we used {\it F350LP} to identify Cepheid light curves, additional photometry was obtained over a few epochs in {\it F555W} and {\it F814W}. These data were phase-corrected to mean-light values using empirical relations based on light curves in both {\it F555W} and {\it F350LP} from Cepheids in NGC$\,$5584.  Figure~\ref{fg:cmplc} shows composite Cepheid light curves in $F350LP/F555W$ for each galaxy. Despite limited sampling of the individual light curves, the composites clearly display the characteristic ``saw-toothed'' light curves of Population~I fundamental-mode Cepheids, with a rise twice as fast as the decline and similar mean amplitudes across all hosts. 

For every host, optical data in {\it F555W} and {\it F814W} from WFC3 were uniformly calibrated using the latest reference files from STScI and aperture corrections derived from isolated stars in deep images to provide uniform flux measurements for all Cepheids. In a few cases, {\it F555W} and {\it F814W} data from ACS and WFC3 were used in concert with their well-defined cross-calibration to obtain photometry with a higher signal-to-noise ratio (S/N). The cross-calibration between these two cameras has been stable to $<\!0.01$ mag over their respective lifetimes. 

\begin{figure}[ht]
\vspace*{150mm}
\includegraphics{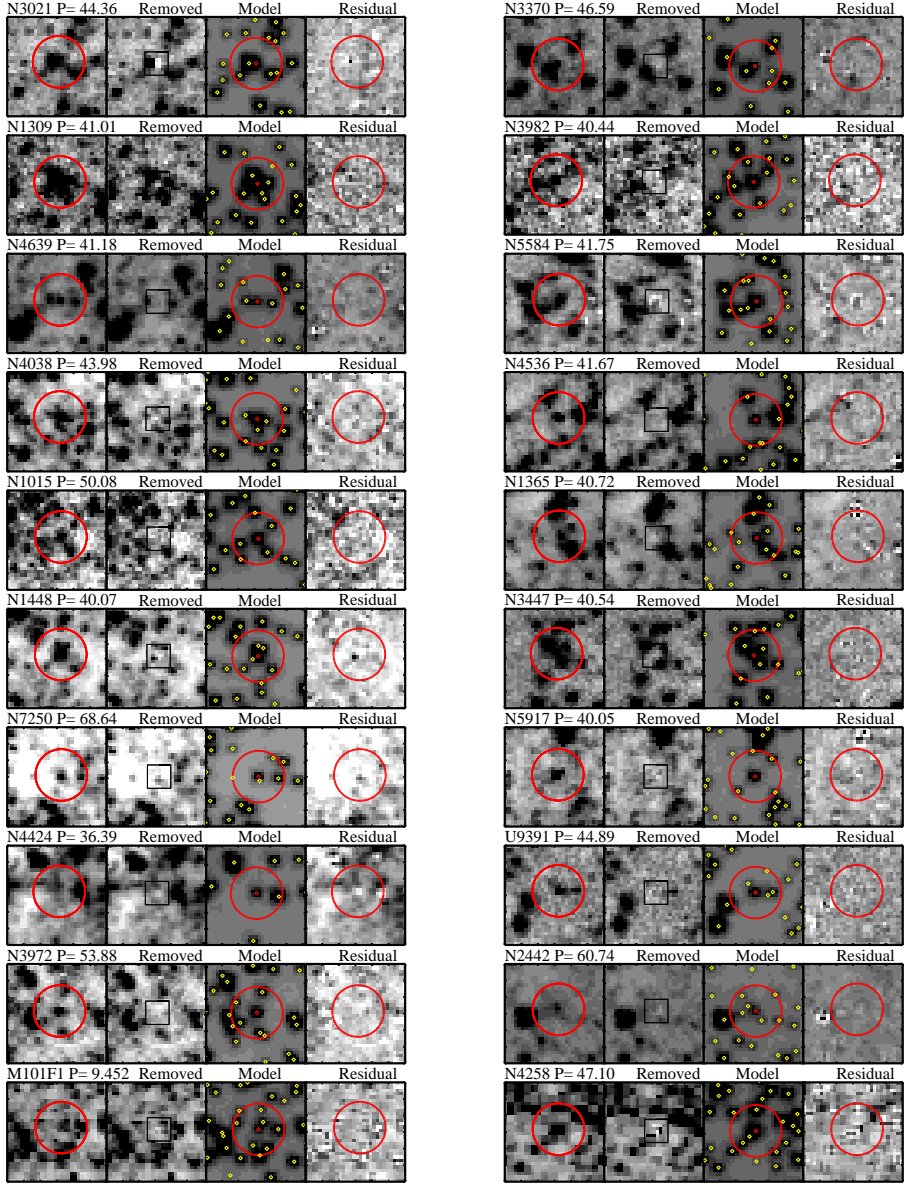}
\caption{\label{fg:stamps} Example WFC3 {\it F160W} Cepheid scene model for each host. A random Cepheid in the period range of $30<P<70$~d was selected.  The four panels of each host show a $1\arcsec$ region of the scene around each known Cepheid, the region after the Cepheid is fit and subtracted, the model of all detected sources, and the model residuals.}
\end{figure}

As in R11, we calculated the positions of Cepheids in the WFC3 {\it F160W} images using a geometric transformation derived from the optical images using bright and isolated stars, with resulting mean position uncertainties for the variables $<\!0.03$~pix. We used the same scene-modeling approach to {\it F160W} NIR photometry developed in R09 and R11. The procedure is to build a model of the Cepheid and sources in its vicinity using the superposition of point-spread functions (PSFs). {\it The position of the Cepheid is fixed at its predicted location to avoid measurement bias.} We model and subtract a single PSF at that location and then produce a list of all unresolved sources within 50 pixels. A scene model is constructed with three parameters per source ($x$, $y$, and flux), one for the Cepheid (flux) and a local sky level in the absence of blending; the best-fit parameters are determined simultaneously using a Levenberg-Marquardt-based algorithm. Example NIR scene models for each of the 19 SN~Ia hosts are shown in Figure~\ref{fg:stamps}. 

Care must be taken when measuring photometry of visible stellar sources in crowded regions as source blending can alter the statistics of the Cepheid background \citep{Stetson:1987}. Typically the mean flux of pixels in an annulus around the Cepheid is subtracted from the measured flux at the position of the Cepheid to produce unbiased photometry of the Cepheid.  This mean background or sky would include unresolved sources and diffuse background. However, we can improve the precision of Cepheid photometry by correctly attributing some flux to the other sources in the scene, especially those visibly overlapping with the Cepheid.  The consequence of differentiating the mean sky into individual source contributions plus a lower constant sky level is that the new sky level will underestimate the true mix of unresolved sources and diffuse background superimposed with the Cepheid flux (in sparse regions without blending, the original and new sky levels would approach the same value). This effect may usefully be called the sky bias or the photometric difference due to blending and is statistically easily rectified. To retrieve the unbiased Cepheid photometry from the result of the scene model we could either recalculate the Cepheid photometry using the original mean sky or correct the overestimate of Cepheid flux based on the measured photometry of artificial stars added to the scenes.  The advantage of the artificial star approach is that the same analysis also produces an empirical error estimate and can provide an estimate of outlier frequency. 

Following this approach, we measure the mean difference between input and recovered photometry of artificial Cepheids added to the local scenes in the {\it F160W} images and fit with the same algorithms.  As in R09 and R11, we added and fitted 100 artificial stars, placed one at a time, at random positions within 5 arcseconds of (but not coincident with) each Cepheid to measure and account for this difference.   To avoid a bias in this procedure, we initially estimate the input flux for the artificial stars from the Cepheid period and an assumed \PL\ relation (iteratively determined), measure the difference caused by blending, refine the \PL\ relation, and iterate until convergence. Additionally, we use the offset in the predicted and measured location of the Cepheid, a visible consequence of blending, to select similarly affected artificial stars to customize the difference measurements for each Cepheid.  The median difference for the Cepheids in all SN hosts hosts observed with {\it HST} is 0.18 mag, mostly due to red-giant blends, and it approaches zero for Cepheids in lower-density regions such as the outskirts of hosts. The Cepheid photometry presented in this paper already accounts for the sky bias.  We also estimate the uncertainty in the Cepheid flux from the dispersion of the measured artificial-star photometry around the 2.7$\sigma$-clipped mean.  The NIR Cepheid \PL\ relations for all hosts and anchors are shown in Figure~\ref{fg:plr}.
  
Likewise, in the optical images, we used as many as 200 measurements of randomly placed stars in the vicinity of each Cepheid in {\it F555W} and {\it F814W} images to measure and account for the photometric difference due to the process of estimating the sky in the presence of blending. Only 10 stars at a time were added to each simulated image to avoid increasing the stellar density. These tests show that similarly to the NIR measurements, uncertainty in the Cepheid background is the leading source of scatter in the observed \PL\ relations of the SN hosts. The mean dispersions at {\it F555W} and {\it F814W}, with values for each host listed in Table~\ref{tb:optbias} in columns 6 and 7, are 0.19 and 0.17~mag, respectively.  All SN hosts and NGC$\,$4258 display some difference in their optical magnitudes due to blending, with mean values of $0.05$ and $0.06$~mag (bright) in {\it F555W} and {\it F814W}, respectively. The most crowded case ($\Delta V=0.32$ and $\Delta I=0.26$~mag) is NGC$\,$4424, a galaxy whose Cepheids are located in a circumnuclear starburst region with prominent dust lanes. We tabulate the mean photometric differences due to blending for each host in Table~\ref{tb:optbias}, columns 2 and 3. However, the effect of blending largely cancels when determining the color {\it F555W}$\!-\!${\it F814W} used to measure Cepheid distances via equation (1) since the blending is highly correlated across these bands. Indeed, the estimated change in color across all hosts given in Table~\ref{tb:optbias}, column 4 has a mean of only $0.005$~mag (blue) and a host-to-host scatter of 0.01~mag, implying no statistically significant difference from the initial measurement and thus we have not applied these to the optical magnitudes in Table~\ref{tb:wfc3irceph}. Even the additional scatter in the $m_H^W$ \PL\ relation owing to blending in the optical color measurement is a relatively minor contribution of 0.07~mag. The small correction due to blending in the optical bands does need to be accounted for when using a conventional {\it optical} Wesenheit magnitude, $m_I^W=${\it F814W}$\!-\!R_I(${\it F555W}$\!-\!${\it F814W}$)$, because (unlike the color) the cancellation in $m_I^W$ is not complete. We find a small mean difference for $m_I^W$ in our SN hosts of $0.025$~mag (bright) with a host-to-host dispersion in this quantity of 0.03 mag. If uncorrected, this would lead to a 1\% underestimate of distances and an overestimate of H$_0$ for studies that rely exclusively on $m_I^W$. The more symmetric effect of blending on $m_I^W$ than $m_H^W$ magnitudes results from the mixture of blue blends (which make $m_I^W$ faint) and red blends (which make $m_I^W$ bright). These results are consistent with those found from simulations by \citet{Ferrarese:2000}, who drew similar conclusions. We will make use of these results for $m_I^W$ in \S4.2.  Although the net effect of blending for $m_I^W$ is typically small, the uncertainty it produces is the dominant source of dispersion with a mean of 0.36~mag for the SN hosts, similar in impact and scatter to what was found for $m^W_H$.


\begin{deluxetable}{lrrrrrrrr}
\tablecaption{Artificial Cepheid Tests in Optical Images\label{tb:optbias}}
\tabletypesize{\small}
\tablewidth{0pc}
\tablehead{\colhead{Host} & \colhead{$\Delta V$} & \colhead{$\Delta I$}& \colhead{$\Delta ct$}& \colhead{$\Delta m_I^W$}& \colhead{$\sigma(V)$} & \colhead{$\sigma(I)$}& \colhead{$\sigma_{\rm ct}$}& \colhead{$\sigma(m_I^W)$}\\ 
& \multicolumn{4}{c}{[mmag]} & \multicolumn{4}{c}{[mag]}}
\startdata
M101   &   6 &   3 &  1  &  -2 & 0.09 & 0.09 & 0.03 & 0.16 \\
N1015  &  41 &  40 &  1  &  27 & 0.13 & 0.13 & 0.06 & 0.31 \\
N1309  & 105 &  63 & 12  &  -1 & 0.35 & 0.26 & 0.10 & 0.48 \\
N1365  &  15 &  19 &  0  &   7 & 0.13 & 0.13 & 0.06 & 0.29 \\
N1448  &  31 &  24 &  1  &   6 & 0.14 & 0.13 & 0.06 & 0.29 \\
N2442  & 141 & 109 &  8  &  23 & 0.24 & 0.21 & 0.10 & 0.48 \\ 
N3021  & 106 & 134 &  0  &  75 & 0.23 & 0.22 & 0.09 & 0.46 \\
N3370  &  69 &  55 &  5  &  26 & 0.23 & 0.19 & 0.07 & 0.37 \\
N3447  &  34 &  23 &  4  &  -1 & 0.14 & 0.12 & 0.06 & 0.29 \\
N3972  &  79 &  68 &  7  &  25 & 0.18 & 0.17 & 0.07 & 0.38 \\
N3982  &  82 &  69 &  0  &  22 & 0.22 & 0.19 & 0.09 & 0.44 \\
N4038  &  38 &  28 &  2  &  12 & 0.19 & 0.15 & 0.07 & 0.34 \\
N4258I &   5 &   7 & -1  &  10 & 0.20 & 0.23 & 0.05 & 0.36 \\
N4258O &  -2 &   1 &  0  &   0 & 0.08 & 0.07 & 0.02 & 0.10 \\
N4424  & 318 & 262 & -2  & 111 & 0.31 & 0.28 & 0.11 & 0.58 \\
N4536  &  12 &  16 & -1  &  10 & 0.11 & 0.10 & 0.05 & 0.24 \\
N4639  &  56 &  85 & -5  &  89 & 0.21 & 0.22 & 0.09 & 0.51 \\
N5584  &  26 &  23 &  2  &   7 & 0.15 & 0.13 & 0.05 & 0.26 \\
N5917  &  54 &  51 & -2  &  32 & 0.20 & 0.19 & 0.08 & 0.42 \\
N7250  & 152 &  91 & 13  &  -1 & 0.24 & 0.20 & 0.08 & 0.42 \\
U9391  &  36 &  42 & -3  &  38 & 0.15 & 0.15 & 0.06 & 0.34 
\enddata
\tablecomments{$\Delta =$ median magnitude or color offset derived from tests; $\sigma =$dispersion around $\Delta$; $V$ stands for {\it F555W}; $I$ stands for {\it F814W}; {\it ct}$=R \times (V\!-\!I)$, with $R=0.39$ for $R_V=3.3$ and the \citet{Fitzpatrick:1999} extinction law; $m_I^W=$defined in text.}
\end{deluxetable}
 
\begin{deluxetable}{lrrrrrrr} 
\tablewidth{0pc}
\tabletypesize{\small}
\tablecaption{Properties of NIR $P$--$L$ Relations\label{tb:nirpl}}
\tablehead{\colhead{Galaxy} & \multicolumn{3}{c}{Number} & \colhead{$\langle P \rangle$} & \colhead{$\Delta T$} & \colhead{$\langle \sigma_{\rm tot}\rangle$} & \colhead{$\sigma_{\rm PL}$} \\
\colhead{} & \colhead{FoV} & \colhead{meas.} & \colhead{fit} & \colhead{(days)} & \colhead{(days)} & \colhead{} & \colhead{}}
\startdata
LMC   &    \ndr &  799 &  785 &   6.6 & \ndr & 0.09 & 0.08 \\
MW    &    \ndr &   15 &   15 &   8.5 & \ndr & 0.21 & 0.12 \\
M31   &    \ndr &  375 &  372 &  11.5 &   0 & 0.15 & 0.15 \\
M101  &     355 &  272 &  251 &  17.0 &   0 & 0.30 & 0.32 \\
N1015 &      27 &   14 &   14 &  59.8 & 100 & 0.32 & 0.36 \\
N1309 &      64 &   45 &   44 &  55.2 &   0 & 0.35 & 0.36 \\
N1365 &      73 &   38 &   32 &  33.6 &  12 & 0.32 & 0.32 \\
N1448 &      85 &   60 &   54 &  30.9 &  54 & 0.30 & 0.36 \\
N2442 &     285 &  143 &  141 &  32.5 &  68 & 0.52 & 0.38 \\
N3021 &      36 &   18 &   18 &  32.8 &   0 & 0.42 & 0.51 \\
N3370 &      86 &   65 &   63 &  42.1 &   0 & 0.33 & 0.33 \\
N3447 &     120 &   86 &   80 &  34.5 &  59 & 0.28 & 0.34 \\
N3972 &      71 &   43 &   42 &  31.5 &  38 & 0.49 & 0.38 \\
N3982 &      22 &   16 &   16 &  40.6 &   0 & 0.30 & 0.32 \\
N4038 &      28 &   13 &   13 &  63.4 &   0 & 0.43 & 0.33 \\
N4258 &     228 &  141 &  139 &  18.8 &   0 & 0.40 & 0.36 \\
N4424 &       8 &    4 &    3 &  28.9 &  33 & 0.56 & \ndr \\
N4536 &      47 &   35 &   33 &  36.5 &   0 & 0.27 & 0.29 \\
N4639 &      35 &   26 &   25 &  40.4 &   0 & 0.36 & 0.45 \\
N5584 &     128 &   85 &   83 &  42.6 &  11 & 0.32 & 0.33 \\
N5917 &      21 &   14 &   13 &  39.8 & 100 & 0.39 & 0.38 \\
N7250 &      39 &   22 &   22 &  31.3 &  60 & 0.44 & 0.43 \\
U9391 &      36 &   29 &   28 &  42.2 & 100 & 0.34 & 0.43 \\
\hline
Total SN & 1566 & 1028 &  975 &  32.5 & \ndr & \ndr & \ndr \\
Total All& \ndr & 2358 & 2286 &  \ndr & \ndr & \ndr & \ndr \\
\enddata
\tablecomments{FoV: located within the WFC3/IR field of view. Meas.: good quality measurement within allowed color range and with period above completeness limit. Fit: after global outlier rejection, see \S 4.1. $\langle P \rangle$: median period of the final NIR sample used in this analysis; $\Delta T=$time interval between first and last NIR epochs; $\langle\sigma_{\rm tot}\rangle=$median value of $\sigma_{\rm tot}$ (uncertainties) for Cepheids in each host (see text for definition); $\sigma_{\rm PL}=$apparent dispersion of NIR $P$--$L$ relation after outlier rejection.}
\end{deluxetable}

\begin{deluxetable}{lrrrrrrrrrr}
\tabletypesize{\scriptsize}
\tablewidth{0pt}
\tablecaption{WFC3-IR Cepheids\label{tb:wfc3irceph}}
\tablenum{4}
\tablehead{\colhead{Field}&\colhead{$\alpha$}&\colhead{$\delta$}&\colhead{ID}&\colhead{$P$}&\colhead{$V\!-\!I$}&\colhead{{\it $H$}}&\colhead{$\sigma_{\rm tot}$}&\colhead{$Z$}&\colhead{Note}\\ 
\colhead{}&\multicolumn{2}{c}{(deg, J2000)}&\colhead{(mag)}&\colhead{(days)}&\multicolumn{3}{c}{(mag)}&\colhead{(dex)}&\colhead{}}
\startdata
N3021 & 147.75035 & 33.547150 & 64252 &  16.18  &  0.92  &  25.72  &  0.578  &  8.831  &  \\
N3021 & 147.74194 & 33.558410 & 97590 &  18.24  &  1.00  &  25.05  &  0.536  &  8.972  &  \\
N3021 & 147.73714 & 33.560090 & 114118 &  20.60  &  1.13  &  26.80  &  0.581  &  8.930  &  \\
N3021 & 147.74692 & 33.556630 & 80760 &  21.01  &  1.17  &  25.79  &  0.596  &  8.914  &  \\
N3021 & 147.72083 & 33.555140 & 155661 &  22.98  &  0.99  &  25.64  &  0.286  &  8.665  &  \\
N3021 & 147.72678 & 33.556140 & 143080 &  23.95  &  1.22  &  25.30  &  0.458  &  8.968  &  \\
N3021 & 147.73210 & 33.548780 & 124526 &  26.78  &  1.18  &  25.49  &  0.365  &  8.875  &  \\
N3021 & 147.73335 & 33.552300 & 122365 &  31.09  &  0.93  &  25.57  &  0.525  &  9.197  &  \\
N3021 & 147.74791 & 33.550320 & 74434 &  31.68  &  0.87  &  24.54  &  0.496  &  9.045  &  \\
N3021 & 147.73688 & 33.559300 & 114576 &  33.18  &  1.06  &  24.83  &  0.308  &  9.007  &  \\
N3021 & 147.73288 & 33.560150 & 127220 &  35.31  &  1.49  &  25.65  &  0.308  &  8.945  &  \\
N3021 & 147.72787 & 33.558920 & 141178 &  36.38  &  1.27  &  25.35  &  0.298  &  8.936  &  \\
N3021 & 147.73387 & 33.551510 & 120418 &  35.34  &  0.84  &  25.23  &  0.432  &  9.166  &  \\
N3021 & 147.73248 & 33.548850 & 123439 &  39.41  &  1.18  &  25.02  &  0.309  &  8.895  &  \\
N3021 & 147.74989 & 33.550530 & 67964 &  39.83  &  1.24  &  26.08  &  0.432  &  8.964  &  \\
N3021 & 147.75172 & 33.549600 & 59565 &  44.28  &  0.58  &  25.06  &  0.235  &  8.869  &  \\
N3021 & 147.73892 & 33.558060 & 107249 &  56.24  &  1.32  &  24.65  &  0.528  &  9.089  &  \\
N3021 & 147.75116 & 33.554140 & 65081 &  58.08  &  0.90  &  24.31  &  0.242  &  8.842  &  
\enddata
\tablecomments{$V\!-\!I$ stands for {\it F555W}$-${\it F814W} and $H$ stands for {\it F160W}. $Z=12+\log(O/H)$}
\end{deluxetable}

Although we quantify and propagate the individual measurement uncertainty for each Cepheid, we conservatively discard the lowest-quality measurements. As in R11, scene models of Cepheids were considered to be useful if our software reported a fitted magnitude for the source with an uncertainty $<\!0.7$~mag, a set of model residual pixels with root-mean square (rms) lower than $3\sigma$ from the other Cepheid scenes, and a measured difference from the artificial star analyses of $<\!1.5$~mag.  In addition, we used a broad (1.2~mag) allowed range of {\it F814W}$\!-\!${\it F160W} colors centered around the median for each host, similar to the $V\!-\!I$ color selection common to optical studies (see H16), to remove any Cepheids strongly blended with redder or bluer stars of comparable brightness.  As simulations in \S4.1 show, most of these result from red giants but also occasionally from blue supergiants.  

1028 of the 1566 Cepheids present in the {\it F160W} images of the SN~Ia hosts with periods above their respective completeness limits yielded a good quality photometric measurement within the allowed color range. Excessive blending in the vicinity of a Cepheid in lower-resolution and lower-contrast NIR images was the leading cause for the failure to derive a useful measurement for the others. The number of Cepheids available at each step in the measurement process is given in Table~\ref{tb:nirpl}.  
 
\subsection{Statistical Uncertainties in Cepheid Distances}

We now quantify the statistical uncertainties that apply to Cepheid-based distance estimates. As described in the previous section, the largest source of measurement uncertainty for $m_H^W$ (defined in equation 1) arises from fluctuations in the NIR sky background due to variations in blending, and it is measured from artificial star tests; we refer to this as $\sigma_{\rm sky}$. For SN~Ia hosts at 20--40~Mpc and for NGC$\,$4258, the mean $\sigma_{\rm sky}$ for Cepheids in the NIR images is $0.28$~mag, but it may be higher or lower depending on the local stellar density. The next term which may contribute uncertainty in equation 1 is $\sigma_{\rm ct}=R\sigma(V-I)$.  While blending does not change the mean measured optical colors  (discussed in \S2.1), it does add a small amount of dispersion.  The artificial star tests in the optical data yield a mean value for $\sigma_{\rm ct}$ of 0.07 mag across all hosts, with values for each host given in Table~\ref{tb:optbias}, column 8). There is also an {\it intrinsic dispersion}, $\sigma_{\rm int}$, resulting from the nonzero temperature width of the Cepheid instability strip. It can be determined empirically using nearby Cepheid samples which have negligible background errors. We find $\sigma_{\rm int}=0.08$~mag for $m_H^W$ (0.12 mag for $m_I^W$) using the LMC Cepheids from \citet{Macri:2015} over a comparable period range (see Figure~\ref{fg:plr}). This agrees well with expectations from the Geneva stellar models (R.~I.~Anderson et al. 2016, in prep.). We use this value as the intrinsic dispersion of mean $m_H^W$ magnitudes. The last contribution comes from our use of random- or limited-phase (rather than mean-phase) {\it F160W} magnitudes. Monte Carlo sampling of complete $H$-band light curves from \citet{persson04} shows that the use of a single random phase adds an error of $\sigma_{\rm ph}=$0.12~mag\footnote{The sum of the intrinsic and random phase errors, 0.14~mag, is smaller than the 0.21~mag assumed by R11; the overestimate of this uncertainty explains why the $\chi^2$ of the \PL\ fits in that paper were low and resulted in the need to rescale parameter errors.}. The relevant fractional contribution of the random-phase uncertainty for a given Cepheid with period $P$ depends on the temporal interval, $\Delta T$, across NIR epochs, a fraction we approximate as $f_{\rm ph}=1-({\Delta T}/ P)$ for $\Delta T\!<\!P$ and $f_{\rm ph}= 1$ for $\Delta T\!>\!P$; the values of $\Delta T$ are given in Table~\ref{tb:nirpl}. The value of this fraction ranges from $\sim\!1$ (NIR observations at every optical epoch) to zero (a single NIR follow-up observation).

Thus, we assign a total statistical uncertainty arising from the quadrature sum of four terms: NIR photometric error, color error, intrinsic width and random-phase:\bq \sigma_{\rm tot}=(\sigma_{\rm sky}^2\!+\!\sigma_{\rm ct}^2\!+\!\sigma_{\rm int}^2\!+\!(f_{\rm ph}\sigma_{\rm ph})^2)^{1 \over 2}. \nonumber \eq
We give the values of $\sigma_{\rm tot}$ for each Cepheid in Table~\ref{tb:wfc3irceph}. These have a median of 0.30 mag (mean of 0.32 mag) across all fields; mean values for each field range from 0.23~mag (NGC$\,$3447) to 0.47~mag (NGC$\,$4424). The mean for NGC$\,$4258 is 0.39~mag.  We also include in Table~\ref{tb:wfc3irceph} an estimate of the metallicity at the position of each Cepheid based on metallicity gradients measured from optical spectra of H~II regions obtained with the Keck-I 10~m telescope and presented by H16.

\section{Measuring the Hubble Constant}

The determination of H$_0$ follows the formalism described in \S3 of R09. To summarize, we perform a single, simultaneous fit to all Cepheid and SN~Ia data to minimize one $\chi^2$ statistic and measure the parameters of the distance ladder. We use the conventional definition of the distance modulus, $\mu=5\, {\rm log}\,D + 25$, with $D$ a luminosity distance in Mpc and measured as the difference in magnitudes of an apparent and absolute flux, $\mu=m-M$.
We express the $j$th Cepheid magnitude in the $i$th host as
\bq m_{H,i,j}^W=(\mu_{0,i}\!-\!\mu_{0,{\rm N4258}})\!+\!{\rm zp}_{W,{\rm N4258}}\!+\!b_W \ {\rm log}\,P_{i,j}\!+\!Z_W \ \Delta \log\,{\rm (O/H)}_{i,j}, \label{eq:cephmag} \eq
\noindent where the individual Cepheid parameters are given in Table~\ref{tb:wfc3irceph} and $m_{H,i,j}^W$ was defined in Equation~\ref{eq:wh}. We determine the values of the nuisance parameters $b_W$ and $Z_W$ --- which define the relation between Cepheid period, metallicity, and luminosity --- by minimizing the $\chi^2$ for the global fit to the sample data. The reddening-free distances for the hosts relative to NGC$\,$4258 are given by the fit parameters $\mu_{0,i}\!-\!\mu_{0,{\rm N4258}}$, while zp$_{W,{\rm N4258}}$ is the intercept of the \PL\ relation simultaneously fit to the Cepheids of NGC$\,$4258.

Uncertainties in the nuisance parameters are due to measurement errors and the limited period and metallicity range spanned by the variables. In R11 we used a prior inferred from external Cepheid datasets to help constrain these parameters. In the present analysis, instead, we explicitly use external data as described below to augment the constraints.

\begin{figure}[ht]
\vspace*{150mm}
\includegraphics{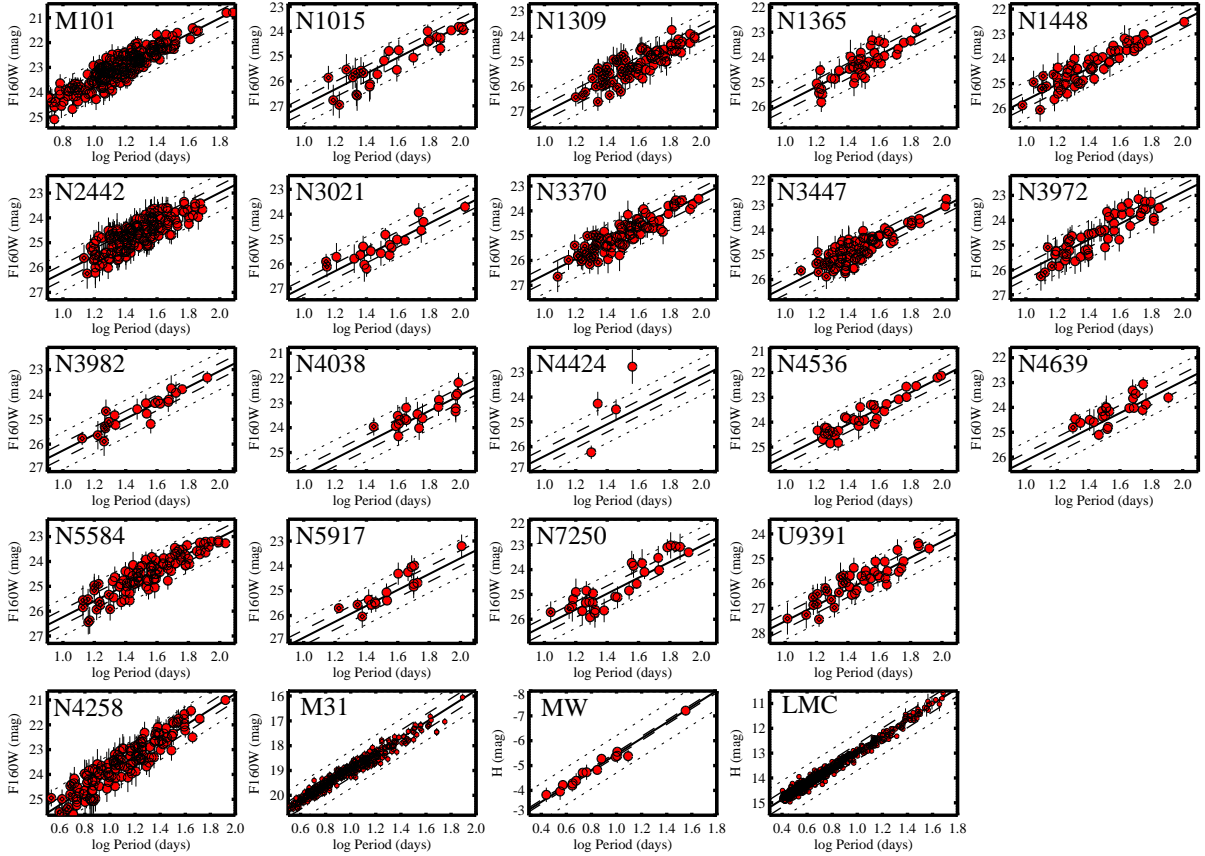}
\vspace*{-2in}
\caption{\label{fg:plr} Near-infrared Cepheid \PL\ relations. The Cepheid magnitudes are shown for the 19 SN hosts and the 4 distance-scale anchors.  Magnitudes labeled as {\it F160W} are all from the same instrument and camera, WFC3 {\it F160W}.  The uniformity of the photometry and metallicity reduces systematic errors along the distance ladder.  A single slope is shown to illustrate the relations, but we also allow for a break (two slopes) as well as limited period ranges.}
\end{figure}

Recent {\it HST} observations of Cepheids in M31 provide a powerful ancillary set of Cepheids at a fixed distance to help characterize NIR \PL\ relations. Analyses of the {\it HST} PHAT Treasury data \citep{Dalcanton:2012} by \citet{Riess:2012}, \citet{Kodric:2015}, and \citet{Wagner-Kaiser:2015} used samples of Cepheids discovered from the ground with NIR and optical magnitudes from {\it HST} to derive low-dispersion \PL\ relations. We used the union set of these samples and their WFC3 photometry in {\it F160W} measured with the same algorithms as the previous hosts to produce a set of 375 Cepheids with $3\!<\!P\!<\!78$~d as shown in Figure~\ref{fg:plr}. We add Equation~\ref{eq:cephmag} (actually, a set of such equations) for these data to those from the other hosts, requiring the addition of one nuisance parameter, the distance to M31, but providing a large range of $\log\,P$ ($\sim\!1.4$~dex) for the determination of the \PL\ relation slopes. These M31 Cepheids alone constrain the slope to an uncertainty of 0.03 mag dex$^{-1}$, a factor of 3 better than the prior used by R11. They also hint at the possible evidence of a break in the $m_H^W$ \PL\ relation at the $2\sigma$ confidence level \citep{Kodric:2015} if the location of a putative break is assumed {\it a priori} to be at 10 days as indicated by optical \PL\ relations \citep{Ngeow:2005}. To allow for a possible break, we include two different slope parameters in Equation~\ref{eq:cephmag} in the primary analysis, one for Cepheids with $P\!>\!10$~d and another for $P\!<\!10$~d. We will consider alternative approaches for dealing with nonlinear \PL\ relations in \S4.1. 

The SN~Ia magnitudes in the calibrator sample are simultaneously expressed as
\bq m_{x,i}^0=( \mu_{0,i}-\mu_{0,{\rm N4258}})+m^0_{x,{\rm N4258}}, \label{eq:snmag} \eq
\noindent where the value $m_{x,i}^0$ is the maximum-light apparent $x$-band brightness of a SN~Ia in the {\it i}th host at the time of $B$-band peak, corrected to the fiducial color and luminosity. This quantity is determined for each SN~Ia from its multiband light curves and a light-curve fitting algorithm. For the primary fits we use SALT-II \citep{Guy:2005,Guy:2010}.  For consistency with  the most recent cosmological fits we use version 2.4 of SALT II as used by \cite{Betoule:2014} and more recently from \cite{Scolnic:2016} \footnote{http://kicp.uchicago.edu/$\sim$dscolnic/Supercal/supercal\textunderscore vH0.fitres} and for which $x=B$.  The fit parameters are discussed in more detail in \S4.2.  In order to compare with R11 and to explore systematics in light-curve fits, we also use MLCS2k2 \citep{jha07} for which $x=V$  (see \S4.2 for further discussion). 

The simultaneous fit to all Cepheid and SN~Ia data via Equations~\ref{eq:cephmag} and \ref{eq:snmag} results in the determination of $m^0_{x,{\rm N4258}}$, which is the expected reddening-free, fiducial, peak magnitude of a SN~Ia appearing in NGC$\,$4258. The individual Cepheid {\PL} relations are shown in Figure~\ref{fg:plr}. Lastly, H$_0$ is determined from
\bq {\rm log}\, {\rm H}_0={(m_{x,{\rm N4258}}^0-\mu_{0,{\rm N4258}})+5a_x+25 \over 5}, \label{eq:h0} \eq
\noindent where $\mu_{0,4258}$ is the independent, geometric distance modulus estimate to NGC$\,$4258 obtained through VLBI observations of water megamasers orbiting its central supermassive black hole \citep[][]{herrnstein99,humphreys05,argon07,Humphreys:2008,Humphreys:2013}. 


Observations of megamasers in Keplerian motion around a supermassive blackhole in NGC 4258 provide one of the best sources of calibration of the absolute distance scale with a total uncertainty given by H13 of 3\%.  However, the leading systematic error in H13 resulted
from limited numerical sampling of the multi-parameter model space of the system, given in H13 as 1.5\%.  The ongoing improvement in computation speed allows us to reduce this error.

\begin{figure}[ht]
\vspace*{150mm}
\includegraphics{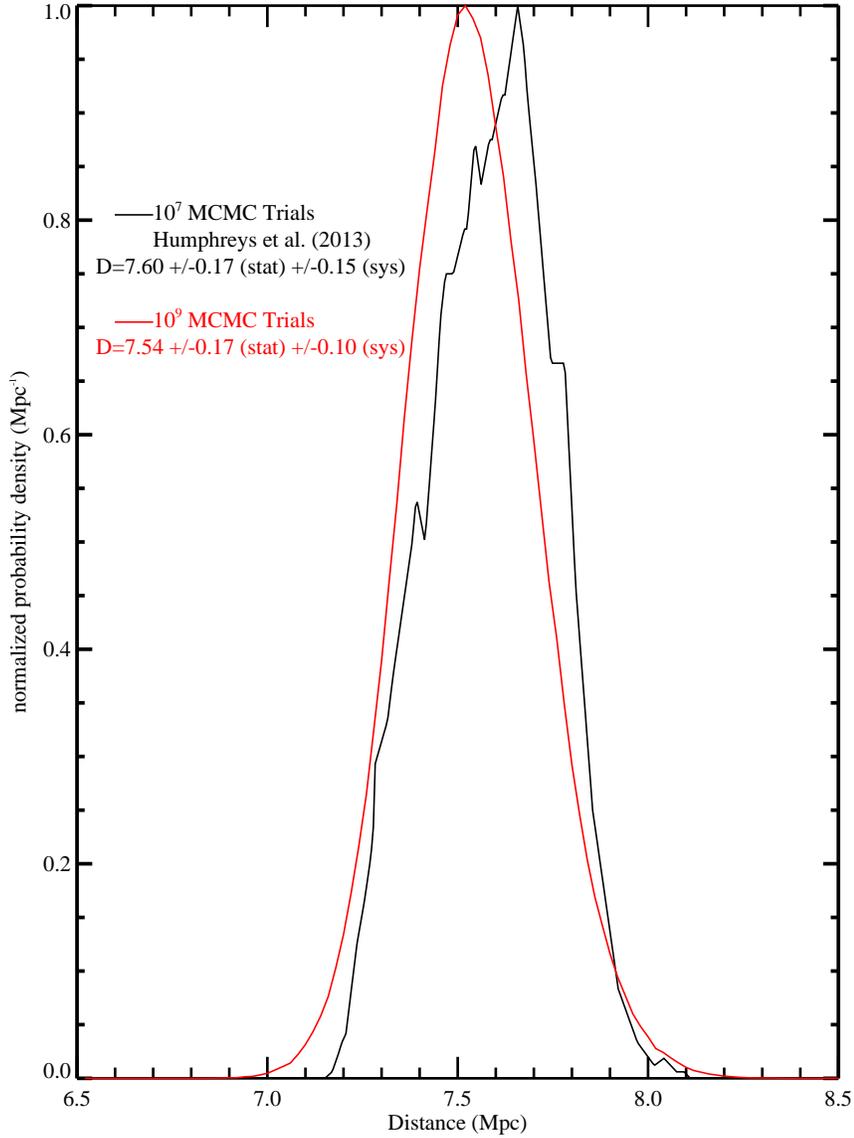}
\caption{\label{fg:maserdist} Normalized probability density function (PDF) for the maser-based distance to NGC$\,$4258.  The curve in black shows the PDF for the distance to NGC$\,$4258 based on the same multiparameter fit of the maser data in NGC$\,$4258 from \citet{Humphreys:2013} with the x-axis expanded by 18\% to match the rescaling used by H13 to account for $\chi^2_\nu=1.4$.  With a 100-fold increase (red curve) in the MCMC sampling, we have reduced the 1.5\% systematic error in distance from \citet{Humphreys:2013}, which reflected different results with differing initial conditions using more-limited MCMC sampling of the parameter space.}
\end{figure}

Here we make use of an {\it improved} distance estimate to NGC$\,$4258 utilizing the same VLBI data and model from H13 but now with a 100-fold increase in the number of Monte Carlo Markov Chain (MCMC) trial values from $10^7$ in that publication to $10^9$ for each of three independent ``strands'' of trials or initial guesses initialized near and at $\pm 10$\% of the H13 distance. By increasing the number of samples, the new simulation averages over many more of the oscillations of trial parameters in an MCMC around their true values. The result is a reduction in the leading systematic error of 1.5\% from H13 caused by ``different initial conditions'' for strands with only $10^7$ MCMC samples to 0.3\% for the differences in strands with $10^9$ MCMC samples. The smoother probability density function (PDF) for the distance to NGC$\,$4258 can be seen in Figure~\ref{fg:maserdist}. The complete uncertainty (statistical and systematic) for the maser distance to NGC$\,$4258 is reduced from 3.0\% to 2.6\%, and the better fit also produces a slight 0.8\% decrease in the distance, yielding
\bq D({\rm NGC}\,4258)=7.54\pm0.17 {\rm (random)} \pm0.10 {\rm (systematic)~Mpc,} \nonumber \eq
\noindent{equivalent to $\mu_{0,{\rm N4258}}=29.387\pm0.0568$~mag.}

\begin{figure}[ht]
\vspace*{150mm}
\includegraphics{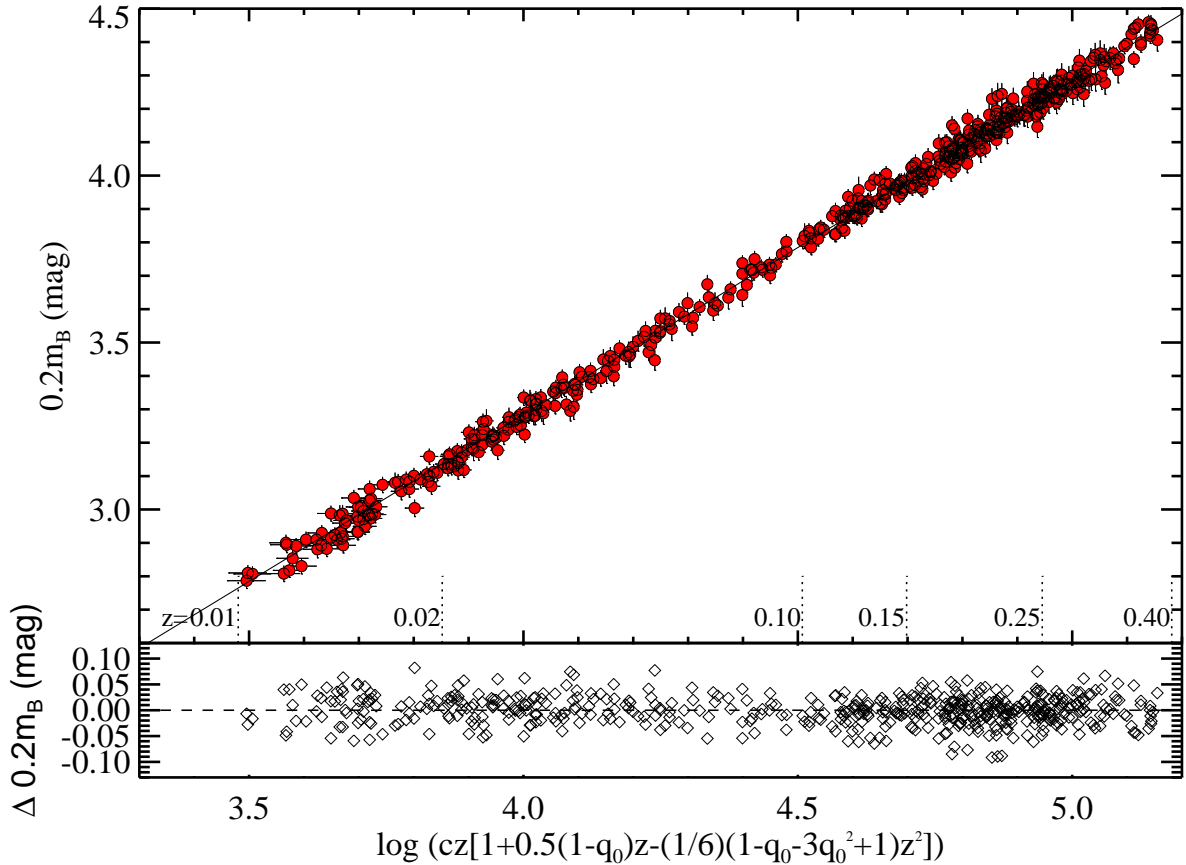}
\caption{\label{fg:snhub} Hubble diagram of more than 600 SNe~Ia at $0.01<z<0.4$ in units of $\log cz$.  Measurements of distance and redshift for a compilation of SN~Ia data as described by \citet{Scolnic:2015}.  These data are used to determine the intercept, $a_X$ (see Equation 5) where $\log cz$=0, which helps measure the value of the Hubble constant as given in Equation 9).  We account for changes in the cosmological parameters empirically by including the kinematic terms, $q_0$ and $j_0$, measured between high- and low-redshift SNe~Ia.  The intercept is measured using variants of this redshift range, as discussed in the text, with the primary fit at $0.0233<z<0.15$.}
\end{figure}

The term $a_x$ in Equation~\ref{eq:h0} is the intercept of the SN~Ia magnitude-redshift relation, approximately $\log\,cz - 0.2m_x^0$ in the low-redshift limit but given for an arbitrary expansion history and for $z\!>\!0$ as
\bq a_x=\log \left (cz \left\{ 1 + {1\over2}\left[1-q_0\right] {z} -{1\over6}\left[1-q_0-3q_0^2+j_0 \right] z^2 + O(z^3) \right \} \right ) - 0.2m_x^0, \label{eq:ax} \eq
\noindent measured from the set of SN~Ia ($z, m_x^0$) independent of any absolute (i.e., luminosity or distance) scale. We determine $a_x$ from a Hubble diagram of up to $281$ SNe~Ia with a light-curve fitter used to find the individual $m_x^0$ as shown in Figure~\ref{fg:snhub}. Limiting the sample to $0.023\!<\!z\!<\!0.15$ (to avoid the possibility of a coherent flow in the more local volume; $z$ is the redshift in the rest frame of the CMB corrected for coherent flows, see \S4.3) leaves 217 SNe~Ia (in the next section we consider a lower cut of $z\!>\!0.01$). Together with the present acceleration $q_0=-0.55$ and prior deceleration $j_0=1$ which can be measured via high-redshift SNe Ia \citep{riess07,Betoule:2014} independently of the CMB or BAO, we find for the primary fit $a_B=0.71273 \pm 0.00176$, with the uncertainty in $q_0$ contributing 0.1\% uncertainty (see \S4.3). Combining the peak SN magnitudes to the intercept of their Hubble diagram as $m_{x,i}^0 + 5 a_x$ provides a measure of distance independent of the choice of light-curve fitter, fiducial source, and measurement filter. These values are provided in Table~\ref{tb:distpar}. 

We use matrix algebra to simultaneously express the over 1500 model equations in Equations~\ref{eq:cephmag} and \ref{eq:snmag}, along with a diagonal correlation matrix containing the uncertainties. We invert the matrices to derive the maximum-likelihood parameters, as in R09 and R11.  

Individual Cepheids may appear as outliers in the $m_H^W$ {\PL} relations owing to (1) a complete blend with a star of comparable brightness and color, (2) a poor model reconstruction of a crowded group when the Cepheid is a small component of the total flux or a resolved cluster is present, (3) objects misidentified as classical Cepheids in the optical (e.g., blended Type II Cepheids), or (4) Cepheids with the wrong period (caused by aliasing or incomplete sampling of a single cycle). For our best fit we identify and remove outliers from the global model fit which exceed $2.7\sigma$ (see \S4.1 for details), comprising $\sim$2\% of all Cepheids (or $\sim$5\% from all SN hosts).  We consider alternative approaches for dealing with these outliers and include their impact into our systematic uncertainty in \S4.1.  

Our best fit using only the maser distance to NGC$\,$4258 in Equation~\ref{eq:h0} to calibrate the Cepheids yields a Hubble constant of \hofnosys\ (statistical uncertainty only; hereafter ``stat''), a \uncfnosys\ determination compared to 4.0\% in R11. The statistical uncertainty is the quadrature sum of the uncertainties in the three independent terms in Equation~\ref{eq:h0}.  We address systematic errors associated with this and other measurements in \S4.

\subsection{Additional Anchors}

We now make use of additional sources for the calibration of Cepheid luminosities, focusing on those which (i) are fundamentally geometric, (ii) have Cepheid photometry available in the $V$, $I$, and $H$ bands, and (iii) offer precision comparable to that of NGC$\,$4258, i.e., less than 5\%. For convenience, the resulting values of H$_0$ are summarized in Table~\ref{tb:h0}.
   
\subsubsection{Milky Way Parallaxes}  

Trigonometric parallaxes to Milky Way Cepheids offer one of the most direct sources of geometric calibration of the luminosity of these variables. As in R11, we use the compilation from \citet{vanleeuwen07}, who combined 10 Cepheid parallax measurements with {\it HST}/FGS from \cite{Benedict07} with those measured at lower precision with {\it Hipparcos}, plus another three measured only with significance by {\it Hipparcos}. We exclude Polaris because it is an overtone pulsator whose ``fundamentalized'' period is an outlier among fundamental-mode Cepheids. In their analysis, \citet{Freedman:2012} further reduced the parallax uncertainties provided by \citet{Benedict07}, attributing the lower-than-expected dispersion of the {\PL} relation of the 10 Cepheids from \cite{Benedict07} as evidence for lower-than-reported measurements errors. However, we think it more likely that this lower scatter is caused by chance (with the odds against $\sim\!2\sigma$) than overestimated parallax uncertainty, as the latter is dominated by the propagation of astrometry errors which were stable and well-characterized through extensive calibration of the {\it HST} FGS. As the sample of parallax measurements expands, we expect that this issue will be resolved, and for now we retain the uncertainties as determined by \cite{Benedict07}.

We add to this sample two more Cepheids with parallaxes measured by \cite{Riess:2014} and \cite{Casertano:2015} using the WFC3 spatial scanning technique. These measurements have similar fractional distance precision as those obtained with FGS despite their factor of 10 greater distance and provide two of only four measured parallaxes for Cepheids with $P\!>\!10$~d. The resulting parallax sample provides an independent anchor of our distance ladder with an error in their mean of 1.6\%, though this effectively increases to 2.2\% after the addition of a conservatively estimated $\sigma_{\rm zp}$=0.03 mag zeropoint uncertainty between the ground and {\it HST} photometric systems (but see discussion in \S5).

We use the parallaxes and the $H$, $V$, and $I$-band photometry of the MW Cepheids by replacing Equation~\ref{eq:cephmag} for the Cepheids in SN~hosts and in M31 with
\bq m_{H,i,j}^W=\mu_{0,i}+M_{H,1}^W+b_W\, \log\,P_{i,j}+Z_W \ \Delta \log\,{\rm (O/H)}_{i,j}, \label{eq:cephmagalt} \eq
\noindent where $M_{H,1}^W$ is the absolute $m_H^W$ magnitude for a Cepheid with $P=1$~d, and simultaneously fitting the MW Cepheids with the relation
\bq M_{H,i,j}^W=M_{H,1}^W\!+\!b_W \ \log\,P_{i,j}\!+\!Z_W \ \Delta \log\,{\rm (O/H)}_{i,j}, \label{eq:cephmagmw} \eq 
\noindent where $M_{H,i,j}^W=m_{H,i,j}^W\!-\!\mu_\pi$ and $\mu_\pi$ is the distance modulus derived from parallaxes, including standard corrections for bias (often referred to as Lutz-Kelker bias) arising from the finite S/N of parallax measurements with an assumed uncertainty of 0.01 mag \citep{Hanson:1979}.   The $H$, $V$, and $I$-band photometry, measured from the ground, are transformed to match the WFC3 {\it F160W}, {\it F555W}, and {\it F814W} as discussed in the next subsection. Equation~\ref{eq:snmag} for the SNe~Ia is replaced with
\bq m_{x,i}^0=\mu_{0,i}\!-\!M_x^0. \label{eq:snmagalt} \eq

The determination of $M^0_x$ for SNe~Ia together with the previous term $a_x$ then determines H$_0$,
\bq \log\,{\rm H}_0={M_x^0\!+\!5a_x\!+\!25 \over 5}. \label{eq:h0alt} \eq

The statistical uncertainty in $H_0$ is now derived from the quadrature sum of the two independent terms in equation \ref{eq:h0alt}, $M_x^0$ and $5a_x$.  

For $m_H^W$ Cepheid photometry not derived directly from {\it HST} WFC3, we assume a fully-correlated uncertainty of 0.03~mag included as an additional, simultaneous constraint equation, $0=\Delta {\rm zp} \pm \sigma_{\rm zp}$, to the global constraints with $\sigma_{\rm zp}=0.03$ mag. The free parameter, $\Delta {\rm zp}$, which expresses the zeropoint difference between {\it HST} WFC3 and ground-based data, is now added to Equation~\ref{eq:cephmagmw} for all of the MW Cepheids.  This is a convenience for tracking the correlation in the zeropoints between ground-based data and providing an estimate of its size.  In future work we intend to eliminate $\Delta {\rm zp}$ and its uncertainty by replacing the ground-based photometry with measurements from {\it HST} WFC3 enabled by spatial scanning \citep{Riess:2014}.

Using these 15 MW parallaxes as the only anchor, we find H$_0$ = \homwnosys (stat). In order to use the parallaxes together with the maser distance to NGC$\,$4258, we recast the equations for the Cepheids in NGC$\,$4258 in the form of Equation~\ref{eq:cephmagmw} with $\mu_{0,{\rm N4258}}$ in place of $\mu_\pi$ 
and the addition of the residual term $\Delta \mu_{\rm N4258}$ to these as a convenience for keeping track of the correlation among these Cepheids and the prior external constraint on the geometric distance of NGC 4258.  We then add the simultaneous constraint equation $0=\Delta \mu_{\rm N4258} \pm \sigma_{\mu 0,{\rm N4258}}$ with $\sigma_{\mu 0,{\rm N4258}}=0.0568$ mag. Compared to the use of the maser-based distance in \S3, $\sigma_{\mu 0,{\rm N4258}}$ has moved from Equation~\ref{eq:h0} to the {\it a priori} constraint on $\Delta \mu_{\rm N4258}$. This combination gives H$_0$ = \homwfnosys\ (stat), a  {\uncmwfnosys} measurement that is consistent with the value from NGC$\,$4258 to $1.2 \sigma$ considering only the distance uncertainty in the geometric anchors.

\subsubsection{LMC Detached Eclipsing Binaries}
In R11 we also used photometry of Cepheids in the LMC and estimates of the distance to this galaxy based on detached eclipsing binaries (DEBs) to augment the set of calibrators of Cepheid luminosities. DEBs provide the means to measure geometric distances \citep{Paczynski:1997} through the ability to determine the physical sizes of the member stars via their photometric light curves and radial velocities. The distance to the LMC has been measured with both early-type and late-type stars in DEBs. \citet{guinan98}, \citet{fitzpatrick02}, and \citet{ribas02} studied three B-type systems (HV\,2274, HV\,982, EROS\,1044) which lie close to the bar of the LMC and therefore provide a good match to the Cepheid sample of \citet{Macri:2015}. In R11 we used an average distance modulus for these of $18.486\pm0.065$~mag\footnote{A fourth system \citep[HV\,5936;][]{fitzpatrick03} is located several degrees away from the bar and yields a distance that is closer by 3$\sigma$. Additional lines of evidence presented in that paper suggest this system lies above the disk of the LMC, closer to the Galaxy.}. However, for early-type stars it is necessary to estimate their surface brightness via non-LTE (local thermodynamic equilibrium) model atmospheres, introducing an uncertainty that is difficult to quantify.

The approach using DEBs composed of late-type stars is more reliable and fully empirical because their surface brightness can be estimated from empirical relations between this quantity and color, using interferometric measurements of stellar angular sizes to derive surface brightnesses \citep{Di-Benedetto:2005}. \cite{Pietrzynski:2013} estimated the distance to the center of the LMC to 2\% precision using 8 DEBs composed of late-type giants in a quiet evolutionary phase on the helium burning loop, located near the center of the galaxy and along its line of nodes. The individual measurements are internally consistent and yield $\mu_{\rm LMC}=18.493\pm0.008\ {\rm (stat)} \pm 0.047$~(sys) mag, with the uncertainty dominated by the accuracy of the surface brightness vs. color relation. 
  
Recently, \citet{Macri:2015} presented NIR photometry for LMC Cepheids discovered by the OGLE-III project \citep{Soszynski:2008}, greatly expanding the sample size relative to that of \citet{persson04} from 92 to 785, although the number of Cepheids with $P\!>\!10$~d increased more modestly from 39 to 110. Similarly to the M31 Cepheids, the LMC Cepheids provide greater precision for characterizing the {\PL} relations than those in the SN~Ia hosts, and independently hint at a change in slope at $P\approx\!10$~d \citep{Bhardwaj:2016}.  

We transform the ground-based $V$, $I$ and $H$-band Vega-system photometry of \citet{Macri:2015} into the Vega-based HST/WFC3 photometric system in {\it F555W}, {\it F814W} and {\it F160W}, respectively, using the following equations:
\begin{eqnarray}
m_{555} & = & V + 0.034 + 0.11 (V-I) \\
m_{814} & =&  I + 0.02 -0.018 (V-I) \\
m_{160} & = & H  + 0.16 (J-H)
\end{eqnarray}
\noindent where the color terms were derived from synthetic stellar photometry for the two systems using SYNPHOT \citep{Laidler:2008}. To determine any zeropoint offsets (aside from the potentially different definitions of Vega) for the optical bands we compared photometry of 97 stars in the LMC observed in $V$ and $I$ by OGLE-III and in WFC3/F555W and F814W as part of HST-GO program \#13010 (P.I.: Bresolin). The latter was calibrated following the exact same procedures as H16, which uses the UVIS 2.0 WFC3 Vegamag zeropoints. The uncertainties of the zeropoints in the optical transformations were found to be only 4~mmag. The change in color, $V-I$ is quite small, at 0.014 mag or a change (decrease) in H$_0$ of 0.3 \% for a value determined solely from an anchor with ground-based Cepheid photometry (LMC or MW).  For $H$-band transformed to {\it F160W}, the net offset besides the aformentioned color term is zero after cancellation of an 0.02 mag offset measured between HST and 2MASS NIR photometry \citep{Riess:2011a} and the same in the reverse direction from the very small count-rate non-linearity of WFC3 at the brightness level of extragalactic Cepheids \citep{Riess:2010}. The mean metallicity of the LMC Cepheids is taken from their spectra by \citet{Romaniello:2008} to be $\oh=-0.25$~dex.
  
Using the late-type DEB distance to the LMC as the sole anchor and the Cepheid sample of \citet{Macri:2015} for a set of constraints in the form of Equation~\ref{eq:cephmagmw} yields  H$_0$ = \holmcnosys\ (stat). As in the prior section, these fits include free parameters $\Delta \mu_{\rm LMC}$ and $\Delta {\rm zp}$, with additional constraint $0=\Delta \mu_{\rm LMC} \pm \sigma_{\mu,{\rm LMC}}$.  The Appendix shows how the system of equations is arranged for this fit.  The last few equations (see Appendix) express the independent constraints on the external distances (i.e., for NGC$\,$4258 and the LMC) with uncertainties contained in the error matrix.  Using the anchor combination of NGC 4258 and the LMC, the optimal set for TRGB calibration,
gives H$_0$=\hotrgbnosys\ (stat).

Using all three anchors, the same set used by R11 and by \cite{Efstathiou:2014}, results in H$_0$ = \hoallthreenosys (stat), a \uncallthreenosys\ determination. The fitted parameters which would indicate consistency within the anchor sample are $\Delta \mu_{\rm N4258}=-0.043$~mag, within the range of its 0.0568 mag prior, and $\Delta \mu_{\rm LMC}=-0.042$~mag, within range of its 0.0452~mag prior.  The metallicity term for the NIR-based Wesenheit has the same sign but only about half the size as in the optical \citep{Sakai:2004} and is not well-detected with $Z_W=-0.14 \pm 0.06$ mag dex$^{-1}$ including systematic uncertainties.  

\subsubsection{DEBs in M31}

As discussed in \S3, we make use of a sample of 375 Cepheids in M31 in order to help characterize the Cepheid {\PL} relations. In principle, we can also use M31 as an anchor in the determination of H$_0$ by taking advantage of the two DEB-based distance estimates to the galaxy \citep{Ribas:2005,Vilardell:2010} which have a mean of $\mu_0=24.36 \pm 0.08$~mag.

Yet, there are several obstacles with the use of M31 as an anchor. The PHAT {\it HST} program \citep{Dalcanton:2012}, which obtained the {\it HST} data, did not use the {\it F555W} filter, nor did it include time-series data, so we cannot use the same individual, mean-light {\it F555W}$\!-\!${\it F814W} colors to deredden the Cepheids in {\it F160W} as for other SH0ES galaxies (or the individual mean $V\!-\!I$ colors to deredden $H$-band data with a 0.03~mag uncertainty as for LMC and MW Cepheids as individual ground-based colors are too noisy). The best available color for measuring the individual reddenings of the M31 Cepheids is {\it F110W}$\!-\!${\it F160W} so we must recalibrate these colors to match the reddening in the $V\!-\!I$ data.  Following \cite{Riess:2012}, we add a constant to these colors so that their mean measured {\it F160W} extinction is the same as derived from the mean $V\!-\!I$ Cepheid colors in M31 based on data from the ground-based DIRECT program \citep{Kaluzny:1998}\footnote{By equating the mean $V\!-\!I$ dereddening with that for {\it F110W}$\!-\!${\it F160W}, we can solve for a color offset to ensure they yield the same result. That is, $0.40\langle V\!-\!I\rangle = 1.49 \langle ${\it F110W}$\!-\!${\it F160W}$-X\rangle$, where $\langle V\!-\!I\rangle=1.23$~mag from DIRECT gives $X=0.22$~mag. Note that the reddening parameters \citep[now adopted from][]{Fitzpatrick:1999} and the Cepheid samples differ from those used by \cite{Riess:2012}, leading to a different value of $X$.}.  The advantage of the latter approach is that it can account for {\it differential} reddening along the line of sight while providing a reddening correction which is consistent with that used for Cepheids in all other targets. We adopt an 0.02~mag systematic uncertainty, $\sigma_{\rm zp, opt}$, between the ground-based optical colors of Cepheids and those measured from space. With the same formalism used for the LMC but with M31 as the sole anchor we find H$_0$ = \hoandnosys (stat), consistent with the value derived from the other three anchors.

On the other hand, as previously discussed, DEB distances for early-type stars (the only ones currently measured in M31) include significant inputs from non-LTE stellar model atmospheres with systematic uncertainties that are hard to assess. It is somewhat reassuring to note that in the LMC, where both types of DEBs have been measured, the difference in the distance moduli obtained from either type is only $0.01\pm 0.08$~mag, a test with the same precision as the early-type DEB distance to M31. Future measurements of late-type DEBs or water masers in M31 \citep{Darling:2011} would place M31 as an anchor on equal footing with the others.

To be conservative, we use as our primary determination of H$_0$ the result from the combination of NGC$\,$4258 masers, MW parallaxes, and LMC late-type DEBs (the same set of anchors used by R11): H$_0$ = \hoallthreenosys\ (stat). Note, however, the consistency of our primary result with the result using M31 alone. If M31 were included together with the other anchors, the resulting value of H$_0$ would be \hoallfournosys\ (stat).

\begin{figure}[ht]
\vspace*{150mm}
\includegraphics{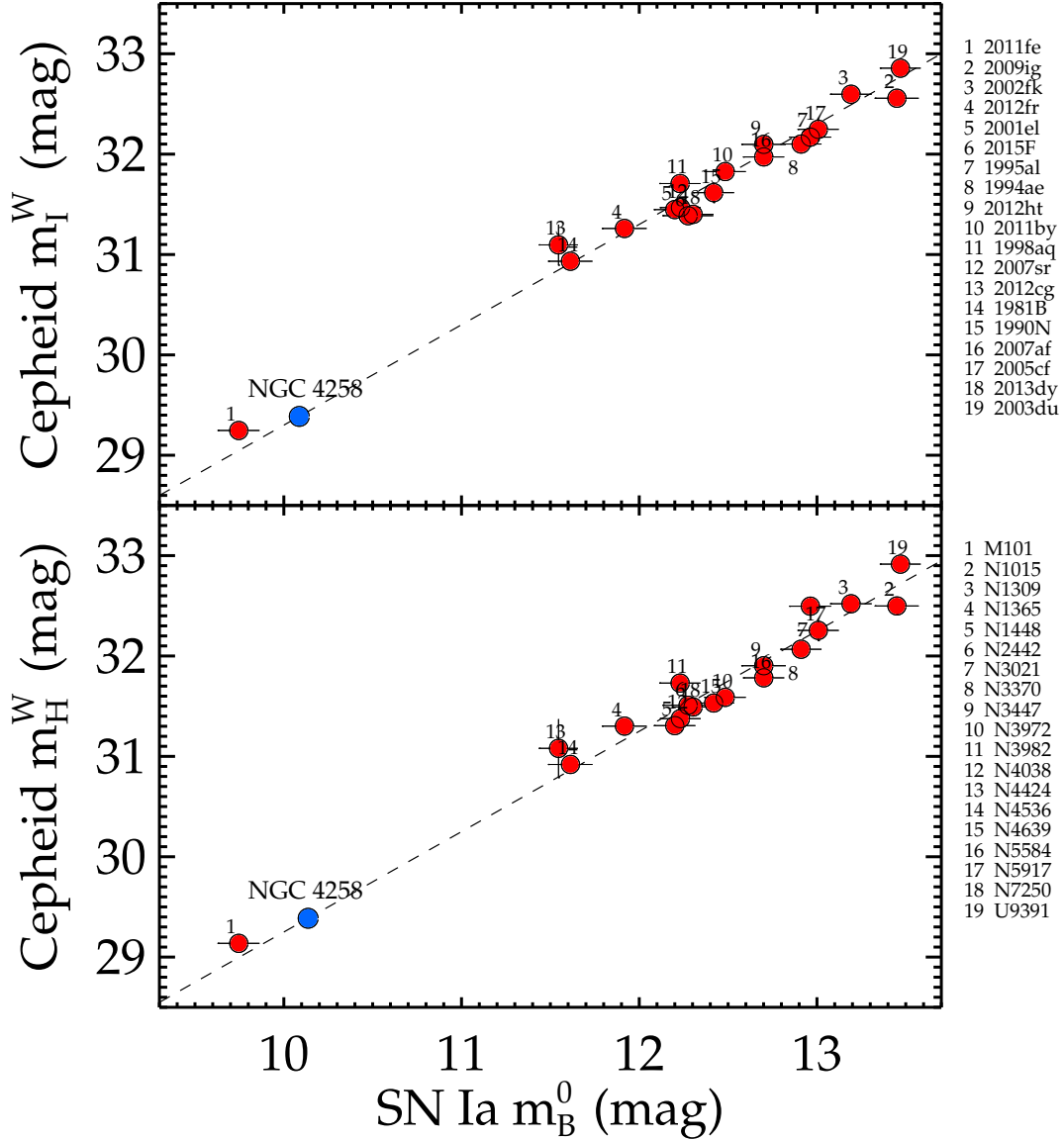}
\caption{\label{fg:cephsndist} Relative distances from Cepheids and SNe~Ia.  The top and bottom panels show relative distances for 19 hosts determined from their SNe~Ia and Cepheid Wesenheit optical and NIR magnitudes, respectively.  The Cepheid result for each host is an approximated distance derived after removing that host's SN~Ia data from the full global fit for H$_0$. The relative dispersions are 0.12 mag (top) and 0.15 mag (bottom).  The maser-calibrated Cepheid distance to NGC$\,$4258 is indicated as well as the model-fit SN~Ia magnitude it would host.}
\end{figure}

\begin{figure}[ht]
\vspace*{180mm}
\includegraphics{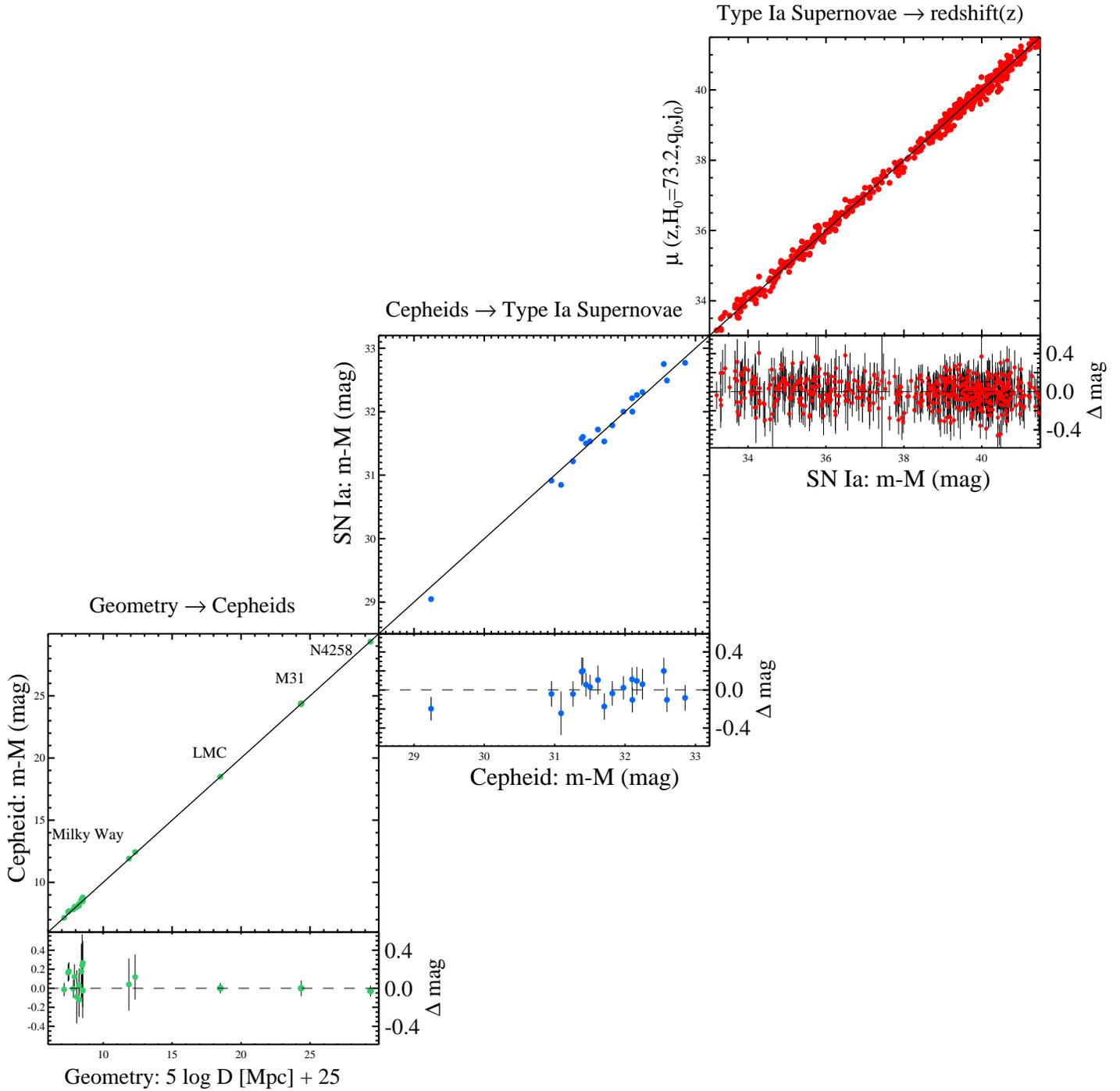}
\caption{\label{fg:ladder} Complete distance ladder.  The simultaneous agreement of pairs of geometric and Cepheid-based distances (lower left), Cepheid and SN Ia-based distances (middle panel) and SN and redshift-based distances provides the measurement of the Hubble constant.  For each step, geometric or calibrated distances on the x-axis serve to calibrate a relative distance indicator on the y-axis through the determination of $M$ or H$_0$.  Results shown are an approximation to the global fit as discussed in the text.}
\end{figure}

While the global model accounts for the covariance between all distances and model parameters, we can explore the internal agreement of the Cepheid and SN distance estimates by deriving {\it approximate} Cepheid-only distances for the 19 hosts. For each host, we remove only its SN distance from the global fit and derive its Cepheid distance, $\mu_{0,i}$ based on the remaining data. The result is a set of Cepheid distances to each host which are independent of their SN distances (although these distances are slightly correlated with each other and thus do not provide a substitute for the full analysis which accounts for such covariance). The results are listed in Table~\ref{tb:distpar}, column 5 as approximate Cepheid distances (i.e., ignoring the covariance) and Figure~\ref{fg:cephsndist} shows the SN distances versus those from Cepheid optical and NIR magnitudes. Figure~\ref{fg:ladder} shows an {\it approximation} to the full distance-ladder fit to provide a sense of the sampling using the previously described approximations. These approximations should be good to $\sim$ 0.01--0.02 mag.  The resulting relation between the SN and Cepheid-based distances will be considered in the next section.  The Cepheid-based distances for 7 of the 8 hosts used in R11 have a mean difference of 0.01~mag and a dispersion of 0.12~mag. The eight host, N4038, shifted from $-1.6\sigma$ to $+1.7\sigma$ relative to the SN-inferred distances ($\Delta\mu=-0.37~mag$, closer in R16). The shift primarily arises because we conservatively excluded a unique set of 10 variables from R11 with ultra-long periods ($P>100$~days) due to very sparse phase coverage and the poorly constrained properties of the P-L relation for these intrinsically rare objects \citep{Bird:2009,Fiorentino:2012}.


\begin{deluxetable}{llrrrrrr}
\tabletypesize{\small}
\tablewidth{0pc}
\tablenum{5}
\tablecaption{Approximations for Distance Parameters\label{tb:distpar}}
\tablehead{\colhead{Host} & \colhead{SN} & \colhead{$m_{B,i}^0$} & \colhead{$\sigma^a$} & \colhead{$\mu_{\rm Ceph}^b$} & \colhead{$\sigma$} & \colhead{$M_{B,i}^0$} & \colhead{$\sigma$} \\
\colhead{} & \colhead{} & \colhead{$+5a_B$} & \colhead{} & \colhead{} & \colhead{} & \colhead{} & \colhead{}\\
\colhead{} & \colhead{} & \multicolumn{6}{c}{(mag)}}
\startdata
M101  & 2011fe& 13.310  &  0.117  &  29.135  &  0.045  &  -19.389  &  0.125  \\
N1015 & 2009ig& 17.015  &  0.123  &  32.497  &  0.081  &  -19.047  &  0.147  \\
N1309 & 2002fk& 16.756  &  0.116  &  32.523  &  0.055  &  -19.331  &  0.128  \\
N1365 & 2012fr& 15.482  &  0.125  &  31.307  &  0.057  &  -19.390  &  0.137  \\
N1448 & 2001el& 15.765  &  0.116  &  31.311  &  0.045  &  -19.111  &  0.125  \\
N2442 & 2015F & 15.840  &  0.142  &  31.511  &  0.053  &  -19.236  &  0.152  \\
N3021 & 1995al& 16.527  &  0.117  &  32.498  &  0.090  &  -19.535  &  0.147  \\
N3370 & 1994ae& 16.476  &  0.115  &  32.072  &  0.049  &  -19.161  &  0.125  \\
N3447 & 2012ht& 16.265  &  0.124  &  31.908  &  0.043  &  -19.207  &  0.131  \\
N3972 & 2011by& 16.048  &  0.116  &  31.587  &  0.070  &  -19.103  &  0.136  \\
N3982 & 1998aq& 15.795  &  0.115  &  31.737  &  0.069  &  -19.507  &  0.134  \\
N4038 & 2007sr& 15.797  &  0.114  &  31.290  &  0.112  &  -19.058  &  0.160  \\
N4424 & 2012cg& 15.110  &  0.109  &  31.080  &  0.292  &  -19.534  &  0.311  \\
N4536 & 1981B & 15.177  &  0.124  &  30.906  &  0.053  &  -19.293  &  0.135  \\
N4639 & 1990N & 15.983  &  0.115  &  31.532  &  0.071  &  -19.113  &  0.135  \\
N5584 & 2007af& 16.265  &  0.115  &  31.786  &  0.046  &  -19.085  &  0.124  \\
N5917 & 2005cf& 16.572  &  0.115  &  32.263  &  0.102  &  -19.255  &  0.154  \\
N7250 & 2013dy& 15.867  &  0.115  &  31.499  &  0.078  &  -19.196  &  0.139  \\
U9391 & 2003du& 17.034  &  0.114  &  32.919  &  0.063  &  -19.449  &  0.130  
\enddata
\tablecomments{(a) For SALT-II, 0.1~mag added in quadrature to fitting error. (b) Approximate, SN-independent Cepheid-based distances as described at the end of \S3.}
\end{deluxetable}

\section{Analysis Systematics}

The statistical uncertainties quoted thus far include the full propagation of all known contributions as well as the degeneracies resulting from simultaneous modeling and characterization of the whole dataset of $>\!2200$ Cepheids ($\sim$ 1000 in SN hosts), 19 SNe Ia, 15 MW parallaxes, the DEB-based distance to the LMC, and the maser distance to NGC$\,$4258. Our model formally contains parameters used to propagate what were considered sources of systematic uncertainties in other analyses \citep{Freedman:2001,Freedman:2012,Sandage:2006} such as zeropoint errors, metallicity dependences, and the slopes and breaks in the \PL\ relation, therefore our statistical uncertainties incorporate many effects that others consider among systematics (see Appendix).  

Following the approach of R09 and R11, we therefore explore reasonable alternatives to the global determination of H$_0$ which are not easily parameterized for inclusion in the framework of \S3, and we use these to determine an additional systematic error component.  While truly unknown systematic errors can never be ruled out, we address this possibility in \S4 by comparing our measurement to independent measurements of H$_0$ which do not utilize SN-based distance measurements.

\subsection{Cepheid Systematics}

The Cepheid outlier fraction in \S3 is $\sim$ 2\% for all hosts (or $\sim$5\% across all SN hosts), smaller than the 15\%--20\% in R11.  This reduction in the outlier fraction results largely from the use of a color selection in {\it F814W}$\!-\!${\it F160W} around the median color in each host to remove blends with unresolved sources of comparable luminosity and different color (e.g., red giants, blue supergiants, unresolved star clusters). This is a useful criterion as it is distance- and period-independent, insensitive to reddening, and anchored to the physical properties of Cepheids (i.e., stars with spectral types F--K). The well-characterized LMC Cepheids from \cite{Macri:2015} have a mean $I\!-\!H$ of 0.96~mag with a dispersion of just 0.10~mag, much smaller than the allowed 1.2~mag breadth which alone would exclude only stars hotter than early-F or cooler than late-K (i.e., colors which cannot result from Cepheids).  Because measurement errors owing to blending are correlated across bands, the uncertainty in this color is smaller than either band and a factor of $\sim$ 6 smaller than the allowed range, so colors outside the range primarily result from color blends rather than noise.  Doubling the breadth of the color cut decreased H$_0$ by 0.9 \hunit and removing a color cut altogether lowered H$_0$ by an additional 0.2 \hunit, both shifts much smaller than the statistical uncertainty.  

We further tested the use of our color cut by simulating the appearance of a distribution of Cepheids in a galaxy at $D\approx\!30$~Mpc using star catalogues of the LMC. Cepheids with low optical blending (hence identifiable by amplitude and allowed range in {\it F555W}$\!-\!${\it F814W}; see H16 and \citealt{Ferrarese:2000}) but with significant NIR blending are most often blended with red giants. This shifts their colors redward in {\it F814W}$\!-\!${\it F160W} to a degree, on average, that is proportional to their local surface brightness. While we account for this mean, blended sky level in our photometry, the ``direct hits'' by red or blue sources are removed by the color cut. However, blending may {\it still} occur with stars of a similar color, such as the (less common) yellow supergiants, or the sample may include a small number of objects erroneously identified as Cepheids. For these reasons we still identify and remove a small fraction of the sample ($\sim 2$\%) as outliers from the {\PL} relations. 

A number of reasonable approaches would likely suffice for identifying these outliers as demonstrated for the R11 sample \citep{Becker:2015, Efstathiou:2014, Kodric:2015}. R11 used a $2.5\sigma$ threshold to identify outliers from the {\it individual} $H$-band {\PL} relations for their primary H$_0$ analysis, while evaluating the impact of no outlier rejection to determine the sensitivity of H$_0$ to this step. \citet{Efstathiou:2014} used a similar threshold but applied to outliers of the final, {\it global} fit. \citet{Kodric:2015} used a global rejection as well but recalculated the global fit after removing the single most deviant point until none remained above the threshold. \citet{Becker:2015} applied a Bayesian characterization of outliers, attributing them to a second, contaminating distribution with uniform properties.  However, the artificial-star tests and LMC analysis indicate that the outliers are well described by the tails of the blending distribution.  For our primary fit we use a {\it global} rejection of $2.7\sigma$, the threshold where the $\chi^2_\nu=0.95$ of our global fit matches that of a normal distribution with the same rejection applied.  Following \citet{Kodric:2015} we recalculated the global fit after removing the single most deviant point until none remain above the $2.7\sigma$ threshold. We also performed as variants a single-pass, global rejection and a rejection from individual {\PL} relations, both applied at the aforementioned threshold and a larger $3.5\sigma$ threshold, as well as {\it no outlier rejection}. The results of all these variants are presented in Table~\ref{tb:h0var}. These variants of outlier rejection changed H$_0$ by less than 0.6 \hunit.  Because the outlier fraction of 2\% is quite small here and the Cepheid slope is better constrained relative to that of R11, we conclude that the outlier analysis does not warrant further consideration. The Cepheids in Table~\ref{tb:wfc3irceph} are those that passed the best-fit, global $2.7\sigma$ outlier rejection.  
    
We consider a number of variants related to the Cepheid reddening law. Besides the primary fits, which use a \citet{Fitzpatrick:1999} law with $R_V=3.3$, we also use $R_V=2.5$ and alternative formulations of the reddening law from \citet{Cardelli:1989} and \citet{Nataf:2016}. We also explore variants related to a possible break in the Cepheid {\PL} relation near 10 days. Our primary fit allows for a break or discontinuity (while not requiring one) by providing two independent slope parameters: one for Cepheids at $P\!>\!10$~d and one for $P\!<\!10$~d. The allowance for a break only increases the uncertainty in H$_0$ by 0.01 \hunit which is negligible.  We also evaluate changes in H$_0$ arising from a single-slope formulation for all periods, as well as from removing all Cepheids with $P\!<\!10$~d, or removing those with $P\!>\!60$~d as shown in Table~\ref{tb:h0var}.  Interestingly, we see no evidence of a change in slope at $P=10$~d in the $M_H^W$ \PL\ relation to a precision of 0.02 mag dex$^{-1}$ in the global fit to all Cepheids.  Hints of an increasing (LMC) or decreasing (M31) slope with period are not confirmed in this broader analysis with many more hosts. We further included variants that ignored the possibility of a Cepheid metallicity dependence and another based on a $T_e$ recalibration of nebular oxygen abundances \citep{Bresolin:2011}. We also included a variant foregoing the use of optical colors to correct for NIR reddening as it tends to be low. The results of all these variants are presented in Table~\ref{tb:h0var}.

Comparing the individual SN distances to the previously discussed approximate, independent Cepheid distances, we find none of the hosts to be an outlier. There is also no evidence ($<\!1\sigma$) for a trend between SN and Cepheid NIR distances over a 3.8~mag range in distance modulus (equivalent to a factor of 5.8 in distance). This suggests that Cepheids are not associated with significant unresolved luminosity overdensities across the range of 7--38~Mpc spanned by our sample of SN hosts and one of our anchors (NGC$\,$4258). This agrees well with \citet{Senchyna:2015}, who used {\it HST} to determine that only $\sim$ 3\% of Cepheids in M31 are in parsec-scale clusters.  Further, only a small fraction of these would alter Cepheid photometry at the resolution available from the ground or the similar resolution of {\it HST} at the distance of the SN hosts.

Lastly, we test for a dependence of the measured Cepheid distance with the level of blending by comparing the six hosts with blending higher than the inner region of NGC$\,$4258 to the remaining 13. The difference in the mean model residual distances of these two subsamples is $0.02 \pm 0.07$~mag, providing no evidence of such a dependence.

\subsection{Optical Wesenheit Period-Luminosity Relation}

The SH0ES program was designed to identify Cepheids from optical images and to observe them in the NIR with {\it F160W} to reduce systematic uncertainties related to the reddening law, its free parameters, sensitivity to metallicity, and breaks in the \PL\ relation.  However, some insights into these systematics may be garnered by replacing the NIR-based Wesenheit magnitude, $m_H^W$, with the optical version used in past studies \citep{Freedman:2001}, $m_I^W=I - R(V\!-\!I)$, where $R\equiv A_I/(A_V\!-\!A_I)$ and the value of $R$ here is $\sim 4$ times larger than in the NIR.  The advantage of this change is the increase in the sample by a little over 600 Cepheids in {\it HST} hosts owing to the greater field of view (FoV) of WFC3/UVIS.  Of these additional Cepheids, 250 come from M101,  140 from NGC$\,$4258, and the rest from the other SN hosts. In Table~\ref{tb:h0var} we give results based on Cepheid measurements of $m_I^W$ instead of $m_H^W$ for the primary fit variant with all 4 anchors, the primary fit anchor set of NGC$\,$4258, MW and LMC and for NGC$\,$4258 as the sole anchor.

The fits for all Cepheids with $m_I^W$ data generally show a significantly steeper slope for $P\!<\!10$~d than for $P\!>\!10$~d, with our preferred variant giving a highly significant slope change of $0.22\pm0.03$~mag dex$^{-1}$.  We also see strong evidence of a metallicity term with a value of $-0.20\pm0.05$~mag dex$^{-1}$ for our preferred fit, also highly significant and consistent with the value from \citet{Sakai:2004} of $-0.24\pm0.05$~mag dex$^{-1}$.  The constraint on the metallicity term is nearly unchanged when using NGC$\,$4258 as the sole anchor, $-0.19 \pm0.05$ mag dex$^{-1}$, demonstrating that the metallicity constraint comes from the metallicity gradients and SN host-to-host distance variations and {\it not} from improving the consistency in the distance scale of different anchors.

The dispersion between the individual SN and Cepheid distances (see Figure~\ref{fg:cephsndist} and the next subsection) is $\sigma = 0.12$ mag for $m_I^W$, somewhat smaller than $\sigma = 0.15$ mag from $m_H^W$.  Some reduction may be expected because a larger number of Cepheids are available in the optical relative to the NIR.  However, the SNe have a mean distance uncertainty of 0.12 mag and the sets of $m_H^W$ magnitudes in each host have a typical mean uncertainty of 0.06 mag, indicating that the dispersion between SN and Cepheid distances is already dominated by the SN error and leaving little room for improvement with additional Cepheids. The one exception is NGC$\,$4424, where the paucity of variables with valid NIR measurements results in a Cepheid-dominated calibration error which is reduced by a third by adding Cepheids only available in the optical. Based on the good agreement between the relative SN and Cepheid distances and uncertainties, we conclude that the intrinsic SN dispersion of 0.1 mag from SALT-II is reasonable.

Using the three primary anchors and the optical Wesenheit {\PL} relation, we find {H$_0$ = \hoallthreenosyswvi} (stat), extremely similar to the NIR-based result and with a {\it statistical} error just 0.05 \hunit smaller. We determined the systematic error for the optical Wesenheit from the dispersion of its variants after eliminating those expected to perform especially poorly in the optical: no allowance for reddening, no metallicity term, and no lower-period cutoff.  Even without these variants, the systematic error in the optical of {\sysuncallthreewvi} is still considerably worse than its NIR counterpart and is also larger than the statistical error.  The reason is that changes to the treatment of reddening, metallicity, {\PL} relation breaks, and outlier rejection cause larger changes in H$_0$ for the optical Wesenheit magnitudes than for the NIR counterparts.  This is a fairly uniform result, not driven by any one or two variants.  For example, changing from the preferred \citet{Fitzpatrick:1999} reddening law to the alternative formulations by \citet{Cardelli:1989} or \citet{Nataf:2016} changes H$_0$ by $0.10$ and 0.15 {\hunit} for $m_H^W$, respectively.  These same variants change H$_0$ by -2.15 and 3.82 {\hunit} for the $m_I^W$ data.  This increased sensitivity to the reddening law is a natural consequence of the larger value of $R$.  Changing the two-slope {\PL} formulation to a single slope or restricting the period range to $P\!>\!10$ or $P\!<\!60$~d changes H$_0$ by $-1.64$, $-1.24$, and $1.79$ {\hunit}, respectively, for the optical  formulation. These changes are generally smaller for the NIR Wesenheit at 0.03, -1.59, and -0.18 {\hunit}, respectively.  Finally, changing the outlier clipping from one-at-a-time to a single pass changes H$_0$ by 0.01 and $-0.90$ {\hunit} for the NIR and optical approaches, respectively.

Using the three primary anchors with the optical Wesenheit and now including systematic errors, we find H$_0$ = {\hoallthreewvi}, equivalent to an uncertainty of \uncallthreewvi. This result is somewhat less precise than the 3.3\% total error of R11, which used the NIR Wesenheit but only 8 SN-Cepheid hosts instead of the present 19. Until or unless additional studies can improve our understanding of Cepheid reddening, metallicity sensitivity, and the scale of {\PL} breaks at optical wavelengths, our analysis shows that improvements in the determination of H$_0$ via Cepheids must primarily rely on the inclusion of NIR observations.

Similar conclusions are reached when using only NGC$\,$4258 as an anchor: H$_0$ = {\hoallfnosyswvi} without systematic errors, so the statistical error is slightly better than the equivalent NIR result at \uncallfnosyswvi. However, the systematic error of {\sysuncallfwvi} is considerably worse, leading to a combined value of H$_0$ = {\hoallfwvi}. While the use of strictly optical Wesenheit magnitudes can be informative, our best results for H$_0$ with lowest systematics consistently come from using the NIR data in concert with optical observations.

\subsection{Supernova Systematics}

The SALT-II SN light-curve fits, the composition of the Hubble-flow sample, and sources of SN photometry used to determine $a_X$ in Equation~\ref{eq:ax} are described in \citet{Scolnic:2015,Scolnic:2016}.  These take advantage of the ``Supercal'' procedure \citep{Scolnic:2015} which uses reference stars in the fields of the SNe and the homogeneous set of star photometry over 3$\pi$ steradians from Pan-STARRS to remove small photometric inconsistencies between SN photometry obtained across multiple observatories and systems.  As is common in recent analyses of SN~Ia distances \citep[e.g.,][]{Betoule:2014}, to determine $a_X$ we use ``quality cuts'' to include only SN~Ia light curves for which the SALT color parameter ($c$) is within $\pm0.3$, the light-curve parameter ($x_1$) is within $\pm 3.0$ (error $<\!1.5$), the $\chi^2$ of the light-curve fit is ``good'' (fitprob $>\!0.001$), the peak time of the light curve is constrained to better than 2 days, and the uncertainty in the corrected peak magnitude is $<\!0.2$ mag.  All of the 19 calibrators pass these quality cuts as well. The SN redshifts are corrected for coherent (peculiar) flows based on density maps \citep{Scolnic:2014b,Scolnic:2015} which reduces correlated deviations from expansion caused by visible large-scale structure and empirical residuals determined from simulations \citep{Scolnic:2016}.  A residual velocity (peculiar) error of 250\,km\,s$^{-1}$ is assumed.  As a final step, we exclude SNe~Ia which deviate from the form of Equation~\ref{eq:ax} by more than $3\sigma$; this excludes 3\% of the sample for the primary fit with $0.0233\!<\!z\!<\!0.15$, leaving 217 SNe~Ia (or 281 SNe Ia for variants with $0.01\!<\!z\!<\!0.15$).  These have a dispersion of 0.128 mag around Equation~\ref{eq:ax} with a mean error of 0.129 mag and a $\chi^2$ per degree of freedom of 0.91, and yield $a_B=0.71273 \pm 0.00176$ for SALT-II ($a_V=0.7005$ for MLCS2k2). As an alternative to the SALT-II light-curve fitter, we used the MLCS2k2 fitter \citep{jha07} with a value of $R_V=2.5$ for the SN host galaxy, the same as the primary fits of R11.  The resulting value of H$_0$ is higher by 1.9 \hunit or $1.1\sigma$ of the total error as given in Table~\ref{tb:h0var}.

As in R11, we make use of several studies \citep{hicken09b,kelly10,lampeitl10,sullivan10} which have shown the existence of a small step {\it brighter} for the corrected SN magnitude for hosts more massive than $\log M_{\rm stellar} \sim 10$.  We use the same value of 0.06 mag used by \citet{Betoule:2014} for the size of the mass step to account for this effect. The net effect on H$_0$ is a small decrease of 0.7\% because of the modest difference in masses of the nearby hosts (mean $\log M_{\rm stellar}=9.8$) and of those that define the magnitude-redshift relation \citep[mean $\log M_{\rm stellar}=10.5$]{sullivan10}. We include these corrections based on host-galaxy mass in our present determination of $m_{B,i}^0$ given in Table~\ref{tb:distpar} and for $a_x$, correcting those with hosts above and below $\log M_{\rm stellar} \sim 10$ by 0.03 fainter and brighter, respectively.

An alternative host dependence on SN~Ia distance has been proposed by \citet{Rigault:2013, Rigault:2015} based on the local star-formation rate (LSFR) measured at the site of the SN.  The results from \citet{Rigault:2015} suggested a $\sim$3$\sigma$ correlation between SN distance residual and LSFR inferred from ultraviolet photometry measured with {\it GALEX} for a set of 82 SNe~Ia from \citet{Hicken:2009}, with somewhat higher significance for distances from MLCS2k2 and somewhat lower for SALT-II.  \citet{Jones:2015} repeated the LSFR analysis using a larger sample of SNe~Ia which better matched the samples and light-curve quality selection used in the cosmological analyses of R11 and \citet{Betoule:2014} as well as the more recent version of SALT II.  Using 179 {\it GALEX}-imaged SN~Ia hosts from the JLA SN sample \citep{Betoule:2014} and the Pan-STARRS sample \citep{Scolnic:2015}, or 157 used by R11, the significance of a LSFR effect diminished to $\lesssim 1\sigma$ due to two differences from \citet{Rigault:2015}: (1) the increase in the sample statistics, and (2) use of the JLA or R11 quality criteria.  Because we employ both the larger local SN sample as well as the quality cuts used by \citet{Jones:2015}, we include only the mass-based correction whose significance has remained in cosmological SN samples.  

Nevertheless, if we were to assume the existence of a LSFR (despite the preceding lack of significance), we can select a Hubble-flow sample to match the LSF of the calibrator sample and thus nullify the possible impact on H$_0$. In the calibrator sample, 17 of 19 hosts (or 89\%) are above the LSFR threshold adopted by \citet{Rigault:2013, Rigault:2015}, which is a larger fraction than the 50--60\% in the Hubble-flow sample \citep{Jones:2015}.  To determine an upper limit on a LSFR mismatch and thus H$_0$ across sets, we selected all Hubble-flow SNe significantly above the LSFR threshold (i.e., a purely LSFR sample), requiring these SNe~Ia to have good {\it GALEX} detections. By changing the Hubble flow selection, only the term $a_X$ is affected.  For this all-LSFR sample, $a_B$ is higher than for the primary fit by 0.00446 at $z\!>\!0.0233$ and lower by 0.0010 at $z\!>\!0.01$. Thus the Hubble constant from the primary fit increased by 0.8 or 0.2 {\hunit}, respectively (see Table~\ref{tb:h0var}). Thus even if a relation existed, we find that a LSFR in SN hosts would have no significant impact on the determination of H$_0$ here.

However, to address the possibility of host-galaxy dependence that arises from sample selection, we also recalculated the intercept of the Hubble-flow SNe ($a_X$) using only those found in spiral hosts. Because the 19 hosts selected for Cepheid observations were chosen on the basis of their appearance as spirals (as well as their proximity and modest inclination), this selection would be expected to match the two samples if global star formation or its history had an impact on the measured SN distance. Because the Hubble-flow sample is so much larger than the nearby sample, such a cut has a modest effect on the uncertainty in H$_0$. Doing so raised H$_0$ for the SALT-II fitter and lowered H$_0$ for MLCS2k2 each by $\sim\!0.5$\,km\,s$^{-1}$\,Mpc$^{-1}$. We note that the spiral-host sample has a mean LSFR of $-2.21$~dex, similar to the mean of the 19 calibrators at $-2.23$~dex and higher than the full Hubble-flow set of $-2.58$~dex.

We also changed the lower redshift cutoff of the Hubble diagram from $z=0.023$ to $z=0.01$, originally adopted to mitigate the impact of a possible local, coherent flow.  This raised H$_0$ by 0.2\,km\,s$^{-1}$\,Mpc$^{-1}$ for the primary fit.  Changing the deceleration parameter used to fit the SNe~Ia at $0.0233\!<\!z\!<\!0.15$ from $q_0 = -0.55$ (as expected for $\Omega_M=0.3$, $\Omega_\Lambda=0.7$) to $-0.60$ ($\Omega_M=0.27$, $\Omega_\Lambda=0.73$; or $\Omega_M=0.3$, $\Omega_{DE}=0.7$, $w=-1.05$) decreases H$_0$ by 0.2\%. More generally, an uncertainty in $\Omega_M$ of 0.02 \citep{Betoule:2014} produces an uncertainty in $q_0$ of 0.03 resulting in an uncertainty in $H_0$ of 0.1\%. As expected, the sensitivity of $H_0$ to knowledge of $q_0$ is very low as the mean SN redshift is only 0.07. As a further test, we reduced the upper range of redshifts used to measure the intercept from 0.15 to 0.07. This reduces the sample by nearly half, increases the uncertainty in the intercept by 40\% and increases the intercept and $H_0$ by 0.7\%.  We do not use this more limited redshift range because it introduces the potential for larger peculiar flows and the sensitivity to knowledge of $q_0$ is already very low.  

Two of the SNe in the calibrator sample (SN 1981B and SN 1990N) were measured before the Hubble-flow sample was acquired. Relative to the global fit, SN 1990N is faint by $0.15\pm0.14$~mag and SN 1981B is bright by $0.08\pm 0.14$~mag, so this older digital photometry does not appear to bias the value of H$_0$ in a significant way.

\begin{deluxetable}{lr}
\tablewidth{0pc}
\tablecaption{Best Estimates of H$_0$ Including Systematics\label{tb:h0}}
\tablenum{6}
\tabletypesize{\small}
\tablewidth{0pc}
\tablehead{\colhead{Anchor(s)} & \colhead{Value} \\
\multicolumn{2}{r}{[km s$^{-1}$ Mpc$^{-1}$]}}
\startdata
\multicolumn{2}{l}{One anchor} \\
\hline
NGC$\,$4258: Masers & \hofnu \\
MW: 15 Cepheid Parallaxes & \homwnu \\
LMC: 8 Late-type DEBs & \holmcnu \\
M31: 2 Early-type DEBs & \hoandnu \\
\hline
\multicolumn{2}{l}{Two anchors} \\
\hline
NGC$\,$4258 + MW & \homwfnu \\
NGC$\,$4258 + LMC & \hotrgbnu \\
\hline
\multicolumn{2}{l}{\bf Three anchors (preferred)} \\
\hline
{\bf NGC$\,$4258 + MW + LMC} & \hoallthreebfnu \\
\hline
\multicolumn{2}{l}{Four anchors} \\
\hline
NGC$\,$4258 + MW + LMC + M31 & \hoallfournu \\
\hline
\multicolumn{2}{l}{Optical only (no NIR), three anchors} \\
\hline
NGC$\,$4258 + MW + LMC & \hoallthreewvinu
\enddata
\end{deluxetable}

\begin{deluxetable}{llrrrrr}
\tablecaption{H$_0$ Error Budget for Cepheid and SN~Ia Distance Ladders$^*$\label{tb:h0unc}}
\tabletypesize{\small}
\tablewidth{0pc}
\tablenum{7}
\tablehead{\colhead{Term} & \colhead{Description} & \multicolumn{1}{c}{Prev.} & \multicolumn{1}{c}{R09}  & \multicolumn{1}{c}{R11} & \multicolumn{2}{c}{This work}\\
\colhead{} & \colhead{} & \colhead{LMC} & \colhead{N4258} & \colhead{All 3} & \colhead{N4258} & \colhead{All 3}}
\startdata
$\sigma_{\rm anchor}$  &   Anchor distance, mean  &  5\%  & 3\% & 1.3\% & 2.6\% &  1.3\% \\
$\sigma_{{\rm anchor PL}}^a$  &  Mean of {\PL} in anchor & 2.5\%  &  1.5\% & 0.8\% & 1.2\% & 0.7\% \\
$\sigma_{{\rm host PL}}/\sqrt{n}$  &  Mean of {\PL} values in SN~Ia hosts  & 1.5\% & 1.5\% &  0.6\% & 0.4\% & 0.4\% \\
$\sigma_{\rm SN}/\sqrt{n}$  &  Mean of SN~Ia calibrators &  2.5\% & 2.5\% & 1.9\% & 1.2\% &  1.2\% \\  
$\sigma_{m-z}$  &  SN~Ia $m$--$z$ relation & 1\% & 0.5\%  & 0.5\% & 0.4\% & 0.4\% \\
$R\sigma_{\rm zp}$  & Cepheid reddening \& colors, anchor-to-hosts & 4.5\% & 0.3\% & 1.4\% & 0\% & 0.3\% \\
$\sigma_{Z}$ & Cepheid metallicity, anchor-to-hosts  & 3\% & 1.1\%   & 1.0\%  & 0.0\% & 0.5\% \\
$\sigma_{\rm PL}$ & {\PL} slope, $\Delta$ log\,$P$, anchor-to-hosts & 4\% & 0.5\%  &  0.6\% & 0.2\% & 0.5\% \\
$\sigma_{\rm WFPC2}$ & WFPC2 CTE, long-short & 3\% & N/A & N/A & N/A & N/A \\
\hline
\multicolumn{2}{l}{subtotal, $\sigma_{{\rm H}_0}^b$} &     10\%  &  4.7\% & 2.9\% & 3.3\%$^c$ & 2.2\% \\
\hline
\multicolumn{2}{l}{Analysis Systematics} & N/A & 1.3\% & 1.0\% & 1.2\% & 1.0\% \\
 \hline
\multicolumn{2}{l}{Total, $\sigma_{{\rm H}_0}$} &     10\%  &  4.8\% & 3.3\% & 3.5\% & 2.4\% 
\enddata
\tablecomments{(*) Derived from diagonal elements of the covariance matrix propagated via the error matrices associated with Equations 1, 3, 7, and 8. (a) For MW parallax, this term is already included with the term above. (b) For R09, R11, and this work, calculated with covariance included. (c) One anchor not included in R11 estimate of $\sigma_{{\rm H}_0}$ to provide a crosscheck.}
\end{deluxetable}

A budget for the sources of uncertainty in the determination of H$_0$ is given in Table~\ref{tb:h0unc}. These are necessarily marginalized approximations, as they do not show the (small) covariance between terms included in the full global fit. 

\begin{figure}[ht]
\vspace*{150mm}
\includegraphics{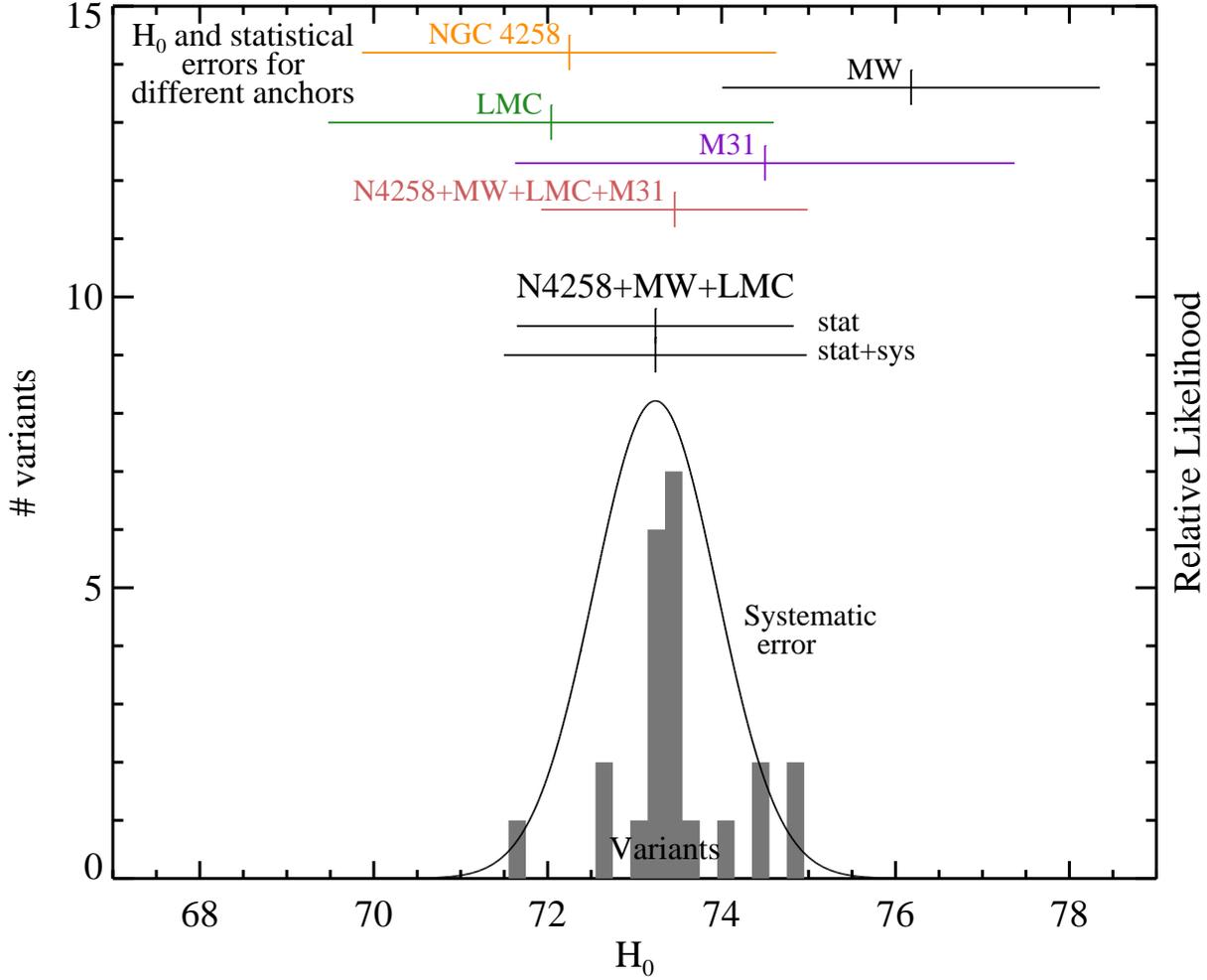}
\caption{\label{fg:h0hist} Determination of systematic errors in H$_0$ for the set of anchors used in the primary fit (N4258, MW \& LMC).  By varying factors outside the global fit and its parameters such as the assumed reddening law, its parameters, the presence of a metallicity dependence, the presence of breaks in the \PL\ relations, selection of SN light curve fitter, morphology or local star formation rate of hosts, etc.  We derive a systematic error from a Gaussian fit to the variants.  This error is smaller than the indicated statistical errors.}
\end{figure}

Our systematic error is estimated based on the variations in H$_0$ resulting from the reasonable, alternative fits.  These alternatives are, by their nature, difficult to formally include in the global fit. All of the discussed NIR variants, 207 in total including combinations of anchors, are listed in Table~\ref{tb:h0var}. As shown in Figure~\ref{fg:h0hist}, the histogram for the primary-fit anchors (NGC$\,$4258, MW, and LMC) is well fit by a Gaussian distribution with $\sigma =$ {\sysuncallthreenop} {\hunit}, a systematic uncertainty that is a little less than half of the statistical error. None of the variants is a noteworthy outlier from this distribution. The complete error in $H_0$ using multiple anchors can be traced to the quadrature sum of three terms, the two independent terms in equation 9 and the systematic error.  The error in $M^0_X$ for any variant, derived from the global fit, is given in third column of Table 8 and the error in the intercept, $a_X$, was given in \S 3.

Including the systematic error, we arrive at a complete result of H$_0$ = {\hoallthree}, corresponding to a total uncertainty, combining statistical and systematic contributions, of \uncallthree. The two largest remaining uncertainties are the mean geometric distance calibration (1.3\%) and the mean of the 19 SN Ia calibrators (1.2\%).
    
\section{Discussion}

Our primary fit of H$_0$ = {\hoallthree} is \PlanckdiffTTTEEE $\sigma$ higher than the value of $66.93\pm0.62$~{\hunit} predicted by \citet{Planck:2016} based on $\Lambda$CDM with 3 neutrino flavors having a mass of 0.06\,eV and the {\it Planck} CMB data (TT,TE,EE+SIMlow; \Planckdiff $\sigma$ for TT+SIMlow).  Assuming the \PlanckdiffTTTEEE $\sigma$ difference is not a fluke (99.9\% confidence), possible explanations include systematic errors in the local H$_0$ or CMB measurements, or an unexpected feature in the cosmological model that connects them.  Previous indications of $\sim$ $2\sigma$ tension from the less-precise measurements of H$_0$ and the CMB \citep{Riess:2011,Planck:2014} elicited a number of new studies, many of which were addressed above and helped improve the present analysis.

The analysis of the R11 dataset by \citet{Efstathiou:2014} yielded a value of H$_0 = 72.5 \pm 2.5$ \hunit, similar to the primary result of $73.0 \pm 2.4$ \hunit found by R11 using the same three anchors (MW, LMC, and NGC$\,$4258, including the same H13 distance for NGC$\,$4258 for both) and resulting in a 1.9$\sigma$ tension with {\it Planck} and $\Lambda$CDM.  \citet{Efstathiou:2014} also found H$_0 = 70.6 \pm 3.3$ \hunit with NGC$\,$4258 as the {\it only} anchor, and the {\it Planck} team adopted this value instead of the three anchor result with its reduced precision and tension.  The main difference in the analysis with R11 was the use of a global instead of \PL-specific outlier rejection.  Our use here of {\it F814W}$\!-\!${\it F160W} colors to identify blends as discussed in \S4.1 has significantly reduced the need for outlier rejection, and we have adopted a global outlier rejection for the 2\% that remain.  The internal model constraints on the slope and metallicity parameters have also improved substantially over the R11 dataset with no need for the priors set by R11 or \citet{Efstathiou:2014}.   We find the difference in H$_0$ between the use of three anchors and just NGC$\,$4258 to be 1.0 \hunit, less than the 1.9 \hunit found by \citet{Efstathiou:2014} with the R11 dataset, a consequence of the tightened constraints on the Cepheid relations, and we conclude that use of the three anchors provides our best determination of H$_0$.   

In the previous section we addressed systematic errors related to Cepheids and SNe used in our determination of H$_0$.  A third component comes from our use of geometric distances to calibrate Cepheids.  We used four sets: masers in NGC$\,$4258, parallaxes to MW Cepheids, DEBs in the LMC, and DEBs in M31.  The four values of H$_0$ using each as the sole anchor (see Table~\ref{tb:h0}) are in good relative agreement, with none more than $1.5\sigma$ from the primary fit considering only their mean geometric distance error of 2.8\%.  Thus we see no basis for excluding any of these four as outliers.  Among the four, NGC$\,$4258 has the advantage of a Cepheid sample with mean period closer to those in the SN hosts and with all their photometry on the same {\it HST} system.

However, in our analyses we parameterize the difference in zeropoints for non-{\it HST} data, and the {\it a posteriori} result of 0.013 mag for the primary fit is well below the estimated {\it a priori} constraint of $\sigma_{\rm zp}=0.03$~mag, indicating no unexpected inconsistency with zeropoints.  Our use of Cepheid samples in M31 and the LMC, which sample the short- and long-period range as well as the allowance in the fits for a {\PL} break, strictly limits the impact of a difference in sample mean periods on H$_0$. The residuals among the anchor distances for our primary fit are $\Delta \mu_{\rm N4258}=-0.043$, within the range of its 0.0568 mag prior, and $\Delta \mu_{\rm LMC}=-0.042$, within range of its 0.0452 mag prior. To be conservative, we removed M31 from the anchor set of our primary fit for the reasons discussed in \S3.1.3 ---  but we discourage any {\it additional} winnowing or editing of the anchor set as it is unwarranted by the data and is likely to give a false sense of reducing the tension merely by inflating the present uncertainties.

\begin{figure}[ht]
\vspace*{150mm}
\includegraphics{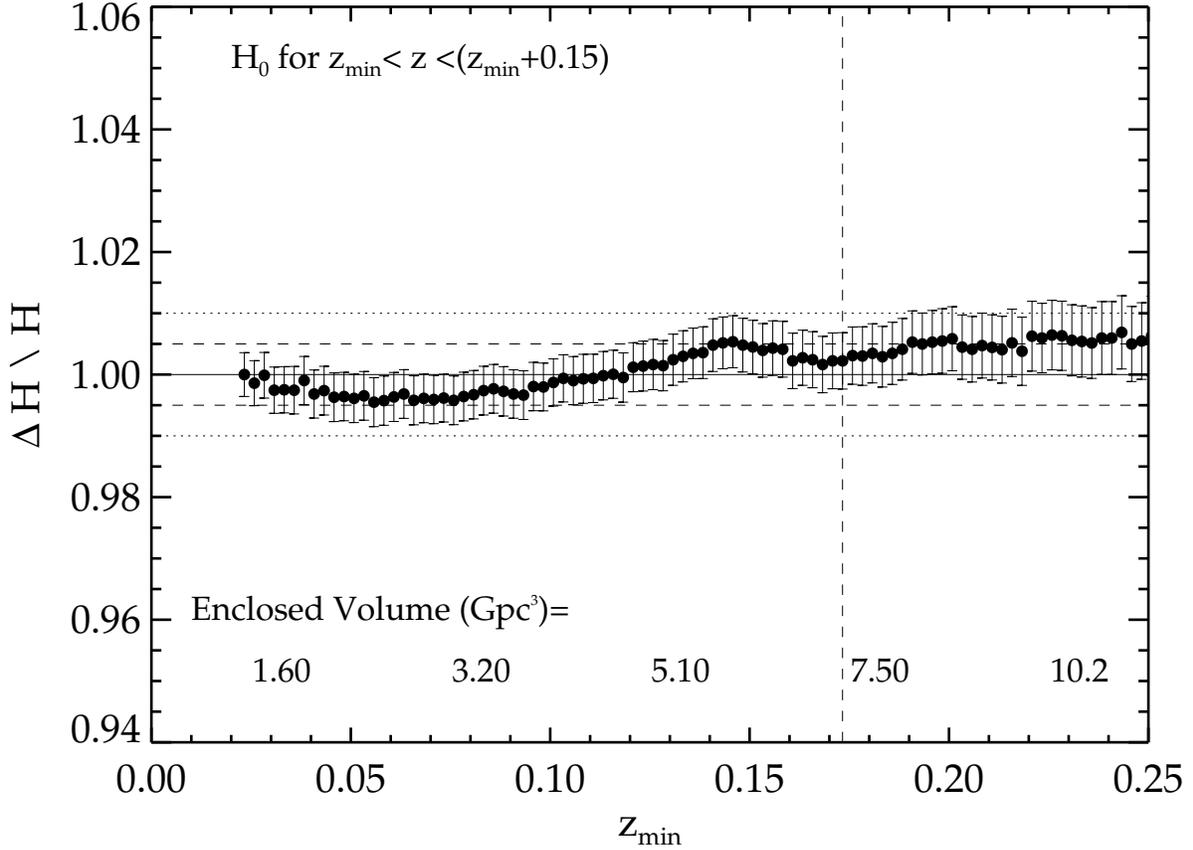}
\caption{\label{fg:h0zmin} Fractional variation in H$_0$ resulting from a progressively higher redshift (lower cosmic variance) range used to measure the Hubble expansion, $z_{\rm min} < z < z_{\rm min}+0.15$.  Empirically increasing $z_{\rm min}$ from 0.0233 (primary fit) to 0.25 and the maximum redshift from 0.15 (primary fit) to 0.40 produces variations consistent with the measurement uncertainty of $\pm$ 0.004-0.006 and the simulated uncertainty of $\pm0.0027$ (intrinsic) from \citet{Odderskov:2016}.  Thus a difference between the local and global H$_0$ of even $\sim 1$\%  is exceedingly unlikely.}
\end{figure}

We may consider whether the local determination of H$_0$ is different than the global (i.e., cosmological) value. In a homogeneous and isotropic Universe the two have the same expectation value.  However, we live in an unusual place (a dark matter halo), and the inhomogeneity of matter on our measurement scale could lead to important variations in H$_0$.  We currently account for flows induced by visible structures using host redshift corrections derived from a map of the matter density field calibrated by the 2M++ catalogue (with a light-to-matter bias parameter of $\beta=0.43$ and a dipole from \citealt{Carrick:2015}).  This produces a small net increase in H$_0$ of a few tenths of a percent over the case of uncorrelated velocities at rest with respect to the CMB as discussed by R11.  We also account for the cosmological change in expansion rate using $q_0$ and $j_0$ as discussed in \S4.2. Because the Hubble diagram of SNe~Ia is continuously sampled from $z=0.01$ to $z=2$, a percent-level change in the local expansion rate at $z\!>\!0.15$ would be empirically evident in the distance residuals. In Figure~\ref{fg:h0zmin} we show the relative change in H$_0$ starting at $0.0233\!<\!z\!<\!0.15$ and decreasing the influence of the local volume by gradually increasing the redshift cutoffs for determining H$_0$.  As shown, the value of H$_0$ never changes by more than 1.3 times the statistical uncertainty in the fit of the intercept over a $\Delta z$ range of 0.2 and a factor of 5 increase in volume.  

\citet{Odderskov:2016} {\it simulated} the effect of inhomogeneities on the local value of H$_0$ using mock sources in N-body simulations using the GADGET code with a box size of 700 Mpc and $512^3$ dark matter particles with cosmological parameters in agreement with \citet{Planck:2014} from $z=50$ to the present.  In the simulation, halos are resolved using the halo-finder ROCKSTAR and realistic SN sampling is obtained from the redshift distribution of the samples with $0.01\!<\!z\!<\!0.1$.  Cosmic variance is taken into account by varying the location of the observer.  The uncertainty in the local measurement of H$_0$ is found to be 0.27\% for the case of a typical SN sample, observer in a Local Group halo and the maximum redshift of $z=0.15$ for our primary fit (I.~Odderskov, priv. comm.). This analysis is in good agreement with our empirical result of 0.4\% uncertainty in $H_0$, which shows that such convergence to the sub-percent level has occurred within the SN sample at $z\!<\!0.15$.  We conclude that the uncertainty in H$_0$ owing to inhomogeneities is adequately taken into account by the procedure of empirically correcting the redshifts for expected flows, testing for convergence of H$_0$ on large scales, and comparing the propagated uncertainty to simulations.  A difference in H$_0$ at even the $>\!1$\% level caused by inhomogeneities would be triple the empirical or theoretical uncertainty and thus appears exceedingly unlikely.

Could the difference result from a systematic error in the {\it Planck} measurement? To explore this possibility, we consider an independent set of CMB data from the combination of WMAP9, ACT, and SPT observations.  Based on the analysis by \citet{Calabrese:2013} using $\Lambda$CDM but including the same neutrino mass of 0.06\,eV used in the {\it Planck} analysis yields H$_0 = 70.9\pm1.6$ \hunit, a difference from our local measurement of \WMAPdiff $\sigma$ and thus quite consistent.  While some of the improved agreement comes from the lower precision of this CMB dataset, most comes from a change in the central value of H$_0$ itself; the WMAP9+ACT+SPT value, even with the uncertainty of the {\it Planck} data, would still be consistent at the 1.3$\sigma$ level.  The difference in CMB datasets appears to play some role in the perceived tension with the local value of H$_0$.  \citet{Addison:2015} has reported a parallel 2.5$\sigma$ tension (size and significance) internally within the {\it Planck} data based on H$_0$ parameters determined from multipoles with $l\!<\!1000$ and $l\!>\!1000$, with the two halves of the data producing H$_0 = 69.7 \pm 1.7$ and $64.1 \pm 1.7$ \hunit, respectively.  Considering the two {\it Planck} halves with the R11 measurement of H$_0$, BAO, WMAP9, and SPT, \citet{Addison:2015} finds 5 of the 6 consistent with H$_0 \approx 70$ \hunit, with only the {\it Planck} $l\!>\!1000$ data pulling toward significantly lower values. Because the SPT and {\it Planck} $l\!>\!1000$ data cover similar ranges in $l$, their disagreement should be independent of the cosmological model and thus could indicate the presence of a systematic error and a role in the present tension with local H$_0$ measurements. 

However, some degree of the previous tension remains, even without {\it Planck}, after including other datasets explicitly to constrain the cosmological model. \citet{Bennett:2014} used WMAP9+ACT+SPT with BAO from BOSS DR11 and 6dFGS (their Table 2, column G) and find H$_0 = 69.3 \pm 0.7$ {\hunit}, which has a \WMAPBAOdiff$\sigma$ tension with our determination of H$_0$ (and a $2\sigma$ tension with {\it Planck} in the other direction). A lower value of $68.1\pm0.7$ is given by \citet{Aubourg:2015} for WMAP9, BAO, and high-redshift SNe --- but this neglects SPT, which pulls toward higher H$_0$ \citep{Addison:2015,Story:2013}. More direct comparisons and analyses of CMB data may be expected to resolve the tensions between them and the local value of H$_0$.  

\begin{figure}[ht]
\vspace*{150mm}
\includegraphics{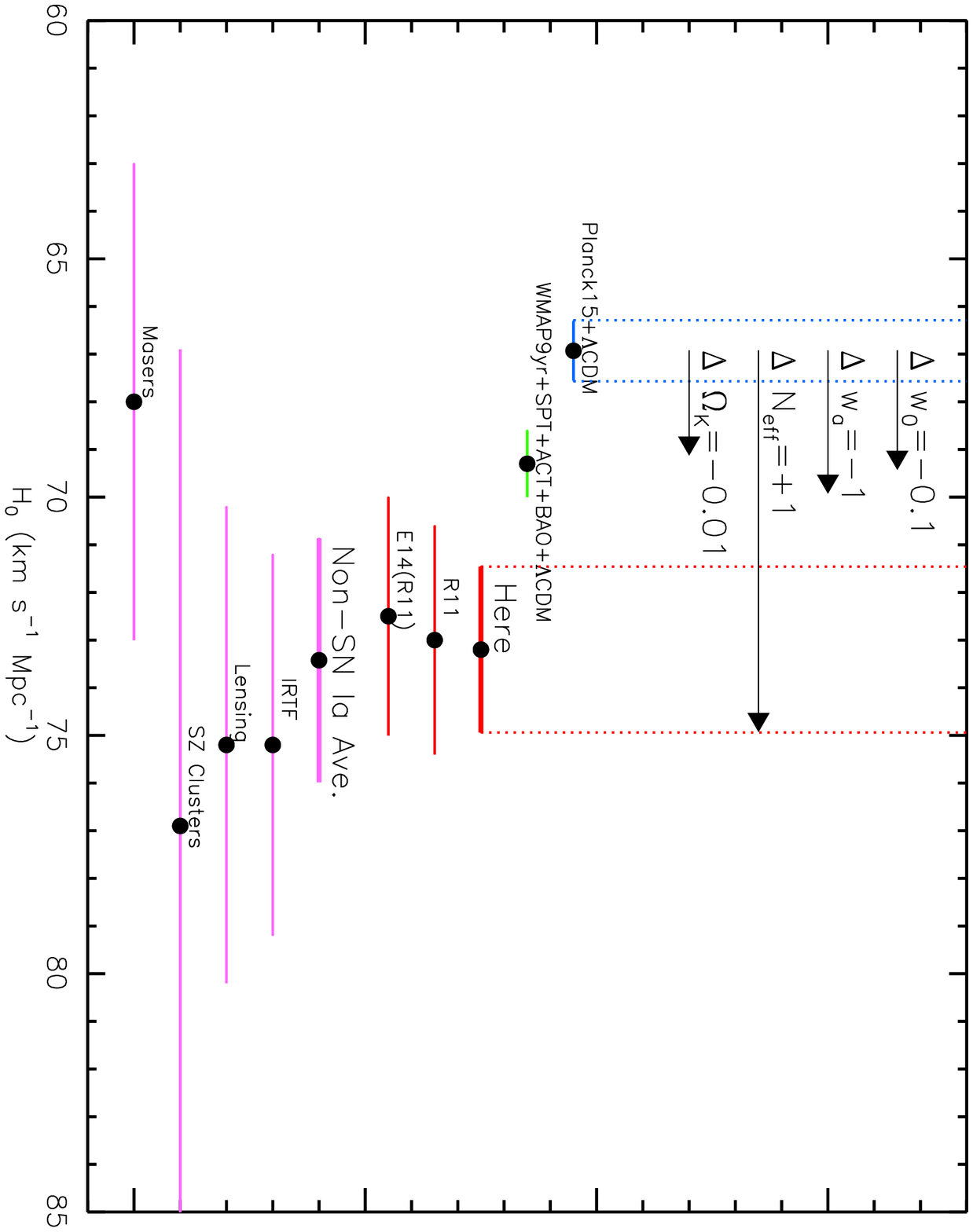}
\caption{\label{fg:h0oth} Local measurements of H$_0$ compared to values predicted by CMB data in conjunction with $\Lambda$CDM. We show 4 SN Ia-independent values selected for comparison by \citet{Planck:2014} and their average, the primary fit from R11, its reanalysis by \citet{Efstathiou:2014} and the results presented here. The \PlanckdiffTTTEEE $\sigma$ difference between {\it Planck}+$\Lambda$CDM \citep{Planck:2016} and our result motivates the exploration of extensions to $\Lambda$CDM.}
\end{figure}

It is useful to compare our result with recent measurements of the local Hubble constant which are {\it independent of SNe Ia}\footnote{Other measurements of H$_0$ which also utilize SNe~Ia do not provide a very meaningful comparison to ours because they are based on far fewer reliable SN Ia calibrators than the 19 presented here as discussed in \S1.1.} and which appear to support a $\sim 5\%$ measurement.  To avoid our own biases in identifying these we use current results from the four SN-independent projects shown in Figure 16 of \citet{Planck:2014}: IR Tully-Fisher from \citet{Sorce:2012}, 2 strong lenses from \citet{Suyu:2013}, 4 distant maser systems from \citet{Gao:2016}, and 38 SZ clusters from \citet{Bonamente:2006}. These are plotted in Figure~\ref{fg:h0oth}. A simple weighted average of these SN-independent measurements gives H$_0=73.4 \pm 2.6$ \hunit, nearly the same as our primary fit though with a 45\% larger uncertainty.  The most precise of these is from the analysis of two strong gravitational lenses and yields H$_0 = 75\pm 4$ {\hunit} \citep{Suyu:2013}, a result that is both independent of ours and has been reaffirmed by an independent lensing analysis \citep{Birrer:2015}. However, we note that while lensing provides an independent, absolute scale, the transformation to $H_0$ depends on knowledge of $H(z)$ between $z=0$ and the redshifts of the two lenses (z=0.295 and z=0.631) which may be gathered from parameter constraints from the CMB or from an empirical distance ladder across this redshift range.  Either approach will add significantly to the overall uncertainty. Given the breadth of evidence that the local measurement of H$_0$ is higher than that inferred from the CMB and $\Lambda$CDM it is worthwhile to explore possible cosmological origins for the discrepancy. 
  
\begin{figure}[ht]
\vspace*{150mm}
\includegraphics{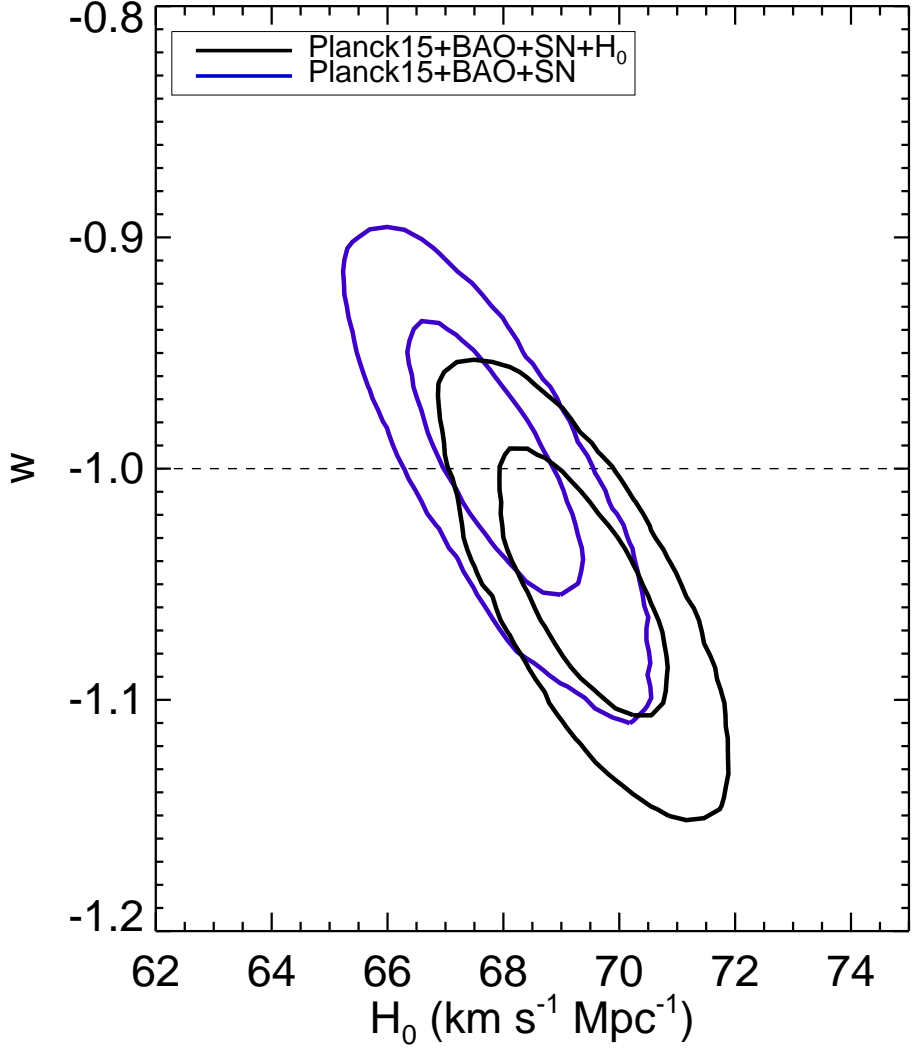}
\caption{\label{fg:h0w} Confidence regions determined with CosmoMC based on the data from Planck (TT+TEB+lensing), BAO including Ly$\alpha$ QSOs, the JLA SN sample \citep{Betoule:2014} and with and without our determination of H$_0$ for the $w$CDM cosmological model.  As shown there is a degeneracy between $w$ and H$_0$ and the local measurement of H$_0$ pulls the solution to a lower value of $w$ though it is still consistent with -1.}
\end{figure}

We may consider the simplest extensions of $\Lambda$CDM which could explain a difference between a local and cosmological Hubble constant of $\sim 4$--6 {\hunit}. We are not the first to look for such a resolution, though the roster of datasets examined has varied substantially and evolves as measurements improve \citep{Wyman:2014,Leistedt:2014, Aubourg:2015, Cuesta:2015,Dvorkin:2014}. The simplest parameterizations of dark energy with $w_0\!<\!-1$ or with $w_0\!>\!-1$ and $w_a\!<\!0$ can alleviate but not fully remove tension with H$_0$ (see Figure~\ref{fg:h0oth}) due to support for $w(z) \sim -1$  signal from high-redshift SNe Ia and BAO \citep[][see Figure~\ref{fg:h0w}]{Cuesta:2015, Aubourg:2015}. A very recent ($z\!<\!0.03$) and dramatic decrease in $w$ or an episode of strong dark energy at $3\!<\!z\!<\!1000$ may evade detection and still produce a high value of H$_0$.  Whether such a model creates additional tensions will depend on its prescription and still, if empirically motivated, is likely to suffer from extreme fine-tuning.

A synthesis of the studies cited above indicates a more fruitful avenue is found in the ``dark radiation'' sector. An increase in the number of relativistic species (dark radiation; e.g., neutrinos) in the early Universe increases the radiation density and expansion rate during the radiation-dominated era, shifting the epoch of matter-radiation equality to earlier times. The resulting reduction in size of the sound horizon (which is used as a standard ruler for the CMB and BAO) by a few percent for one additional species (N$_{\rm eff}=4$) increases H$_0$ by about 7 {\hunit} for a flat Universe, more than enough to bridge the divide between the local and high-redshift scales.  A fractional increase (i.e., less than unity) is also quite plausible for neutrinos of differing temperatures or massless bosons decoupling before muon annihilation in the early Universe \citep[e.g., Goldstone bosons;][]{Weinberg:2013}, producing $\Delta$N$_{\rm eff}=0.39$ or $0.57$ depending on the decoupling temperature.  An example of such a fit comes from \citet{Aubourg:2015} using a comprehensive set of BAO measurements and {\it Planck} data, finding N$_{\rm eff}=3.43 \pm 0.26$ and H$_0 = 71 \pm 1.7$ {\hunit}.  A similar result from WMAP9+SPT+ACT+SN+BAO gives N$_{\rm eff}=3.61 \pm 0.6$ and H$_0 = 71.8 \pm 3.1$ {\hunit} \citep{Hinshaw:2013}. Thus, a value of $\Delta$N$_{\rm eff}$ in the range 0.4--1.0 would relieve some or all of the tension.  Although fits to the {\it Planck} dataset \citep{Planck:2015} do not indicate the presence of such additional radiation, they do not exclude this full range either.  

Allowing the N$_{\rm eff}$ degree of freedom triples the uncertainty in the cosmological value of H$_0$ from \citet{Planck:2015}, BAO and high-redshift SNe and modestly raises its value to H$_0=68\pm1.6$ \hunit, reducing the tension to 2.1$\sigma$ and demonstrating that a local measurement of H$_0$ appears to offer a powerful aid to determining N$_{\rm eff}$. A cosmologically constrained value of $\Delta N_{\rm eff}$ can be used to diagnose the nature of the new particle and its decoupling temperature \citep{Brust:2013}.

\begin{figure}[ht]
\vspace*{150mm}
\includegraphics{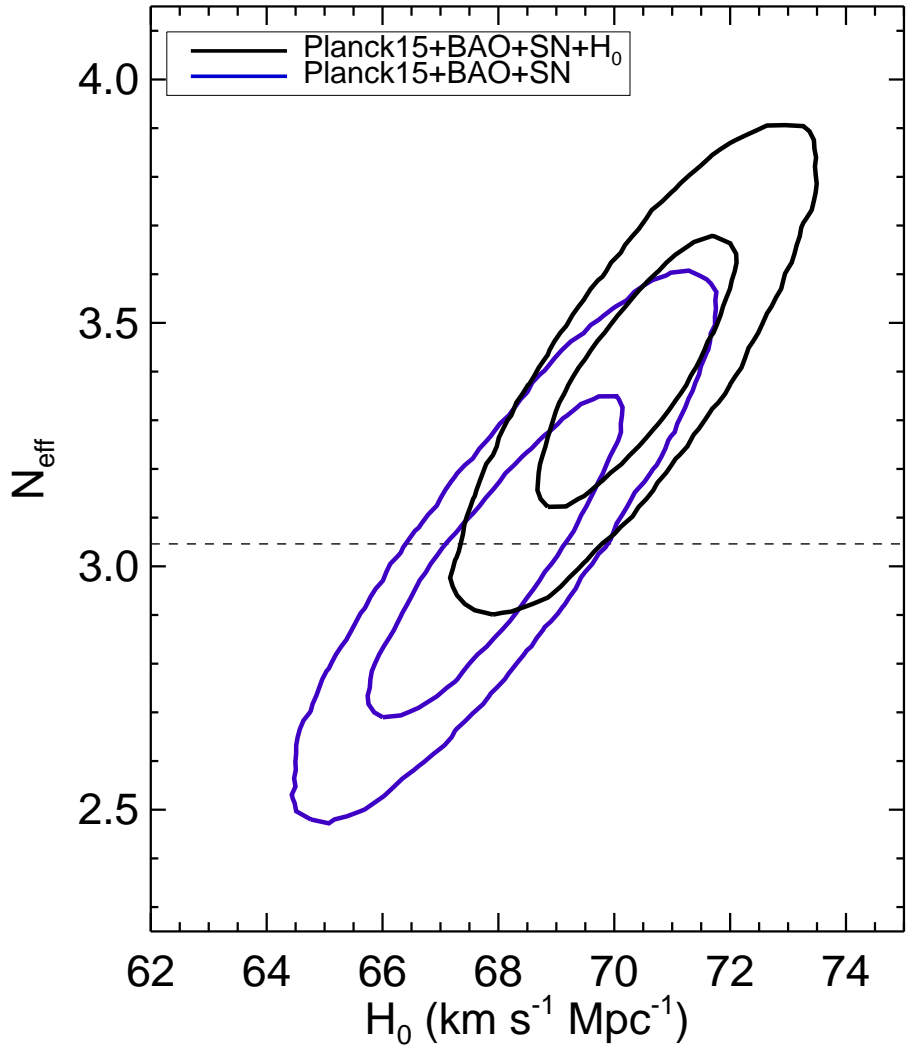}
\caption{\label{fg:h0neff} Same as Figure~\ref{fg:h0w} but for the N$_{\rm eff}$CDM model.  The local measurement of H$_0$ pulls the solution towards N$_{\rm eff}>3.046$ which also provides a marginally better fit to the full data set than $\Lambda$CDM.}
\end{figure}

Including the present measurement of H$_0$ with the {\it Planck} \citep{Planck:2015} data (including lensing), the full BAO set of measurements (including the Lyman-alpha QSO's) and the \citet{Betoule:2014} SN sample pulls N$_{\rm eff}$ higher to a value of $3.41 \pm 0.22$ (and H$_0=70.4 \pm 1.2$ \hunit), a result favoring (though not requiring) additional dark radiation. This fit provides the lowest value of the best-fit log likelihood among standard extensions to $\Lambda$CDM we considered (lower than $\Lambda$CDM by $\sim$ 2) and the result is shown in Figure~\ref{fg:h0neff}. If {\it Planck} CMB, BAO, SN, and H$_0$ data are taken at face value, this extension of $\Lambda$CDM remains an intriguing avenue toward their resolution and highlights the need for additional improvements in local determinations of H$_0$.  More broadly, the present discrepancy in the measured Hubble constant may provide a clue to one of the many enigmas contained in the 95\% of the Universe within the dark sector.

\begin{figure}[ht]
\vspace*{150mm}
\includegraphics{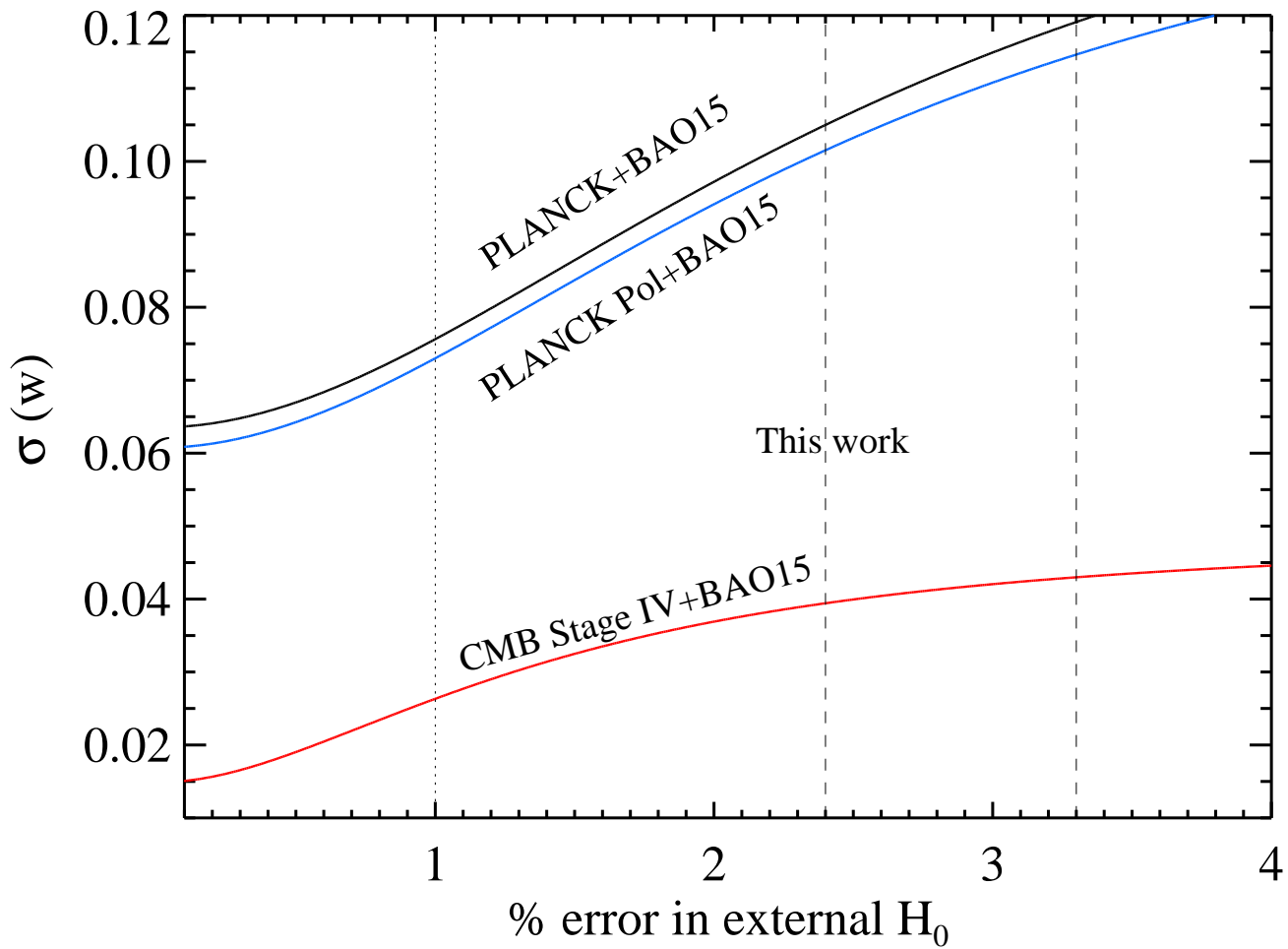}
\caption{\label{fg:h0fut} Constraint in the dark energy equation of state as a function of the precision of the local determination of the Hubble constant.  Past and current precision is indicated as well as a future goal of 1\%.}
\end{figure}

Fortunately, the prospects for near-term improvements in the local determination of the Hubble constant are quite promising.  We have begun obtaining a new sample of parallax measurements of long-period MW Cepheids using the spatial scanning technique with WFC3 on {\it HST} \citep{Riess:2014,Casertano:2015}.  These improvements alone would reduce the total uncertainty in H$_0$ to $\sim$ 1.8\% based on the terms in Table~\ref{tb:h0unc}. In a parallel effort, we are obtaining spatial-scan photometry of a larger sample of MW Cepheids slated for even higher-precision {\it Gaia} parallax determinations in a few years.   With additional progress from this route and others, the goal of 1\% \citep{Suyu:2012} is not far-fetched and has the potential, in concert with Stage-IV CMB experiments (see Figure~\ref{fg:h0fut}), to provide new leverage on the dark Universe.
 
\section{Acknowledgements}

We thank Alessandro Manzotti, Ariel Goobar, Mike Hudson, WeiKang Zheng, Bill Januszewski, Robert Kirshner, Licia Verde, Liz Humphreys, Dan Shafer, the PHAT collaboration, and Peter Stetson for valuable discussions and other contributions, as well as Doug Welch for providing the optimal temporal spacings used in the optical Cepheid searches. We are grateful to Melissa L.~Graham for assistance with the Keck observations. 

This research was supported by NASA/{\it HST} grants GO-12879, 12880, 13334, 13335 \& 13344 from the Space Telescope Science Institute, which is operated by the Association of Universities for Research in Astronomy, Inc., under NASA contract NAS5-26555. L.M.M.'s group at Texas A\&M University acknowledges additional support from the Mitchell Institute for Fundamental Physics \& Astronomy. D.S.~acknowledges support from the Kavli Institute for Cosmological Physics at the University of Chicago through grant NSF PHY-1125897 and an endowment from the Kavli Foundation and its founder Fred Kavli. A.V.F.'s group at UC Berkeley is also grateful for financial assistance from NSF grant AST-1211916, the TABASGO Foundation, Gary and Cynthia Bengier, and the Christopher R. Redlich Fund. P.C. is supported by NSF grant AST-1516854 to the Harvard College Observatory. J.M.S.~is supported by an NSF Astronomy and Astrophysics Postdoctoral Fellowship under under award AST-1302771. R.J.F.~gratefully acknowledges support from the Alfred P.~Sloan Foundation.

Some of the results presented herein are based on data obtained at the W.~M.~Keck Observatory, which is operated as a scientific partnership among the California Institute of Technology, the University of California, and NASA; the observatory was made possible by the generous financial support of the W.~M.~Keck Foundation. We thank the Keck staff for their expert help.

\appendix

\section{Setup of System of Equations}

Equations~\ref{eq:wh} through \ref{eq:snmagalt} describe the relationships between the measurements and parameters with additional constraint equations given in \S3. To improve clarity we explicitly show the system of equations we solve to derive the value of $M^0_B$ which together with the independent determination of $a_B$ provides the measurement of H$_0$ via Equation~\ref{eq:h0alt}. Here we refer to the vector of measurements as ${\bf y}$, the free parameters as ${\bf q}$, the equation (or design) matrix as ${\bf L}$, and the error matrix as ${\bf C}$ with $\chi^2=(y-Lq)^TC^{-1}(y-Lq)$ and maximum likelihood parameters given as $q_{best}=(L^TC^{-1}L)^{-1}L^TC^{-1}y$ and covariance matrix $(L^TC^{-1}L)^{-1}$. For the primary fit which uses 3 anchors, NGC$\,$4258, Milky Way parallaxes, and LMC DEBs we arrange ${\bf L}$, ${\bf C}$ and ${\bf q}$ as given below so that some terms are fully correlated across a set of measurements like the anchor distances for NGC$\,$4258 and the LMC and ground-to-HST zeropoint errors are fully correlated and others like the MW parallax distances are not.  

$$y=\begin{pmatrix} m^W_{H,1,j} \\ .. \\  m^W_{H,19,j} \\  m^W_{H,j,N4258}-\mu_{0,N4258} \\ m^W_{H,M31,j} \\ m^W_{H,MW,j}-\mu_{\pi,j} \\  m^W_{H,LMC,j}-\mu_{0,LMC}  \\ 
m_{B,1}^0 \\ .. \\ m_{B,19}^0 \\ 0 \\ 0 \\ 0 \end{pmatrix}$$

$$l=\begin{pmatrix}
\sy{1} &\td & \sy{0} & \sy{0} & \sy{1} & \sy{0} & \sy{0} &\lp^h_{19,1}/\sy{0}    & \sy{0} & \oh_{19,1}    & \sy{0} & \lp^l_{19,1}/\sy{0}    \\  
   \td &\td & \td    & \td    & \td    & \td    & \td    &\td                    & \td    & \td           & \td    & \td                    \\
\sy{0} &\td & \sy{1} & \sy{0} & \sy{1} & \sy{0} & \sy{0} &\lp^h_{19,j}/\sy{0}    & \sy{0} & \oh_{19,j}    & \sy{0} & \lp^l_{19,j}/\sy{0}    \\  
\sy{0} &\td & \sy{0} & \sy{1} & \sy{1} & \sy{0} & \sy{0} &\lp^h_{N4258,j}/\sy{0} & \sy{0} & \oh_{N4258,j} & \sy{0} & \lp^l_{N4258,j}/\sy{0} \\  
\sy{0} &\td & \sy{0} & \sy{0} & \sy{1} & \sy{0} & \sy{1} &\lp^h_{M31,j}/\sy{0}   & \sy{0} & \oh_{M31,j}   & \sy{0} & \lp^l_{M31,j}/\sy{0}   \\  
\sy{0} &\td & \sy{0} & \sy{0} & \sy{1} & \sy{0} & \sy{0} &\lp^h_{MW,j}/\sy{0}    & \sy{0} & \oh_{MW,j}    & \sy{1} & \lp^l_{MW,j}/\sy{0}    \\
\sy{0} &\td & \sy{0} & \sy{0} & \sy{1} & \sy{1} & \sy{0} &\lp^h_{LMC,j}/\sy{0}   & \sy{0} & \oh_{MW,j}    & \sy{1} & \lp^l_{LMC,j}/\sy{0}   \\
\sy{1} &\td & \sy{0} & \sy{0} & \sy{0} & \sy{0} & \sy{0} &\sy{0}                 & \sy{1} & \sy{0}        & \sy{0} & \sy{0}                 \\
\sy{0} &\td & \sy{1} & \sy{0} & \sy{0} & \sy{0} & \sy{0} &\sy{0}                 & \sy{1} & \sy{0}        & \sy{0} & \sy{0}                 \\
\sy{0} &\td & \sy{0} & \sy{0} & \sy{0} & \sy{0} & \sy{0} &\sy{0}                 & \sy{0} & \sy{0}        & \sy{1} & \sy{0}                 \\ 
\sy{0} &\td & \sy{0} & \sy{1} & \sy{0} & \sy{0} & \sy{0} &\sy{0}                 & \sy{0} & \sy{0}        & \sy{0} & \sy{0}                 \\ 
\sy{0} &\td & \sy{0} & \sy{0} & \sy{0} & \sy{1} & \sy{0} &\sy{0}                 & \sy{0} & \sy{0}        & \sy{0} & \sy{0}                 
\end{pmatrix}$$
        
$$q=\begin{pmatrix} \mu_{0,1} \\\td \\  \mu_{0,19} \\  \Delta \mu_{\rm N4258} \\ M_{H,1}^W\! \\ \Delta \mu_{\rm LMC} \\ \mu_{M31} \\ b \\ M_B^0\! \\ Z_W \\ \Delta {\rm zp} \\ b_l \end{pmatrix}$$
          
$$C\!=\!\begin{pmatrix}
\sy{\sigma_{{\rm tot},1,j}^2}\!\!\!\! &\td & \sy{0} & \sy{0} & \sy{0} & \sy{0} & \sy{0} & \sy{0} &\td & \sy{0} & \sy{0} & \sy{0} & \sy{0} \\
\td &\td & \td &\td &\td &\td &\td &\td &\td &\td &\td &\td &\td \\
\sy{0} &\td & \sy{\sigma_{{\rm tot},19,j}^2}\!\!\!\! & \sy{0} & \sy{0} & \sy{0} & \sy{0} & \sy{0} &\td & \sy{0} & \sy{0} & \sy{0} & \sy{0} \\
\sy{0} &\td & \sy{0} & \sy{\sigma_{{\rm tot},N4258,j}^2}\!\!\!\! & \sy{0} & \sy{0} & \sy{0} & \sy{0} &\td & \sy{0} & \sy{0} & \sy{0} & \sy{0} \\
\sy{0} &\td & \sy{0} & \sy{0} & \sy{\sigma_{{\rm tot},M31,j}^2}\!\!\!\! & \sy{0} & \sy{0} & \sy{0} &\td & \sy{0} & \sy{0} & \sy{0} & \sy{0} \\
\sy{0} &\td & \sy{0} & \sy{0} & \sy{0} & \sy{\sigma_{{\rm tot},MW,j}^2\!+\sigma_{\pi,j}^2}\!\!\!\!\!\! & \sy{0} & \sy{0} &\td & \sy{0} & \sy{0} & \sy{0} & \sy{0} \\
\sy{0} &\td & \sy{0} & \sy{0} & \sy{0} & \sy{0} & \sy{\sigma_{{\rm tot},{\rm LMC},j}^2}\!\!\!\! & \sy{0} &\td & \sy{0} & \sy{0} & \sy{0} & \sy{0} \\
\sy{0} &\td & \sy{0} & \sy{0} & \sy{0} & \sy{0} & \sy{0} & \sy{\sigma_{m_B,1}^2}\!\!\!\! &\td & \sy{0} & \sy{0} & \sy{0} & \sy{0} \\
\td    &\td & \td    & \td    &\td     & \td    & \td    & \td    &\td & \td    &\td     &\td &\td \\
\sy{0} &\td & \sy{0} & \sy{0} & \sy{0} & \sy{0} & \sy{0} & \sy{0} &\td & \sy{\sigma_{m_B,19}^2}\!\!\!\! & \sy{0} & \sy{0} & \sy{0} \\
\sy{0} &\td & \sy{0} & \sy{0} & \sy{0} & \sy{0} & \sy{0} & \sy{0} &\td & \sy{0} & \sy{\sigma_{\rm zp}^2}\!\!\!\! & \sy{0} & \sy{0} \\
\sy{0} &\td & \sy{0} & \sy{0} & \sy{0} & \sy{0} & \sy{0} & \sy{0} &\td & \sy{0} & \sy{0} & \sy{\sigma_{\mu,N4258}^2}\!\!\!\! & \sy{0} \\
\sy{0} &\td & \sy{0} & \sy{0} & \sy{0} & \sy{0} & \sy{0} & \sy{0} &\td & \sy{0} & \sy{0} & \sy{0} & \sy{\sigma_{\mu,{\rm LMC}}^2}\end{pmatrix}$$

\noindent Note: The term $\lp^h_{19,1}/0$ equals $\lp_{19,1}$ if $P\!>\!10$~d or 0 if $P\!<\!10$~d. The term $\lp^l_{19,1}/0$ equals $\lp_{19,1}$ if $P\!<\!10$~d or 0 if $P\!>\!10$~d.

\vspace*{1in}

\begin{deluxetable}{cccccccccccrrrrccccc}
\tabletypesize{\scriptsize}
\tablewidth{0pc}
\setlength{\tabcolsep}{0.33em}
\tablenum{8}
\tablecaption{Fits for H$_0$\label{tb:h0var}}
\tablehead{\colhead{$\chi^2_{dof}$}&\colhead{$H_0$}&\colhead{Anc}&\colhead{Brk}&\colhead{Clp}&\colhead{$\sigma$}&\colhead{Opt}&\colhead{PL}&\colhead{R}&\colhead{$R_V$}&\colhead{N}&\colhead{Z}&\colhead{$\gamma$}&\colhead{b}& \colhead{bl} &\colhead{SN}&\colhead{$z_m$}&\colhead{$M_V^0$}&\colhead{$a_v$}&\colhead{Gal}}
\startdata
 0.92  & 73.46 1.53  & All & Y & 1 &  2.7  & Y & $W_H$ & F & 3.3 &         2277 & Z & -0.13 0.07  & -3.26 0.02  & -3.25 0.02  & S & 0.02 &  -19.24  &  0.71273 & A\\
 0.93  & 73.48 1.53  & All & Y & G &  2.7  & Y & $W_H$ & F & 3.3 &         2280 & Z & -0.12 0.07  & -3.25 0.02  & -3.25 0.02  & S & 0.02 &  -19.24  &  0.71273 & A\\
 0.88  & 72.73 1.49  & All & Y & I &  2.7  & Y & $W_H$ & F & 3.3 &         2280 & Z & -0.10 0.07  & -3.31 0.02  & -3.22 0.02  & S & 0.02 &  -19.27  &  0.71273 & A\\
 1.28  & 73.68 1.79  & All & Y & G & No & Y & $W_H$ & F & 3.3 &         2345 & Z & -0.10 0.08  & -3.26 0.02  & -3.24 0.02  & S & 0.02 &  -19.24  &  0.71273 & A\\
 1.10  & 73.67 1.66  & All & Y & G &  3.5  & Y & $W_H$ & F & 3.3 &         2321 & Z & -0.14 0.07  & -3.25 0.02  & -3.25 0.02  & S & 0.02 &  -19.24  &  0.71273 & A\\
 1.09  & 73.64 1.66  & All & Y & 1 &  3.5  & Y & $W_H$ & F & 3.3 &         2320 & Z & -0.14 0.07  & -3.25 0.02  & -3.25 0.02  & S & 0.02 &  -19.24  &  0.71273 & A\\
 0.92  & 72.69 1.52  & All & Y & I &  3.5  & Y & $W_H$ & F & 3.3 &         2320 & Z & -0.09 0.07  & -3.30 0.02  & -3.22 0.02  & S & 0.02 &  -19.27  &  0.71273 & A\\
 1.11  & 73.21 1.66  & All & Y & G &  3.5  & Y & $W_H$ & F & 2.5 &         2322 & Z & -0.14 0.07  & -3.25 0.02  & -3.24 0.02  & S & 0.02 &  -19.25  &  0.71273 & A\\
 0.92  & 74.04 1.54  & All & Y & G &  2.7  & Y & $W_H$ & C & 3.3 &         2278 & Z & -0.11 0.07  & -3.27 0.02  & -3.27 0.02  & S & 0.02 &  -19.23  &  0.71273 & A\\
 0.94  & 73.14 1.54  & All & Y & G &  2.7  & Y & $W_H$ & N & 3.3 &         2278 & Z & -0.11 0.07  & -3.24 0.02  & -3.24 0.02  & S & 0.02 &  -19.25  &  0.71273 & A\\
 0.93  & 73.49 1.52  & All & N & G &  2.7  & Y & $W_H$ & F & 3.3 &         2280 & Z & -0.12 0.07  & -3.25 0.01  & \nd & S & 0.02 &  -19.24  &  0.71273 & A\\
 1.12  & 72.19 1.72  & All & 10 & G &  2.7  & Y & $W_H$ & F & 3.3 &         1318 & Z & -0.15 0.08  & -3.26 0.03  & \nd & S & 0.02 &  -19.28  &  0.71273 & A\\
 0.91  & 73.32 1.51  & All & 60 & G &  2.7  & Y & $W_H$ & F & 3.3 &         2180 & Z & -0.16 0.07  & -3.25 0.01  & \nd & S & 0.02 &  -19.25  &  0.71273 & A\\
 1.06  & 75.55 1.67  & All & Y & G &  2.7  & Y & $H$ & F & 3.3 &         2240 & Z & -0.08 0.07  & -3.07 0.02  & -3.16 0.02  & S & 0.02 &  -19.18  &  0.71273 & A\\
 0.94  & 73.45 1.54  & All & Y & G &  2.7  & Y & $W_H$ & F & 3.3 &         2281 & Z & \nd & -3.26 0.02  & -3.25 0.02  & S & 0.02 &  -19.24  &  0.71273 & A\\
 0.93  & 73.60 1.53  & All & Y & G &  2.7  & Y & $W_H$ & F & 3.3 &         2280 & Z & -0.12 0.07  & -3.25 0.02  & -3.25 0.02  & S & 0.01 &  -19.24  &  0.71347 & A\\
 0.93  & 75.08 1.65  & All & Y & G &  2.7  & Y & $W_H$ & F & 3.3 &         2279 & Z & -0.06 0.07  & -3.27 0.02  & -3.25 0.02  & M & 0.02 &  -19.15  &  0.70326 & A\\
 0.94  & 74.66 1.55  & All & Y & G &  2.7  & N & $W_H$ & F & 3.3 &         2417 & Z & -0.15 0.07  & -3.21 0.02  & -3.26 0.02  & S & 0.02 &  -19.21  &  0.71273 & A\\
 0.93  & 73.85 1.58  & All & Y & G &  2.7  & Y & $W_H$ & F & 3.3 &         2280 & B & -0.10 0.09  & -3.25 0.02  & -3.25 0.02  & S & 0.02 &  -19.23  &  0.71273 & A\\
 0.93  & 73.59 1.60  & All & Y & G &  2.7  & Y & $W_H$ & F & 3.3 &         2280 & Z & -0.12 0.07  & -3.25 0.02  & -3.25 0.02  & S & 0.02 &  -19.24  &  0.71340 & S\\
 0.93  & 74.57 1.70  & All & Y & G &  2.7  & Y & $W_H$ & F & 3.3 &         2279 & Z & -0.06 0.07  & -3.27 0.02  & -3.25 0.02  & M & 0.02 &  -19.15  &  0.70031 & S\\
 0.93  & 73.76 1.66  & All & Y & G &  2.7  & Y & $W_H$ & F & 3.3 &         2280 & Z & -0.12 0.07  & -3.25 0.02  & -3.25 0.02  & S & 0.01 &  -19.24  &  0.71444 & L\\
 0.93  & 74.23 1.71  & All & Y & G &  2.7  & Y & $W_H$ & F & 3.3 &         2280 & Z & -0.12 0.07  & -3.25 0.02  & -3.25 0.02  & S & 0.02 &  -19.24  &  0.71719 & L\\
 0.92  & 73.24 1.59  & NML & Y & 1 &  2.7  & Y & $W_H$ & F & 3.3 &         2276 & Z & -0.13 0.07  & -3.26 0.02  & -3.25 0.02  & S & 0.02 &  -19.25  &  0.71273 & A\\
 0.93  & 73.25 1.60  & NML & Y & G &  2.7  & Y & $W_H$ & F & 3.3 &         2279 & Z & -0.12 0.07  & -3.25 0.02  & -3.25 0.02  & S & 0.02 &  -19.25  &  0.71273 & A\\
 0.88  & 72.67 1.56  & NML & Y & I &  2.7  & Y & $W_H$ & F & 3.3 &         2279 & Z & -0.10 0.07  & -3.31 0.02  & -3.22 0.02  & S & 0.02 &  -19.27  &  0.71273 & A\\
 1.28  & 73.49 1.87  & NML & Y & G & No & Y & $W_H$ & F & 3.3 &         2344 & Z & -0.10 0.08  & -3.26 0.02  & -3.24 0.02  & S & 0.02 &  -19.24  &  0.71273 & A\\
 1.10  & 73.45 1.74  & NML & Y & G &  3.5  & Y & $W_H$ & F & 3.3 &         2320 & Z & -0.14 0.07  & -3.25 0.02  & -3.25 0.02  & S & 0.02 &  -19.24  &  0.71273 & A\\
 1.09  & 73.42 1.73  & NML & Y & 1 &  3.5  & Y & $W_H$ & F & 3.3 &         2319 & Z & -0.14 0.07  & -3.25 0.02  & -3.25 0.02  & S & 0.02 &  -19.25  &  0.71273 & A\\
 0.92  & 72.62 1.59  & NML & Y & I &  3.5  & Y & $W_H$ & F & 3.3 &         2319 & Z & -0.09 0.07  & -3.30 0.02  & -3.22 0.02  & S & 0.02 &  -19.27  &  0.71273 & A\\
 1.11  & 73.15 1.74  & NML & Y & G &  3.5  & Y & $W_H$ & F & 2.5 &         2321 & Z & -0.14 0.07  & -3.25 0.02  & -3.24 0.02  & S & 0.02 &  -19.25  &  0.71273 & A\\
 0.92  & 73.33 1.59  & NML & Y & G &  2.7  & Y & $W_H$ & C & 3.3 &         2277 & Z & -0.12 0.07  & -3.27 0.02  & -3.27 0.02  & S & 0.02 &  -19.25  &  0.71273 & A\\
 0.94  & 73.39 1.61  & NML & Y & G &  2.7  & Y & $W_H$ & N & 3.3 &         2277 & Z & -0.11 0.07  & -3.24 0.02  & -3.24 0.02  & S & 0.02 &  -19.25  &  0.71273 & A\\
 0.93  & 73.26 1.59  & NML & N & G &  2.7  & Y & $W_H$ & F & 3.3 &         2279 & Z & -0.12 0.07  & -3.25 0.01  & \nd & S & 0.02 &  -19.25  &  0.71273 & A\\
 1.12  & 71.64 1.81  & NML & 10 & G &  2.7  & Y & $W_H$ & F & 3.3 &         1317 & Z & -0.16 0.08  & -3.26 0.03  & \nd & S & 0.02 &  -19.30  &  0.71273 & A\\
 0.91  & 73.06 1.58  & NML & 60 & G &  2.7  & Y & $W_H$ & F & 3.3 &         2179 & Z & -0.17 0.07  & -3.25 0.01  & \nd & S & 0.02 &  -19.26  &  0.71273 & A\\
 1.06  & 74.79 1.73  & NML & Y & G &  2.7  & Y & $H$ & F & 3.3 &         2239 & Z & -0.09 0.07  & -3.06 0.02  & -3.17 0.02  & S & 0.02 &  -19.21  &  0.71273 & A\\
 0.94  & 73.30 1.61  & NML & Y & G &  2.7  & Y & $W_H$ & F & 3.3 &         2280 & Z & \nd & -3.26 0.02  & -3.25 0.02  & S & 0.02 &  -19.25  &  0.71273 & A\\
 0.93  & 73.38 1.60  & NML & Y & G &  2.7  & Y & $W_H$ & F & 3.3 &         2279 & Z & -0.12 0.07  & -3.25 0.02  & -3.25 0.02  & S & 0.01 &  -19.25  &  0.71347 & A\\
 0.93  & 74.89 1.72  & NML & Y & G &  2.7  & Y & $W_H$ & F & 3.3 &         2278 & Z & -0.06 0.07  & -3.27 0.02  & -3.25 0.02  & M & 0.02 &  -19.15  &  0.70326 & A\\
 0.94  & 74.39 1.62  & NML & Y & G &  2.7  & N & $W_H$ & F & 3.3 &         2416 & Z & -0.15 0.07  & -3.21 0.02  & -3.26 0.02  & S & 0.02 &  -19.22  &  0.71273 & A\\
 0.93  & 73.64 1.63  & NML & Y & G &  2.7  & Y & $W_H$ & F & 3.3 &         2279 & B & -0.11 0.09  & -3.25 0.02  & -3.25 0.02  & S & 0.02 &  -19.24  &  0.71273 & A\\
 0.93  & 73.37 1.66  & NML & Y & G &  2.7  & Y & $W_H$ & F & 3.3 &         2279 & Z & -0.12 0.07  & -3.25 0.02  & -3.25 0.02  & S & 0.02 &  -19.25  &  0.71340 & S\\
 0.93  & 74.39 1.76  & NML & Y & G &  2.7  & Y & $W_H$ & F & 3.3 &         2278 & Z & -0.06 0.07  & -3.27 0.02  & -3.25 0.02  & M & 0.02 &  -19.15  &  0.70031 & S\\
 0.93  & 73.54 1.73  & NML & Y & G &  2.7  & Y & $W_H$ & F & 3.3 &         2279 & Z & -0.12 0.07  & -3.25 0.02  & -3.25 0.02  & S & 0.01 &  -19.25  &  0.71444 & L\\
 0.93  & 74.01 1.77  & NML & Y & G &  2.7  & Y & $W_H$ & F & 3.3 &         2279 & Z & -0.12 0.07  & -3.25 0.02  & -3.25 0.02  & S & 0.02 &  -19.25  &  0.71719 & L\\
 0.92  & 74.04 1.74  & NM & Y & 1 &  2.7  & Y & $W_H$ & F & 3.3 &         2275 & Z & -0.16 0.07  & -3.26 0.02  & -3.25 0.02  & S & 0.02 &  -19.23  &  0.71273 & A\\
 0.93  & 74.01 1.75  & NM & Y & G &  2.7  & Y & $W_H$ & F & 3.3 &         2278 & Z & -0.14 0.07  & -3.25 0.02  & -3.25 0.02  & S & 0.02 &  -19.23  &  0.71273 & A\\
 0.87  & 73.67 1.71  & NM & Y & I &  2.7  & Y & $W_H$ & F & 3.3 &         2278 & Z & -0.13 0.07  & -3.31 0.02  & -3.22 0.02  & S & 0.02 &  -19.24  &  0.71273 & A\\
 1.28  & 74.20 2.04  & NM & Y & G & No & Y & $W_H$ & F & 3.3 &         2343 & Z & -0.12 0.08  & -3.26 0.02  & -3.24 0.02  & S & 0.02 &  -19.22  &  0.71273 & A\\
 1.10  & 74.30 1.90  & NM & Y & G &  3.5  & Y & $W_H$ & F & 3.3 &         2319 & Z & -0.17 0.08  & -3.25 0.02  & -3.25 0.02  & S & 0.02 &  -19.22  &  0.71273 & A\\
 1.09  & 74.27 1.89  & NM & Y & 1 &  3.5  & Y & $W_H$ & F & 3.3 &         2318 & Z & -0.17 0.08  & -3.25 0.02  & -3.25 0.02  & S & 0.02 &  -19.22  &  0.71273 & A\\
 0.92  & 73.57 1.75  & NM & Y & I &  3.5  & Y & $W_H$ & F & 3.3 &         2318 & Z & -0.12 0.07  & -3.30 0.02  & -3.22 0.02  & S & 0.02 &  -19.24  &  0.71273 & A\\
 1.11  & 74.04 1.90  & NM & Y & G &  3.5  & Y & $W_H$ & F & 2.5 &         2320 & Z & -0.17 0.08  & -3.25 0.02  & -3.24 0.02  & S & 0.02 &  -19.23  &  0.71273 & A\\
 0.92  & 74.03 1.74  & NM & Y & G &  2.7  & Y & $W_H$ & C & 3.3 &         2277 & Z & -0.15 0.07  & -3.28 0.02  & -3.27 0.02  & S & 0.02 &  -19.23  &  0.71273 & A\\
 0.95  & 74.17 1.77  & NM & Y & G &  2.7  & Y & $W_H$ & N & 3.3 &         2278 & Z & -0.14 0.07  & -3.24 0.02  & -3.24 0.02  & S & 0.02 &  -19.22  &  0.71273 & A\\
 0.93  & 74.03 1.73  & NM & N & G &  2.7  & Y & $W_H$ & F & 3.3 &         2278 & Z & -0.14 0.07  & -3.25 0.01  & \nd & S & 0.02 &  -19.23  &  0.71273 & A\\
 1.12  & 71.36 2.17  & NM & 10 & G &  2.7  & Y & $W_H$ & F & 3.3 &         1316 & Z & -0.15 0.08  & -3.26 0.03  & \nd & S & 0.02 &  -19.31  &  0.71273 & A\\
 0.91  & 73.98 1.73  & NM & 60 & G &  2.7  & Y & $W_H$ & F & 3.3 &         2179 & Z & -0.22 0.08  & -3.25 0.01  & \nd & S & 0.02 &  -19.23  &  0.71273 & A\\
 1.06  & 75.57 1.89  & NM & Y & G &  2.7  & Y & $H$ & F & 3.3 &         2239 & Z & -0.12 0.07  & -3.07 0.02  & -3.17 0.02  & S & 0.02 &  -19.18  &  0.71273 & A\\
 0.94  & 73.70 1.74  & NM & Y & G &  2.7  & Y & $W_H$ & F & 3.3 &         2279 & Z & \nd & -3.26 0.02  & -3.25 0.02  & S & 0.02 &  -19.24  &  0.71273 & A\\
 0.93  & 74.14 1.75  & NM & Y & G &  2.7  & Y & $W_H$ & F & 3.3 &         2278 & Z & -0.14 0.07  & -3.25 0.02  & -3.25 0.02  & S & 0.01 &  -19.23  &  0.71347 & A\\
 0.94  & 75.50 1.86  & NM & Y & G &  2.7  & Y & $W_H$ & F & 3.3 &         2278 & Z & -0.10 0.07  & -3.27 0.02  & -3.25 0.02  & M & 0.02 &  -19.14  &  0.70326 & A\\
 0.93  & 75.27 1.77  & NM & Y & G &  2.7  & N & $W_H$ & F & 3.3 &         2415 & Z & -0.19 0.07  & -3.21 0.02  & -3.26 0.02  & S & 0.02 &  -19.19  &  0.71273 & A\\
 0.93  & 74.57 1.83  & NM & Y & G &  2.7  & Y & $W_H$ & F & 3.3 &         2278 & B & -0.16 0.10  & -3.25 0.02  & -3.25 0.02  & S & 0.02 &  -19.21  &  0.71273 & A\\
 0.93  & 74.13 1.80  & NM & Y & G &  2.7  & Y & $W_H$ & F & 3.3 &         2278 & Z & -0.14 0.07  & -3.25 0.02  & -3.25 0.02  & S & 0.02 &  -19.23  &  0.71340 & S\\
 0.94  & 74.99 1.91  & NM & Y & G &  2.7  & Y & $W_H$ & F & 3.3 &         2278 & Z & -0.10 0.07  & -3.27 0.02  & -3.25 0.02  & M & 0.02 &  -19.14  &  0.70031 & S\\
 0.93  & 74.31 1.86  & NM & Y & G &  2.7  & Y & $W_H$ & F & 3.3 &         2278 & Z & -0.14 0.07  & -3.25 0.02  & -3.25 0.02  & S & 0.01 &  -19.23  &  0.71444 & L\\
 0.93  & 74.78 1.91  & NM & Y & G &  2.7  & Y & $W_H$ & F & 3.3 &         2278 & Z & -0.14 0.07  & -3.25 0.02  & -3.25 0.02  & S & 0.02 &  -19.23  &  0.71719 & L\\
 0.92  & 71.62 1.68  & NL & Y & 1 &  2.7  & Y & $W_H$ & F & 3.3 &         2276 & Z & -0.18 0.07  & -3.26 0.02  & -3.25 0.02  & S & 0.02 &  -19.30  &  0.71273 & A\\
 0.93  & 71.86 1.70  & NL & Y & G &  2.7  & Y & $W_H$ & F & 3.3 &         2279 & Z & -0.15 0.07  & -3.25 0.02  & -3.25 0.02  & S & 0.02 &  -19.29  &  0.71273 & A\\
 0.87  & 71.60 1.66  & NL & Y & I &  2.7  & Y & $W_H$ & F & 3.3 &         2279 & Z & -0.13 0.07  & -3.31 0.02  & -3.22 0.02  & S & 0.02 &  -19.30  &  0.71273 & A\\
 1.28  & 72.14 1.98  & NL & Y & G & No & Y & $W_H$ & F & 3.3 &         2344 & Z & -0.13 0.08  & -3.25 0.02  & -3.24 0.02  & S & 0.02 &  -19.28  &  0.71273 & A\\
 1.09  & 72.01 1.84  & NL & Y & G &  3.5  & Y & $W_H$ & F & 3.3 &         2320 & Z & -0.18 0.07  & -3.25 0.02  & -3.25 0.02  & S & 0.02 &  -19.29  &  0.71273 & A\\
 1.09  & 71.99 1.83  & NL & Y & 1 &  3.5  & Y & $W_H$ & F & 3.3 &         2319 & Z & -0.18 0.07  & -3.25 0.02  & -3.25 0.02  & S & 0.02 &  -19.29  &  0.71273 & A\\
 0.92  & 71.51 1.70  & NL & Y & I &  3.5  & Y & $W_H$ & F & 3.3 &         2319 & Z & -0.12 0.07  & -3.29 0.02  & -3.22 0.02  & S & 0.02 &  -19.30  &  0.71273 & A\\
 1.10  & 71.77 1.84  & NL & Y & G &  3.5  & Y & $W_H$ & F & 2.5 &         2321 & Z & -0.17 0.07  & -3.24 0.02  & -3.25 0.02  & S & 0.02 &  -19.29  &  0.71273 & A\\
 0.92  & 71.80 1.69  & NL & Y & G &  2.7  & Y & $W_H$ & C & 3.3 &         2278 & Z & -0.18 0.07  & -3.27 0.02  & -3.27 0.02  & S & 0.02 &  -19.29  &  0.71273 & A\\
 0.94  & 71.86 1.71  & NL & Y & G &  2.7  & Y & $W_H$ & N & 3.3 &         2277 & Z & -0.15 0.07  & -3.24 0.02  & -3.24 0.02  & S & 0.02 &  -19.29  &  0.71273 & A\\
 0.93  & 71.84 1.69  & NL & N & G &  2.7  & Y & $W_H$ & F & 3.3 &         2279 & Z & -0.15 0.07  & -3.25 0.01  & \nd & S & 0.02 &  -19.29  &  0.71273 & A\\
 1.13  & 71.21 1.86  & NL & 10 & G &  2.7  & Y & $W_H$ & F & 3.3 &         1319 & Z & -0.19 0.08  & -3.26 0.03  & \nd & S & 0.02 &  -19.31  &  0.71273 & A\\
 0.91  & 71.42 1.67  & NL & 60 & G &  2.7  & Y & $W_H$ & F & 3.3 &         2180 & Z & -0.24 0.08  & -3.25 0.01  & \nd & S & 0.02 &  -19.31  &  0.71273 & A\\
 1.06  & 73.32 1.83  & NL & Y & G &  2.7  & Y & $H$ & F & 3.3 &         2240 & Z & -0.13 0.07  & -3.06 0.02  & -3.17 0.02  & S & 0.02 &  -19.25  &  0.71273 & A\\
 0.93  & 72.25 1.70  & NL & Y & G &  2.7  & Y & $W_H$ & F & 3.3 &         2280 & Z & \nd & -3.26 0.02  & -3.25 0.02  & S & 0.02 &  -19.28  &  0.71273 & A\\
 0.93  & 71.98 1.70  & NL & Y & G &  2.7  & Y & $W_H$ & F & 3.3 &         2279 & Z & -0.15 0.07  & -3.25 0.02  & -3.25 0.02  & S & 0.01 &  -19.29  &  0.71347 & A\\
 0.93  & 73.53 1.81  & NL & Y & G &  2.7  & Y & $W_H$ & F & 3.3 &         2279 & Z & -0.11 0.07  & -3.26 0.02  & -3.25 0.02  & M & 0.02 &  -19.19  &  0.70326 & A\\
 0.93  & 72.78 1.71  & NL & Y & G &  2.7  & N & $W_H$ & F & 3.3 &         2417 & Z & -0.21 0.07  & -3.21 0.02  & -3.26 0.02  & S & 0.02 &  -19.26  &  0.71273 & A\\
 0.93  & 72.24 1.70  & NL & Y & G &  2.7  & Y & $W_H$ & F & 3.3 &         2279 & B & -0.22 0.10  & -3.25 0.02  & -3.25 0.02  & S & 0.02 &  -19.28  &  0.71273 & A\\
 0.93  & 71.97 1.75  & NL & Y & G &  2.7  & Y & $W_H$ & F & 3.3 &         2279 & Z & -0.15 0.07  & -3.25 0.02  & -3.25 0.02  & S & 0.02 &  -19.29  &  0.71340 & S\\
 0.93  & 73.03 1.85  & NL & Y & G &  2.7  & Y & $W_H$ & F & 3.3 &         2279 & Z & -0.11 0.07  & -3.26 0.02  & -3.25 0.02  & M & 0.02 &  -19.19  &  0.70031 & S\\
 0.93  & 72.15 1.81  & NL & Y & G &  2.7  & Y & $W_H$ & F & 3.3 &         2279 & Z & -0.15 0.07  & -3.25 0.02  & -3.25 0.02  & S & 0.01 &  -19.29  &  0.71444 & L\\
 0.93  & 72.60 1.85  & NL & Y & G &  2.7  & Y & $W_H$ & F & 3.3 &         2279 & Z & -0.15 0.07  & -3.25 0.02  & -3.25 0.02  & S & 0.02 &  -19.29  &  0.71719 & L\\
 0.92  & 74.15 1.82  & M+L & Y & 1 &  2.7  & Y & $W_H$ & F & 3.3 &         2275 & Z & -0.14 0.07  & -3.26 0.02  & -3.25 0.02  & S & 0.02 &  -19.22  &  0.71273 & A\\
 0.93  & 74.27 1.84  & M+L & Y & G &  2.7  & Y & $W_H$ & F & 3.3 &         2278 & Z & -0.11 0.07  & -3.25 0.02  & -3.25 0.02  & S & 0.02 &  -19.22  &  0.71273 & A\\
 0.88  & 72.84 1.75  & M+L & Y & I &  2.7  & Y & $W_H$ & F & 3.3 &         2278 & Z & -0.10 0.07  & -3.31 0.02  & -3.22 0.02  & S & 0.02 &  -19.26  &  0.71273 & A\\
 1.28  & 74.51 2.15  & M+L & Y & G & No & Y & $W_H$ & F & 3.3 &         2343 & Z & -0.10 0.08  & -3.26 0.02  & -3.24 0.02  & S & 0.02 &  -19.21  &  0.71273 & A\\
 1.10  & 74.43 1.99  & M+L & Y & G &  3.5  & Y & $W_H$ & F & 3.3 &         2319 & Z & -0.14 0.07  & -3.25 0.02  & -3.25 0.02  & S & 0.02 &  -19.22  &  0.71273 & A\\
 1.09  & 74.40 1.99  & M+L & Y & 1 &  3.5  & Y & $W_H$ & F & 3.3 &         2318 & Z & -0.14 0.07  & -3.25 0.02  & -3.25 0.02  & S & 0.02 &  -19.22  &  0.71273 & A\\
 0.92  & 72.92 1.80  & M+L & Y & I &  3.5  & Y & $W_H$ & F & 3.3 &         2318 & Z & -0.09 0.07  & -3.30 0.02  & -3.22 0.02  & S & 0.02 &  -19.26  &  0.71273 & A\\
 1.11  & 73.95 1.99  & M+L & Y & G &  3.5  & Y & $W_H$ & F & 2.5 &         2320 & Z & -0.14 0.07  & -3.25 0.02  & -3.24 0.02  & S & 0.02 &  -19.23  &  0.71273 & A\\
 0.92  & 74.45 1.83  & M+L & Y & G &  2.7  & Y & $W_H$ & C & 3.3 &         2276 & Z & -0.12 0.07  & -3.27 0.02  & -3.27 0.02  & S & 0.02 &  -19.22  &  0.71273 & A\\
 0.94  & 74.35 1.85  & M+L & Y & G &  2.7  & Y & $W_H$ & N & 3.3 &         2275 & Z & -0.11 0.07  & -3.24 0.02  & -3.24 0.02  & S & 0.02 &  -19.22  &  0.71273 & A\\
 0.93  & 74.28 1.82  & M+L & N & G &  2.7  & Y & $W_H$ & F & 3.3 &         2278 & Z & -0.11 0.07  & -3.25 0.01  & \nd & S & 0.02 &  -19.22  &  0.71273 & A\\
 1.12  & 72.70 2.18  & M+L & 10 & G &  2.7  & Y & $W_H$ & F & 3.3 &         1316 & Z & -0.15 0.08  & -3.26 0.03  & \nd & S & 0.02 &  -19.27  &  0.71273 & A\\
 0.90  & 74.29 1.81  & M+L & 60 & G &  2.7  & Y & $W_H$ & F & 3.3 &         2176 & Z & -0.14 0.07  & -3.25 0.01  & \nd & S & 0.02 &  -19.22  &  0.71273 & A\\
 1.06  & 75.96 1.99  & M+L & Y & G &  2.7  & Y & $H$ & F & 3.3 &         2237 & Z & -0.09 0.07  & -3.06 0.02  & -3.16 0.02  & S & 0.02 &  -19.17  &  0.71273 & A\\
 0.93  & 74.40 1.84  & M+L & Y & G &  2.7  & Y & $W_H$ & F & 3.3 &         2279 & Z & \nd & -3.26 0.02  & -3.25 0.02  & S & 0.02 &  -19.22  &  0.71273 & A\\
 0.93  & 74.39 1.84  & M+L & Y & G &  2.7  & Y & $W_H$ & F & 3.3 &         2278 & Z & -0.11 0.07  & -3.25 0.02  & -3.25 0.02  & S & 0.01 &  -19.22  &  0.71347 & A\\
 0.93  & 75.96 1.95  & M+L & Y & G &  2.7  & Y & $W_H$ & F & 3.3 &         2277 & Z & -0.06 0.07  & -3.26 0.02  & -3.25 0.02  & M & 0.02 &  -19.12  &  0.70326 & A\\
 0.93  & 75.65 1.86  & M+L & Y & G &  2.7  & N & $W_H$ & F & 3.3 &         2414 & Z & -0.15 0.07  & -3.21 0.02  & -3.26 0.02  & S & 0.02 &  -19.18  &  0.71273 & A\\
 0.93  & 74.94 1.89  & M+L & Y & G &  2.7  & Y & $W_H$ & F & 3.3 &         2278 & B & -0.13 0.09  & -3.25 0.02  & -3.25 0.02  & S & 0.02 &  -19.20  &  0.71273 & A\\
 0.93  & 74.38 1.89  & M+L & Y & G &  2.7  & Y & $W_H$ & F & 3.3 &         2278 & Z & -0.11 0.07  & -3.25 0.02  & -3.25 0.02  & S & 0.02 &  -19.22  &  0.71340 & S\\
 0.93  & 75.44 1.99  & M+L & Y & G &  2.7  & Y & $W_H$ & F & 3.3 &         2277 & Z & -0.06 0.07  & -3.26 0.02  & -3.25 0.02  & M & 0.02 &  -19.12  &  0.70031 & S\\
 0.93  & 74.56 1.95  & M+L & Y & G &  2.7  & Y & $W_H$ & F & 3.3 &         2278 & Z & -0.11 0.07  & -3.25 0.02  & -3.25 0.02  & S & 0.01 &  -19.22  &  0.71444 & L\\
 0.93  & 75.03 1.99  & M+L & Y & G &  2.7  & Y & $W_H$ & F & 3.3 &         2278 & Z & -0.11 0.07  & -3.25 0.02  & -3.25 0.02  & S & 0.02 &  -19.22  &  0.71719 & L\\
 1.04  & 72.25 2.38  & N & Y & 1 &  2.7  & Y & $W_H$ & F & 3.3 &         1485 & Z & -0.16 0.08  & -3.10 0.03  & -3.46 0.05  & S & 0.02 &  -19.28  &  0.71273 & A\\
 1.05  & 72.52 2.39  & N & Y & G &  2.7  & Y & $W_H$ & F & 3.3 &         1486 & Z & -0.14 0.08  & -3.10 0.03  & -3.45 0.05  & S & 0.02 &  -19.27  &  0.71273 & A\\
 0.89  & 73.31 2.43  & N & Y & I &  2.7  & Y & $W_H$ & F & 3.3 &         1486 & Z & -0.14 0.07  & -3.21 0.03  & -3.40 0.05  & S & 0.02 &  -19.25  &  0.71273 & A\\
 1.48  & 72.78 2.56  & N & Y & G & No & Y & $W_H$ & F & 3.3 &         1540 & Z & -0.13 0.09  & -3.11 0.04  & -3.46 0.06  & S & 0.02 &  -19.26  &  0.71273 & A\\
 1.26  & 72.93 2.48  & N & Y & G &  3.5  & Y & $W_H$ & F & 3.3 &         1522 & Z & -0.17 0.08  & -3.09 0.04  & -3.47 0.06  & S & 0.02 &  -19.26  &  0.71273 & A\\
 1.25  & 72.97 2.48  & N & Y & 1 &  3.5  & Y & $W_H$ & F & 3.3 &         1520 & Z & -0.18 0.08  & -3.09 0.04  & -3.47 0.06  & S & 0.02 &  -19.26  &  0.71273 & A\\
 0.96  & 73.12 2.44  & N & Y & I &  3.5  & Y & $W_H$ & F & 3.3 &         1520 & Z & -0.13 0.08  & -3.18 0.04  & -3.42 0.05  & S & 0.02 &  -19.25  &  0.71273 & A\\
 1.26  & 73.03 2.49  & N & Y & G &  3.5  & Y & $W_H$ & F & 2.5 &         1522 & Z & -0.18 0.08  & -3.08 0.04  & -3.47 0.06  & S & 0.02 &  -19.26  &  0.71273 & A\\
 1.06  & 72.20 2.38  & N & Y & G &  2.7  & Y & $W_H$ & C & 3.3 &         1486 & Z & -0.14 0.08  & -3.11 0.03  & -3.44 0.06  & S & 0.02 &  -19.28  &  0.71273 & A\\
 1.04  & 72.61 2.39  & N & Y & G &  2.7  & Y & $W_H$ & N & 3.3 &         1485 & Z & -0.14 0.08  & -3.09 0.03  & -3.45 0.05  & S & 0.02 &  -19.27  &  0.71273 & A\\
 1.06  & 71.57 2.35  & N & N & G &  2.7  & Y & $W_H$ & F & 3.3 &         1485 & Z & -0.15 0.08  & -3.23 0.02  & \nd & S & 0.02 &  -19.30  &  0.71273 & A\\
 1.12  & 70.99 2.39  & N & 10 & G &  2.7  & Y & $W_H$ & F & 3.3 &         1203 & Z & -0.15 0.08  & -3.11 0.04  & \nd & S & 0.02 &  -19.32  &  0.71273 & A\\
 1.04  & 71.09 2.35  & N & 60 & G &  2.7  & Y & $W_H$ & F & 3.3 &         1388 & Z & -0.23 0.08  & -3.25 0.03  & \nd & S & 0.02 &  -19.32  &  0.71273 & A\\
 1.11  & 74.35 2.45  & N & Y & G &  2.7  & Y & $H$ & F & 3.3 &         1470 & Z & -0.12 0.07  & -2.90 0.03  & -3.14 0.06  & S & 0.02 &  -19.22  &  0.71273 & A\\
 1.05  & 72.21 2.38  & N & Y & G &  2.7  & Y & $W_H$ & F & 3.3 &         1487 & Z & \nd & -3.09 0.03  & -3.45 0.05  & S & 0.02 &  -19.28  &  0.71273 & A\\
 1.05  & 72.64 2.39  & N & Y & G &  2.7  & Y & $W_H$ & F & 3.3 &         1486 & Z & -0.14 0.08  & -3.10 0.03  & -3.45 0.05  & S & 0.01 &  -19.27  &  0.71347 & A\\
 1.06  & 74.16 2.50  & N & Y & G &  2.7  & Y & $W_H$ & F & 3.3 &         1487 & Z & -0.10 0.08  & -3.11 0.03  & -3.44 0.06  & M & 0.02 &  -19.18  &  0.70326 & A\\
 1.05  & 72.90 2.39  & N & Y & G &  2.7  & N & $W_H$ & F & 3.3 &         1626 & Z & -0.17 0.07  & -3.05 0.03  & -3.46 0.05  & S & 0.02 &  -19.26  &  0.71273 & A\\
 1.05  & 72.52 2.39  & N & Y & G &  2.7  & Y & $W_H$ & F & 3.3 &         1486 & B & -0.20 0.11  & -3.10 0.03  & -3.45 0.05  & S & 0.02 &  -19.27  &  0.71273 & A\\
 1.05  & 72.63 2.43  & N & Y & G &  2.7  & Y & $W_H$ & F & 3.3 &         1486 & Z & -0.14 0.08  & -3.10 0.03  & -3.45 0.05  & S & 0.02 &  -19.27  &  0.71340 & S\\
 1.06  & 73.66 2.53  & N & Y & G &  2.7  & Y & $W_H$ & F & 3.3 &         1487 & Z & -0.10 0.08  & -3.11 0.03  & -3.44 0.06  & M & 0.02 &  -19.18  &  0.70031 & S\\
 1.05  & 72.80 2.48  & N & Y & G &  2.7  & Y & $W_H$ & F & 3.3 &         1486 & Z & -0.14 0.08  & -3.10 0.03  & -3.45 0.05  & S & 0.01 &  -19.27  &  0.71444 & L\\
 1.05  & 73.26 2.52  & N & Y & G &  2.7  & Y & $W_H$ & F & 3.3 &         1486 & Z & -0.14 0.08  & -3.10 0.03  & -3.45 0.05  & S & 0.02 &  -19.27  &  0.71719 & L\\
 0.92  & 76.18 2.17  & M & Y & 1 &  2.7  & Y & $W_H$ & F & 3.3 &         2274 & Z & -0.18 0.07  & -3.26 0.02  & -3.25 0.02  & S & 0.02 &  -19.17  &  0.71273 & A\\
 0.93  & 76.12 2.18  & M & Y & G &  2.7  & Y & $W_H$ & F & 3.3 &         2277 & Z & -0.15 0.07  & -3.25 0.02  & -3.25 0.02  & S & 0.02 &  -19.17  &  0.71273 & A\\
 0.87  & 74.58 2.08  & M & Y & I &  2.7  & Y & $W_H$ & F & 3.3 &         2277 & Z & -0.14 0.07  & -3.31 0.02  & -3.22 0.02  & S & 0.02 &  -19.21  &  0.71273 & A\\
 1.28  & 76.30 2.55  & M & Y & G & No & Y & $W_H$ & F & 3.3 &         2342 & Z & -0.13 0.08  & -3.26 0.02  & -3.24 0.02  & S & 0.02 &  -19.16  &  0.71273 & A\\
 1.10  & 76.46 2.37  & M & Y & G &  3.5  & Y & $W_H$ & F & 3.3 &         2318 & Z & -0.17 0.08  & -3.25 0.02  & -3.25 0.02  & S & 0.02 &  -19.16  &  0.71273 & A\\
 1.09  & 76.40 2.36  & M & Y & 1 &  3.5  & Y & $W_H$ & F & 3.3 &         2317 & Z & -0.17 0.08  & -3.25 0.02  & -3.25 0.02  & S & 0.02 &  -19.16  &  0.71273 & A\\
 0.92  & 74.64 2.13  & M & Y & I &  3.5  & Y & $W_H$ & F & 3.3 &         2317 & Z & -0.12 0.07  & -3.29 0.02  & -3.22 0.02  & S & 0.02 &  -19.21  &  0.71273 & A\\
 1.10  & 75.94 2.36  & M & Y & G &  3.5  & Y & $W_H$ & F & 2.5 &         2319 & Z & -0.17 0.08  & -3.25 0.02  & -3.24 0.02  & S & 0.02 &  -19.17  &  0.71273 & A\\
 0.92  & 76.27 2.17  & M & Y & G &  2.7  & Y & $W_H$ & C & 3.3 &         2276 & Z & -0.16 0.07  & -3.27 0.02  & -3.27 0.02  & S & 0.02 &  -19.16  &  0.71273 & A\\
 0.94  & 76.21 2.20  & M & Y & G &  2.7  & Y & $W_H$ & N & 3.3 &         2276 & Z & -0.14 0.07  & -3.23 0.02  & -3.24 0.02  & S & 0.02 &  -19.16  &  0.71273 & A\\
 0.93  & 76.15 2.16  & M & N & G &  2.7  & Y & $W_H$ & F & 3.3 &         2277 & Z & -0.15 0.07  & -3.25 0.01  & \nd & S & 0.02 &  -19.17  &  0.71273 & A\\
 1.12  & 74.42 3.70  & M & 10 & G &  2.7  & Y & $W_H$ & F & 3.3 &         1315 & Z & -0.16 0.08  & -3.26 0.03  & \nd & S & 0.02 &  -19.22  &  0.71273 & A\\
 0.91  & 76.38 2.15  & M & 60 & G &  2.7  & Y & $W_H$ & F & 3.3 &         2176 & Z & -0.23 0.08  & -3.25 0.01  & \nd & S & 0.02 &  -19.16  &  0.71273 & A\\
 1.06  & 77.83 2.37  & M & Y & G &  2.7  & Y & $H$ & F & 3.3 &         2238 & Z & -0.12 0.07  & -3.06 0.02  & -3.17 0.02  & S & 0.02 &  -19.12  &  0.71273 & A\\
 0.93  & 75.64 2.16  & M & Y & G &  2.7  & Y & $W_H$ & F & 3.3 &         2278 & Z & \nd & -3.26 0.02  & -3.25 0.02  & S & 0.02 &  -19.18  &  0.71273 & A\\
 0.93  & 76.25 2.18  & M & Y & G &  2.7  & Y & $W_H$ & F & 3.3 &         2277 & Z & -0.15 0.07  & -3.25 0.02  & -3.25 0.02  & S & 0.01 &  -19.17  &  0.71347 & A\\
 0.94  & 77.64 2.29  & M & Y & G &  2.7  & Y & $W_H$ & F & 3.3 &         2277 & Z & -0.10 0.07  & -3.26 0.02  & -3.25 0.02  & M & 0.02 &  -19.08  &  0.70326 & A\\
 0.93  & 77.84 2.22  & M & Y & G &  2.7  & N & $W_H$ & F & 3.3 &         2413 & Z & -0.20 0.07  & -3.21 0.02  & -3.26 0.02  & S & 0.02 &  -19.12  &  0.71273 & A\\
 0.93  & 77.77 2.41  & M & Y & G &  2.7  & Y & $W_H$ & F & 3.3 &         2277 & B & -0.23 0.10  & -3.25 0.02  & -3.25 0.02  & S & 0.02 &  -19.12  &  0.71273 & A\\
 0.93  & 76.24 2.22  & M & Y & G &  2.7  & Y & $W_H$ & F & 3.3 &         2277 & Z & -0.15 0.07  & -3.25 0.02  & -3.25 0.02  & S & 0.02 &  -19.17  &  0.71340 & S\\
 0.94  & 77.11 2.32  & M & Y & G &  2.7  & Y & $W_H$ & F & 3.3 &         2277 & Z & -0.10 0.07  & -3.26 0.02  & -3.25 0.02  & M & 0.02 &  -19.08  &  0.70031 & S\\
 0.93  & 76.42 2.28  & M & Y & G &  2.7  & Y & $W_H$ & F & 3.3 &         2277 & Z & -0.15 0.07  & -3.25 0.02  & -3.25 0.02  & S & 0.01 &  -19.17  &  0.71444 & L\\
 0.93  & 76.91 2.32  & M & Y & G &  2.7  & Y & $W_H$ & F & 3.3 &         2277 & Z & -0.15 0.07  & -3.25 0.02  & -3.25 0.02  & S & 0.02 &  -19.17  &  0.71719 & L\\
 0.92  & 72.04 2.56  & L & Y & 1 &  2.7  & Y & $W_H$ & F & 3.3 &         2275 & Z & -0.18 0.07  & -3.26 0.02  & -3.25 0.02  & S & 0.02 &  -19.29  &  0.71273 & A\\
 0.93  & 72.27 2.58  & L & Y & G &  2.7  & Y & $W_H$ & F & 3.3 &         2278 & Z & -0.15 0.07  & -3.25 0.02  & -3.25 0.02  & S & 0.02 &  -19.28  &  0.71273 & A\\
 0.87  & 70.97 2.49  & L & Y & I &  2.7  & Y & $W_H$ & F & 3.3 &         2278 & Z & -0.14 0.07  & -3.31 0.02  & -3.22 0.02  & S & 0.02 &  -19.32  &  0.71273 & A\\
 1.28  & 72.58 2.88  & L & Y & G & No & Y & $W_H$ & F & 3.3 &         2343 & Z & -0.13 0.08  & -3.25 0.03  & -3.24 0.02  & S & 0.02 &  -19.27  &  0.71273 & A\\
 1.09  & 72.30 2.72  & L & Y & G &  3.5  & Y & $W_H$ & F & 3.3 &         2319 & Z & -0.18 0.08  & -3.25 0.02  & -3.25 0.02  & S & 0.02 &  -19.28  &  0.71273 & A\\
 1.09  & 72.28 2.72  & L & Y & 1 &  3.5  & Y & $W_H$ & F & 3.3 &         2318 & Z & -0.17 0.08  & -3.25 0.02  & -3.25 0.02  & S & 0.02 &  -19.28  &  0.71273 & A\\
 0.92  & 71.09 2.53  & L & Y & I &  3.5  & Y & $W_H$ & F & 3.3 &         2318 & Z & -0.12 0.07  & -3.29 0.02  & -3.23 0.02  & S & 0.02 &  -19.32  &  0.71273 & A\\
 1.10  & 71.82 2.71  & L & Y & G &  3.5  & Y & $W_H$ & F & 2.5 &         2320 & Z & -0.17 0.08  & -3.24 0.02  & -3.25 0.02  & S & 0.02 &  -19.29  &  0.71273 & A\\
 0.91  & 72.34 2.57  & L & Y & G &  2.7  & Y & $W_H$ & C & 3.3 &         2276 & Z & -0.17 0.07  & -3.27 0.02  & -3.27 0.02  & S & 0.02 &  -19.28  &  0.71273 & A\\
 0.94  & 72.31 2.59  & L & Y & G &  2.7  & Y & $W_H$ & N & 3.3 &         2276 & Z & -0.14 0.07  & -3.24 0.02  & -3.24 0.02  & S & 0.02 &  -19.28  &  0.71273 & A\\
 0.93  & 72.25 2.57  & L & N & G &  2.7  & Y & $W_H$ & F & 3.3 &         2278 & Z & -0.15 0.07  & -3.25 0.01  & \nd & S & 0.02 &  -19.28  &  0.71273 & A\\
 1.12  & 72.20 2.75  & L & 10 & G &  2.7  & Y & $W_H$ & F & 3.3 &         1316 & Z & -0.16 0.08  & -3.26 0.03  & \nd & S & 0.02 &  -19.28  &  0.71273 & A\\
 0.91  & 71.74 2.54  & L & 60 & G &  2.7  & Y & $W_H$ & F & 3.3 &         2178 & Z & -0.23 0.08  & -3.25 0.01  & \nd & S & 0.02 &  -19.30  &  0.71273 & A\\
 1.06  & 73.82 2.74  & L & Y & G &  2.7  & Y & $H$ & F & 3.3 &         2238 & Z & -0.12 0.07  & -3.06 0.02  & -3.17 0.02  & S & 0.02 &  -19.23  &  0.71273 & A\\
 0.93  & 73.07 2.58  & L & Y & G &  2.7  & Y & $W_H$ & F & 3.3 &         2279 & Z & \nd & -3.26 0.02  & -3.25 0.02  & S & 0.02 &  -19.26  &  0.71273 & A\\
 0.93  & 72.39 2.58  & L & Y & G &  2.7  & Y & $W_H$ & F & 3.3 &         2278 & Z & -0.15 0.07  & -3.25 0.02  & -3.25 0.02  & S & 0.01 &  -19.28  &  0.71347 & A\\
 0.93  & 74.05 2.69  & L & Y & G &  2.7  & Y & $W_H$ & F & 3.3 &         2278 & Z & -0.11 0.07  & -3.26 0.02  & -3.25 0.02  & M & 0.02 &  -19.18  &  0.70326 & A\\
 0.93  & 73.43 2.61  & L & Y & G &  2.7  & N & $W_H$ & F & 3.3 &         2414 & Z & -0.20 0.07  & -3.21 0.02  & -3.26 0.02  & S & 0.02 &  -19.25  &  0.71273 & A\\
 0.93  & 72.93 2.57  & L & Y & G &  2.7  & Y & $W_H$ & F & 3.3 &         2278 & B & -0.22 0.10  & -3.25 0.02  & -3.25 0.02  & S & 0.02 &  -19.26  &  0.71273 & A\\
 0.93  & 72.38 2.62  & L & Y & G &  2.7  & Y & $W_H$ & F & 3.3 &         2278 & Z & -0.15 0.07  & -3.25 0.02  & -3.25 0.02  & S & 0.02 &  -19.28  &  0.71340 & S\\
 0.93  & 73.55 2.71  & L & Y & G &  2.7  & Y & $W_H$ & F & 3.3 &         2278 & Z & -0.11 0.07  & -3.26 0.02  & -3.25 0.02  & M & 0.02 &  -19.18  &  0.70031 & S\\
 0.93  & 72.56 2.66  & L & Y & G &  2.7  & Y & $W_H$ & F & 3.3 &         2278 & Z & -0.15 0.07  & -3.25 0.02  & -3.25 0.02  & S & 0.01 &  -19.28  &  0.71444 & L\\
 0.93  & 73.02 2.70  & L & Y & G &  2.7  & Y & $W_H$ & F & 3.3 &         2278 & Z & -0.15 0.07  & -3.25 0.02  & -3.25 0.02  & S & 0.02 &  -19.28  &  0.71719 & L\\
 0.92  & 74.50 2.87  & A & Y & 1 &  2.7  & Y & $W_H$ & F & 3.3 &         2275 & Z & -0.18 0.07  & -3.26 0.02  & -3.25 0.02  & S & 0.02 &  -19.21  &  0.71273 & A\\
 0.93  & 74.53 2.89  & A & Y & G &  2.7  & Y & $W_H$ & F & 3.3 &         2278 & Z & -0.15 0.07  & -3.25 0.02  & -3.25 0.02  & S & 0.02 &  -19.21  &  0.71273 & A\\
 0.87  & 73.00 2.75  & A & Y & I &  2.7  & Y & $W_H$ & F & 3.3 &         2278 & Z & -0.14 0.07  & -3.31 0.02  & -3.22 0.02  & S & 0.02 &  -19.26  &  0.71273 & A\\
 1.28  & 74.53 3.39  & A & Y & G & No & Y & $W_H$ & F & 3.3 &         2343 & Z & -0.13 0.08  & -3.25 0.03  & -3.24 0.02  & S & 0.02 &  -19.21  &  0.71273 & A\\
 1.09  & 74.69 3.14  & A & Y & G &  3.5  & Y & $W_H$ & F & 3.3 &         2319 & Z & -0.17 0.08  & -3.25 0.02  & -3.25 0.02  & S & 0.02 &  -19.21  &  0.71273 & A\\
 1.09  & 74.66 3.14  & A & Y & 1 &  3.5  & Y & $W_H$ & F & 3.3 &         2318 & Z & -0.17 0.08  & -3.25 0.02  & -3.25 0.02  & S & 0.02 &  -19.21  &  0.71273 & A\\
 0.92  & 73.00 2.82  & A & Y & I &  3.5  & Y & $W_H$ & F & 3.3 &         2318 & Z & -0.12 0.07  & -3.29 0.02  & -3.23 0.02  & S & 0.02 &  -19.26  &  0.71273 & A\\
 1.10  & 73.38 3.10  & A & Y & G &  3.5  & Y & $W_H$ & F & 2.5 &         2320 & Z & -0.17 0.08  & -3.24 0.02  & -3.25 0.02  & S & 0.02 &  -19.25  &  0.71273 & A\\
 0.92  & 77.66 3.00  & A & Y & G &  2.7  & Y & $W_H$ & C & 3.3 &         2277 & Z & -0.18 0.07  & -3.27 0.02  & -3.27 0.02  & S & 0.02 &  -19.12  &  0.71273 & A\\
 0.94  & 71.80 2.81  & A & Y & G &  2.7  & Y & $W_H$ & N & 3.3 &         2277 & Z & -0.14 0.07  & -3.23 0.02  & -3.24 0.02  & S & 0.02 &  -19.29  &  0.71273 & A\\
 0.93  & 74.50 2.88  & A & N & G &  2.7  & Y & $W_H$ & F & 3.3 &         2278 & Z & -0.15 0.07  & -3.25 0.01  & \nd & S & 0.02 &  -19.21  &  0.71273 & A\\
 1.12  & 74.40 3.19  & A & 10 & G &  2.7  & Y & $W_H$ & F & 3.3 &         1316 & Z & -0.16 0.08  & -3.26 0.03  & \nd & S & 0.02 &  -19.22  &  0.71273 & A\\
 0.91  & 74.70 2.86  & A & 60 & G &  2.7  & Y & $W_H$ & F & 3.3 &         2178 & Z & -0.23 0.08  & -3.25 0.01  & \nd & S & 0.02 &  -19.21  &  0.71273 & A\\
 1.06  & 79.49 3.29  & A & Y & G &  2.7  & Y & $H$ & F & 3.3 &         2239 & Z & -0.12 0.07  & -3.06 0.02  & -3.17 0.02  & S & 0.02 &  -19.07  &  0.71273 & A\\
 0.93  & 74.04 2.87  & A & Y & G &  2.7  & Y & $W_H$ & F & 3.3 &         2279 & Z & \nd & -3.26 0.02  & -3.25 0.02  & S & 0.02 &  -19.23  &  0.71273 & A\\
 0.93  & 74.66 2.89  & A & Y & G &  2.7  & Y & $W_H$ & F & 3.3 &         2278 & Z & -0.15 0.07  & -3.25 0.02  & -3.25 0.02  & S & 0.01 &  -19.21  &  0.71347 & A\\
 0.93  & 75.97 3.00  & A & Y & G &  2.7  & Y & $W_H$ & F & 3.3 &         2278 & Z & -0.10 0.07  & -3.26 0.02  & -3.25 0.02  & M & 0.02 &  -19.12  &  0.70326 & A\\
 0.93  & 75.98 2.94  & A & Y & G &  2.7  & N & $W_H$ & F & 3.3 &         2413 & Z & -0.19 0.07  & -3.21 0.02  & -3.26 0.02  & S & 0.02 &  -19.17  &  0.71273 & A\\
 0.93  & 76.11 3.09  & A & Y & G &  2.7  & Y & $W_H$ & F & 3.3 &         2278 & B & -0.22 0.10  & -3.25 0.02  & -3.25 0.02  & S & 0.02 &  -19.17  &  0.71273 & A\\
 0.93  & 74.65 2.93  & A & Y & G &  2.7  & Y & $W_H$ & F & 3.3 &         2278 & Z & -0.15 0.07  & -3.25 0.02  & -3.25 0.02  & S & 0.02 &  -19.21  &  0.71340 & S\\
 0.93  & 75.46 3.02  & A & Y & G &  2.7  & Y & $W_H$ & F & 3.3 &         2278 & Z & -0.10 0.07  & -3.26 0.02  & -3.25 0.02  & M & 0.02 &  -19.12  &  0.70031 & S\\
 0.93  & 74.83 2.97  & A & Y & G &  2.7  & Y & $W_H$ & F & 3.3 &         2278 & Z & -0.15 0.07  & -3.25 0.02  & -3.25 0.02  & S & 0.01 &  -19.21  &  0.71444 & L\\
 0.93  & 75.30 3.01  & A & Y & G &  2.7  & Y & $W_H$ & F & 3.3 &         2278 & Z & -0.15 0.07  & -3.25 0.02  & -3.25 0.02  & S & 0.02 &  -19.21  &  0.71719 & L\\
 \hline
  0.91  & 71.96 1.47  & All & Y & 1 &  2.7  & Y & $W_I$ & F & 3.3 &         3138 & Z & -0.20 0.05  & -3.17 0.02  & -3.40 0.02  & S & 0.02 &  -19.29  &  0.71273 & A\\
  0.91  & 71.74 1.54  & NML & Y & 1 &  2.7  & Y & $W_I$ & F & 3.3 &         3137 & Z & -0.20 0.05  & -3.17 0.02  & -3.40 0.02  & S & 0.02 &  -19.30  &  0.71273 & A\\
  1.09  & 72.41 2.26  & N & Y & 1 &  2.7  & Y & $W_I$ & F & 3.3 &         2364 & Z & -0.19 0.05  & -3.08 0.03  & -4.14 0.05  & S & 0.02 &  -19.28  &  0.71273 & A\\
\enddata
\tablecomments{$H_0$: error listed from fit for $M_X^0$ in eq 9 or $m_{x,4258}^0$ in eq 4 only.  Anc: Anchors used; N=N4258 Masers, M=MW Parallaxes, L=LMC DEBs, NML=primary fit using all three, A=NML+M31 DEBs. Brk: Break in \PL relation; Y=two-slope solution, N=single-slope solution, 10=single slope restricted to $P\!>\!10$~d, 60=single slope restrcited to $P\!<\!60$~d. Clp: Clipping procedure; G=global, I=individual, 1=Global but removing single largest outlier at a time. $\sigma$: clipping threshold. Opt: optical completeness required, Y=Yes, N=No. PL: Form of \PL relation used; $W_H$: NIR Wesenheit; $H$ NIR without extinction correction; $W_I$: Optical Wesenheit. R: reddening law; F99=\citet{Fitzpatrick:1999}, CCM=\citet{Cardelli:1989}, N=\citet{Nataf:2016}. $R_V$: Extinction-law parameter. N: Number of Cepheids fit. Z: Metallicity scale; Z=traditional $R_{23}$ method \citep{zaritsky94}, B=$T_e$ method \citep{Bresolin:2011}. $\gamma$: change in Wesenheit mag per dex in $\log (O/H)$. b: slope of \PL\ for all $P$ in no-break variants or for $P\!>\!10$~d for two-slope variants. bl: slope of \PL\ for $P\!<\!10$~d (when applicable). SN: light-curve fitter; S=SALT, M=MLCS2k2. $z_m$: minimum $z$ used in SN Hubble diagram (0.02 stands for 0.0233). $M_V^0$: SN absolute magnitude ($X$ stands for $B$ or $V$ depending on the SN fitter, see text). $a_X$: intercept of SN Hubble diagram ($X$=$B$ or $V$). Gal: SN host galaxy sample; A=All, S=Spiral, L=high LSF.}
\end{deluxetable}

\clearpage
\bibliographystyle{apj} %
\bibliography{bibdesk}
\clearpage
\end{document}